\documentclass{aa}

\usepackage[T1]{fontenc} 
\usepackage{siunitx}
\usepackage{amsmath}
\usepackage{amssymb}
\usepackage{graphicx}
\usepackage{longtable}   
\usepackage[para]{footmisc}
\usepackage{lscape}
\usepackage[export]{adjustbox}
\usepackage{subcaption}
\usepackage{perpage}
\usepackage{alphalph}
\usepackage{hyperref}
\usepackage[capitalise]{cleveref} 
\usepackage{float}

\newcommand{\kms}{km.s$^{\rm -1}$}

\newcommand{\Mjup}{M$_{\rm Jup}$}

\newcommand{\vsini}{$v\sin{i}$}
\newcommand{\Msun}{M$_{\sun}$}

\newcommand{\Mstar}{M$_{\rm \star}$}
\newcommand{\bv}{$B-V$}
\newcommand{\msini}{$m_{\rm p}\sin{i}$}
\newcommand{\rhk}{log$R'_{\rm HK}$}
\newcommand{\safir}{S{\small AFIR}}
\newcommand{\sophie}{S{\small OPHIE}}
\newcommand{\harps}{H{\small ARPS}}
\newcommand{\hipp}{H{\small IPPARCOS}}

\newcommand{\sphere}{S{\small PHERE}}

\DeclareSIUnit\year{yr}
\DeclareSIUnit\arcsecond{as}
\DeclareSIUnit\MJ{M_{Jup}}
\DeclareSIUnit\au{au}
\DeclareSIUnit\jdb{d}
\DeclareSIUnit\day{days}
\DeclareSIUnit\hour{h}
\DeclareSIUnit\night{nights}
\DeclareSIUnit\msun{M_{\odot}}
\makeatletter
\newcommand\footnoteref[1]{\protected@xdef\@thefnmark{\ref{#1}}\@footnotemark}
\makeatother

\begin{document} 
   \title{A HARPS RV search for planets around young nearby stars} 

   \author{A. Grandjean
          \inst{1}
          \and
          A.-M. Lagrange\inst{1}
	\and
         M. Keppler\inst{2}
	\and
	N. Meunier\inst{1}
	\and
	L. Mignon\inst{1}
	\and
	S. Borgniet\inst{3}
	\and
	G. Chauvin\inst{1}
	\and
	S. Desidera \inst{4}
	\and
	F. Galland \inst{1}
	\and
	S. Messina \inst{5}
	\and
	M. Sterzik \inst{6}
	\and
	B. Pantoja \inst{7}
	\and
	L. Rodet\inst{1}
	\and
	N. Zicher  \inst{8}
          }

   \institute{
Univ. Grenoble Alpes, CNRS, IPAG, 38000 Grenoble, France
 \\
\email{Antoine.Grandjean1@univ-grenoble-alpes.fr}
\and 
Max Planck Institute for Astronomy, Königstuhl 17, D-69117 Heidelberg, Germany
\and
CNRS Lesia (UMR8109) - Observatoire de Paris, Paris, France
\and
INAF-Osservatorio Astronomico di Padova, Vicolo dell’Osservatorio 5, Padova, Italy, 35122-I
\and
INAF–Osservatorio Astrofisico di Catania, via Santa Sofia, 78 Catania, Italy 
\and
European Southern Observatory (ESO), Alonso de C\'ordova 3107, Vitacura, Casilla 19001, Santiago, Chile
\and
Departamento de Astronomía, Universidad de Chile, Camino al Observatorio, Cerro Calán, Santiago, Chile
\and
Oxford Astrophysics, Department of Physics, Denys Wilkinson Building,UK
             }

   \date{Received 2019 June 11/ Accepted 2019 October 14}

 
  \abstract
   {Young nearby stars are good candidates in the search for planets with both radial velocity (RV) and direct imaging techniques. 
This, in turn, allows for the computation of the giant planet occurrence rates at all separations. 
The RV search around young stars is a challenge as they are generally faster rotators than older stars of similar spectral types and they exhibit signatures of  magnetic activity (spots) or pulsation in their RV time series.
Specific analyses are necessary to characterize, and possibly correct for, this activity.
 }
   {Our aim is to search for planets around young nearby stars and to estimate the giant planet (GP) occurrence rates for periods up to \SI{1000}{\day}.}
   {We used the \harps \ spectrograph on the \SI{3.6}{\meter} telescope at La Silla Observatory to observe $89 \ A-M$ young ($< \SI{600}{\mega\year}$) stars. 
We used our \safir \ (Spectroscopic data via Analysis of the Fourier Interspectrum Radial velocities ) software  to compute the RV and other spectroscopic observables. Then, we computed the companion occurrence rates on this sample.}
   {We confirm the binary nature of HD177171, HD181321 and HD186704.
 We report the detection of a close low mass stellar companion for HIP36985. 
No planetary companion was detected. We obtain upper limits on the GP ($< \SI{13}{\MJ}$) and BD ($\in [13;80] \ \si{\MJ}$) occurrence rates based on $83$ young stars for periods less than \SI{1000}{\day}, which are set, $2_{-2}^{+3} \ \%$ and $1_{-1}^{+3}\ \%$. }
   {}

       \keywords{ Techniques: radial velocities -- stars: activity - (stars:) binaries: spectroscopic -- stars: planetary systems -- stars: starspots -- stars: variables: general
               }

   \maketitle
%

\section{Introduction}

More than four thousand exoplanets have been confirmed and most of them have been found by transit or radial velocity (RV) techniques\footnote{\url{exoplanet.eu}}.
The latter, although very powerful, is limited by the parent star's activity  (spots, plages, convection, and pulsations).
Young stars are generally faster rotators than their older counterparts \citep{Stauffer_2016, Rebull_2016, Gallet_2015}. 
They can also exhibit activity-induced RV jitter with amplitudes up to $\SI{1}{\kilo\meter\per\second}$ \citep{Lagrange}, larger than the planet's induced signal.
False positive detections have been reported around young stars in the past \citep{Huelamo,Figueira,Soto}.

We have carried out an RV survey to search for planets around such young stars with the  High Accuracy Radial velocity Planet Searcher  (\harps) \citep{HARPS} and \sophie \ \citep{SOPHIE} spectrographs with the final aim of coupling RV data with direct imaging (DI) data, which will allow for the computation of detection limits,  for each targets at all separations and then to compute occurrences rates for all separations.
A feasibility study was carried out by \cite{Lagrange} on 26 stars of the survey with \harps, demonstrating that we can probe for giant planets ($1$ to $\SI{13}{\MJ}$, hereafter GP) with semi-major axis up to \SI{2}{\au} and couple the survey data with direct imaging data.
The time baseline of our survey also permits a probe of the hot Jupiter (Hereafter HJ) domain around young stars.
Although GP formation models predict a formation at a few $\si{\au}$ \citep{Pollack}, migration through disc-planet interaction \citep{Migration-disk} or gravitational interaction with a third body can allow the planet to finally orbit close to the star \citep{High_excentricity_migration}.
HJ are common among exoplanets orbiting older main sequence stars as they represent one  detected planet out of five \citep{Wright_2012}.
While previous RV surveys on small sets of young stars showed no evidence for the presence of young HJ \citep{Esposito,Paulson}, two HJ around young stars were recently discovered by \cite{Donati} and \cite{Yu}. 
However, we still need to find out if this kind of object is common at young age or not and we need to compare the migration models with observations in order to constrain the migration timescales.

Here, we report on the results of our large \harps \ survey. 
We describe our survey sample and observations in  \cref{survey_description}. 
In \cref{comp}, we focus on GP, brown dwarf (BD), and stellar companions detections. 
In \cref{occ_rate}, we perform a statistical analysis of our sample and compute the close GP and BD occurrence rates around young stars.
We conclude our findings in  \cref{conc}.

\section{Description of the survey}
\label{survey_description}

\subsection{Sample}

We selected a sample of $89$ stars, chosen to their brightness ($V < 10$), ages as fund in the literature ($< \SI{300}{\mega\year}$ for most of them, see   \cref{tab_carac}), and distance ($< \SI{80}{pc}$) determined from their \hipp \ parallaxes \citep{hipp2}. 
These criteria ensure we will get the best detection limits for both the \harps \ RV and SPHERE DI surveys at, respectively, short (typically $2-5 \ \si{\au}$), and large (further than typically $\SI{5}{\au}$) separations.
 Indeed, observing bright stars allow us to obtain spectra with a better signal to noise ratio (S/N).
Moreover, young planets are still warm from their formation and are then brighter, which lowers the contrast between them and their host stars, while short distances allow for better angular resolutions in direct imaging.
 Most of the targets are part of the \sphere \ GTO The SpHere INfrared survey for Exoplanets (SHINE) survey sample \citep{Chauvin_shine}. 
Binary stars with an angular separation on the sky that is lower than $\SI{2}{\arcsecond}$  were not selected to avoid contamination in the spectra from the companion

\begin{figure*}[ht!]
  \centering
\begin{subfigure}[t]{0.24\textwidth}
\includegraphics[width=1\hsize]{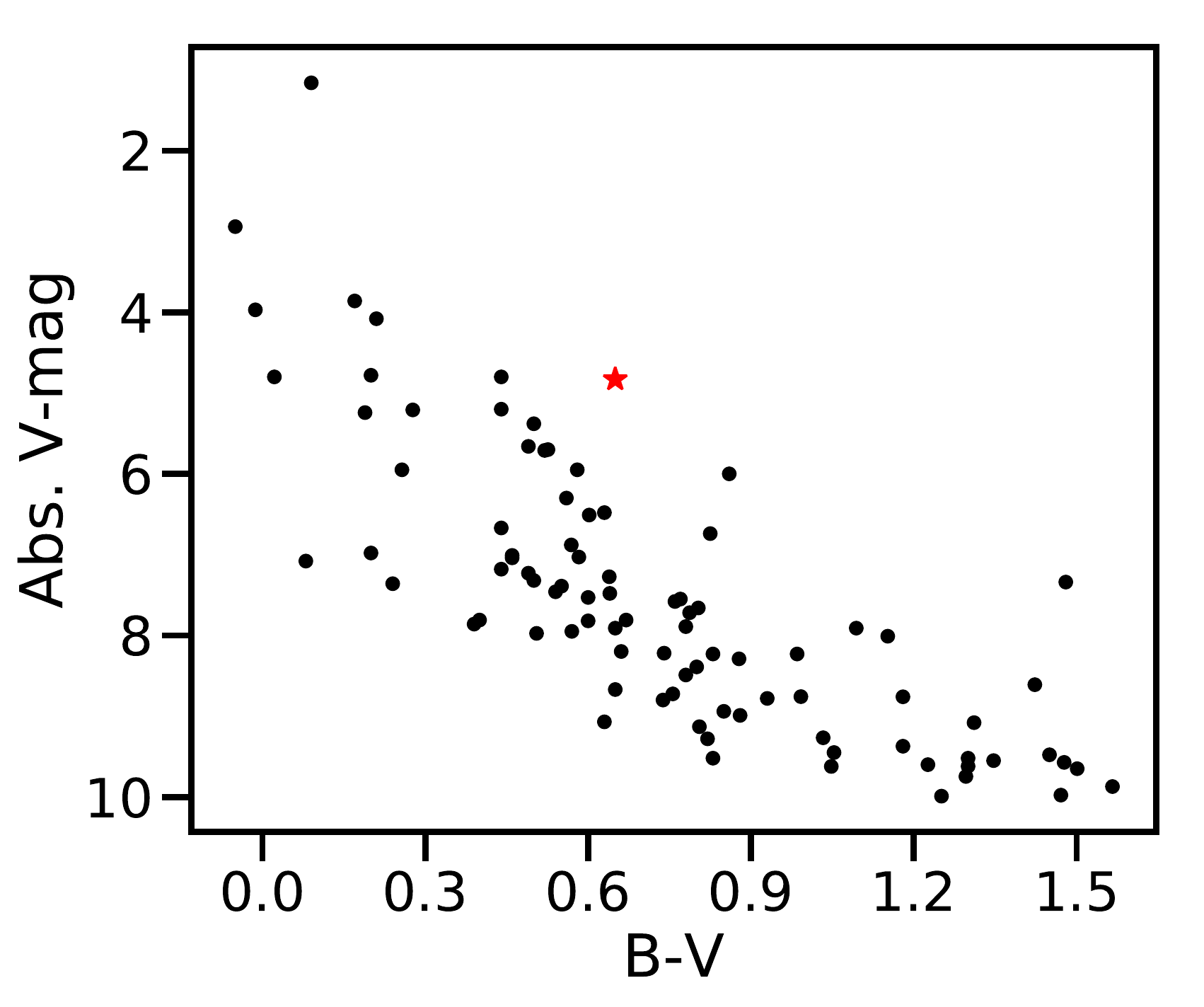}
\caption{\label{HR}}
\end{subfigure}
\begin{subfigure}[t]{0.24\textwidth}
\includegraphics[width=1\hsize]{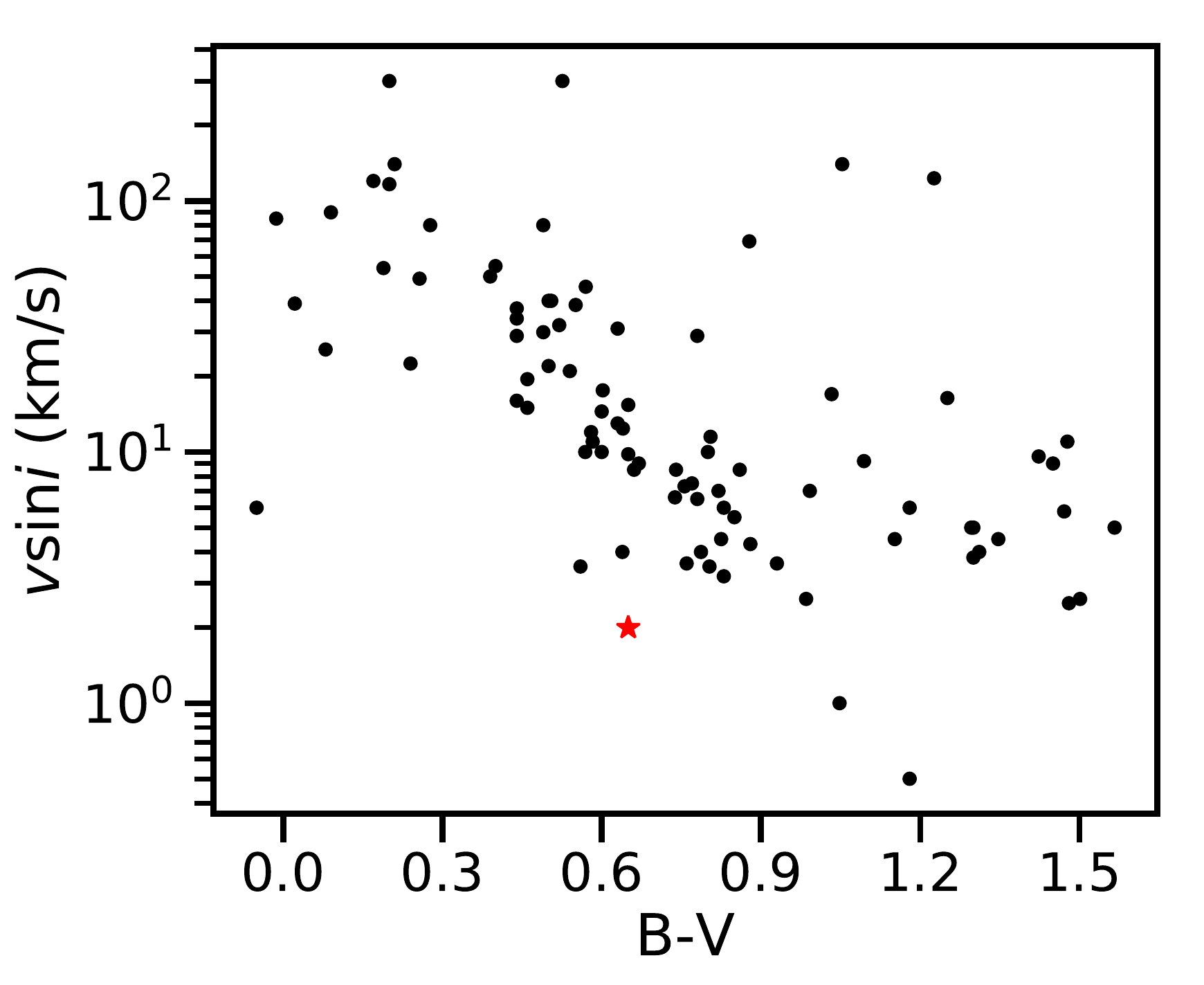}
\caption{\label{vsini}}
\end{subfigure}
\begin{subfigure}[t]{0.24\textwidth}
\includegraphics[width=1\hsize]{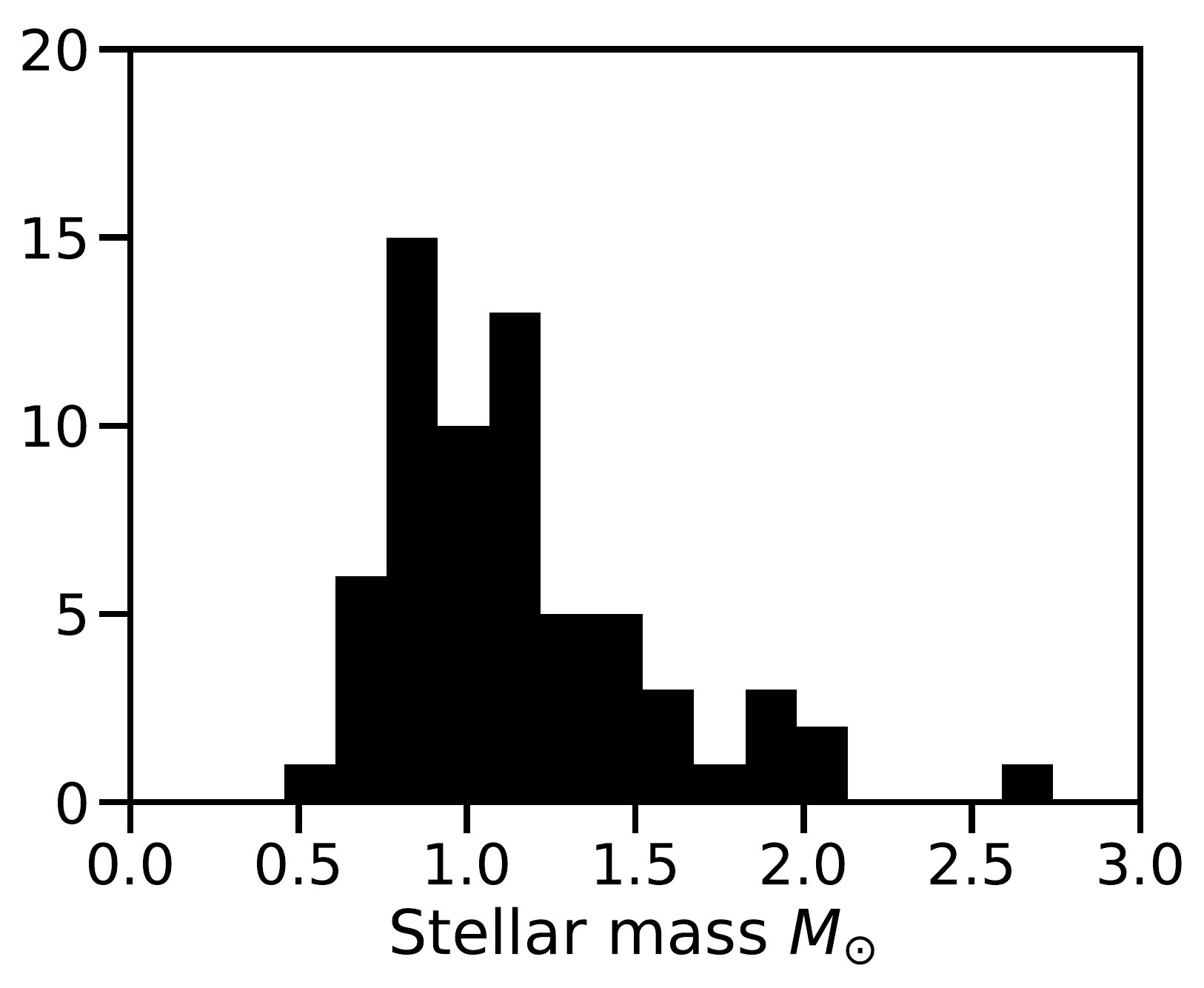}
\caption{\label{mass}}
\end{subfigure}
\begin{subfigure}[t]{0.24\textwidth}
\includegraphics[width=1\hsize]{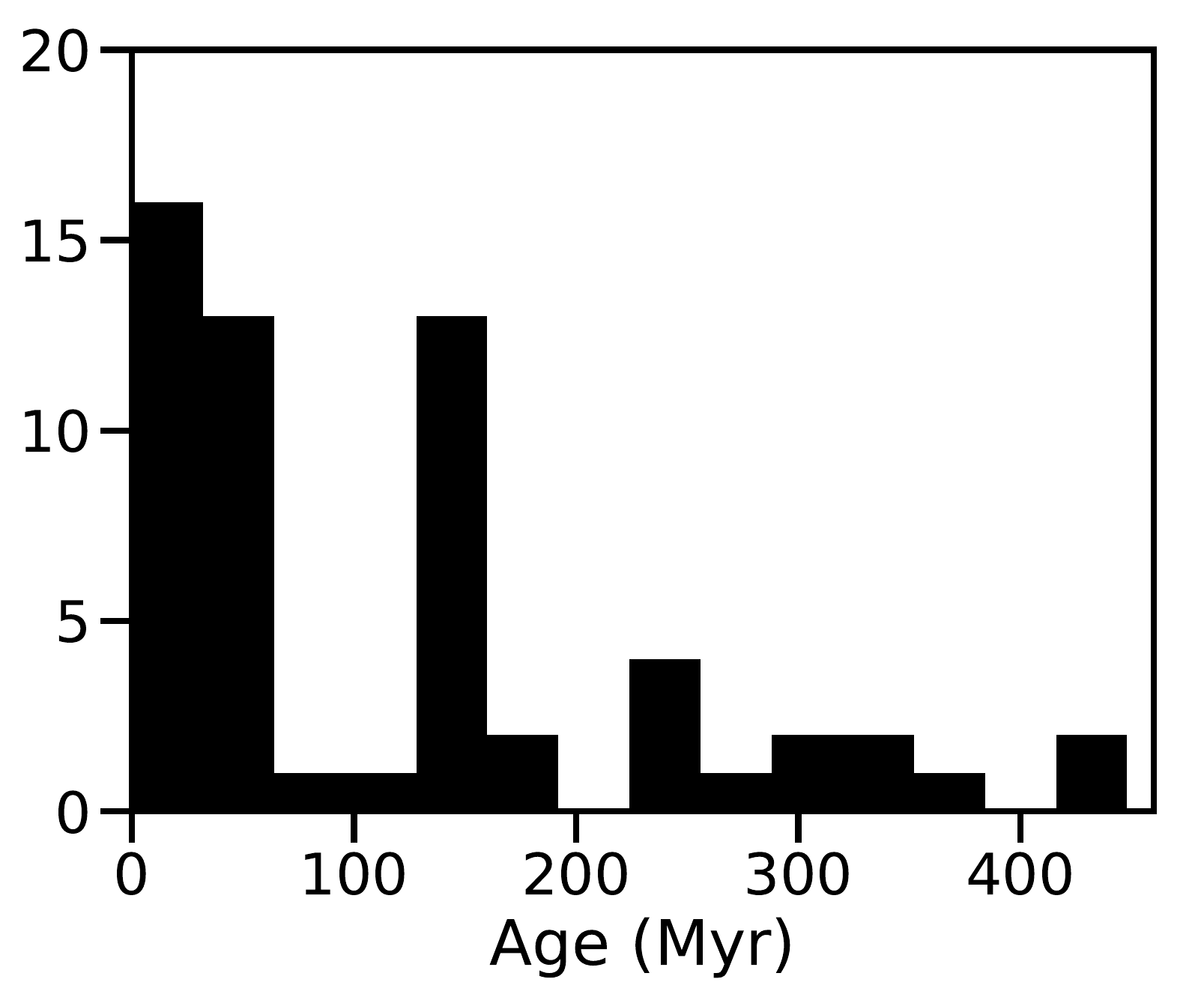}
\caption{\label{age}}
\end{subfigure}
\caption{Main physical properties of our sample.
 \subref{HR})  Absolute $V$-magnitude vs \bv. Each black dot corresponds to one target.
 The Sun is displayed (red star) for comparison. 
\subref{vsini}) \vsini~vs \bv~distribution.
\subref{mass})  Mass histogram (in \Msun).
\subref{age}) Age histogram.}
       \label{survey_carac_1}
\end{figure*}

Their spectral types range from A0V to M5V (\Cref{survey_carac_1}).
 Their projected rotational velocity  (\vsini) range from $0.5$ to $\SI{300}{\kilo\meter\per\second}$ with a median of  $\SI{11}{\kilo\meter\per\second}$,
 their V-band magnitude are mainly between $4$ and $10$ with a median of $7.8$.
 Their masses are between $0.6$ and $\SI{2.74}{\msun}$ with median of  $\SI{1.07}{\msun}$. 
Our sample includes $26$ targets between A0 and F5V ($B-V \in [-0.05:0.52[$), $55$ between F6 and K5   ($B-V \in [-0.05:0.52[$), and $8$ between K6 and M5 ($B-V \geq 1.33$).
Noticeably, our sample includes stars with imaged planetary or substellar companions (among them $\beta$ Pic, AB Pic, HN Peg, GJ 504, HR8799, HD95086, HD106906 or PZ Tel).
We present the main characteristics  of our star sample in \Cref{survey_carac_1} and \cref{tab_carac}.

\begin{figure*}[ht!]
  \centering
\begin{subfigure}[t]{0.32\textwidth}
\includegraphics[width=1\hsize]{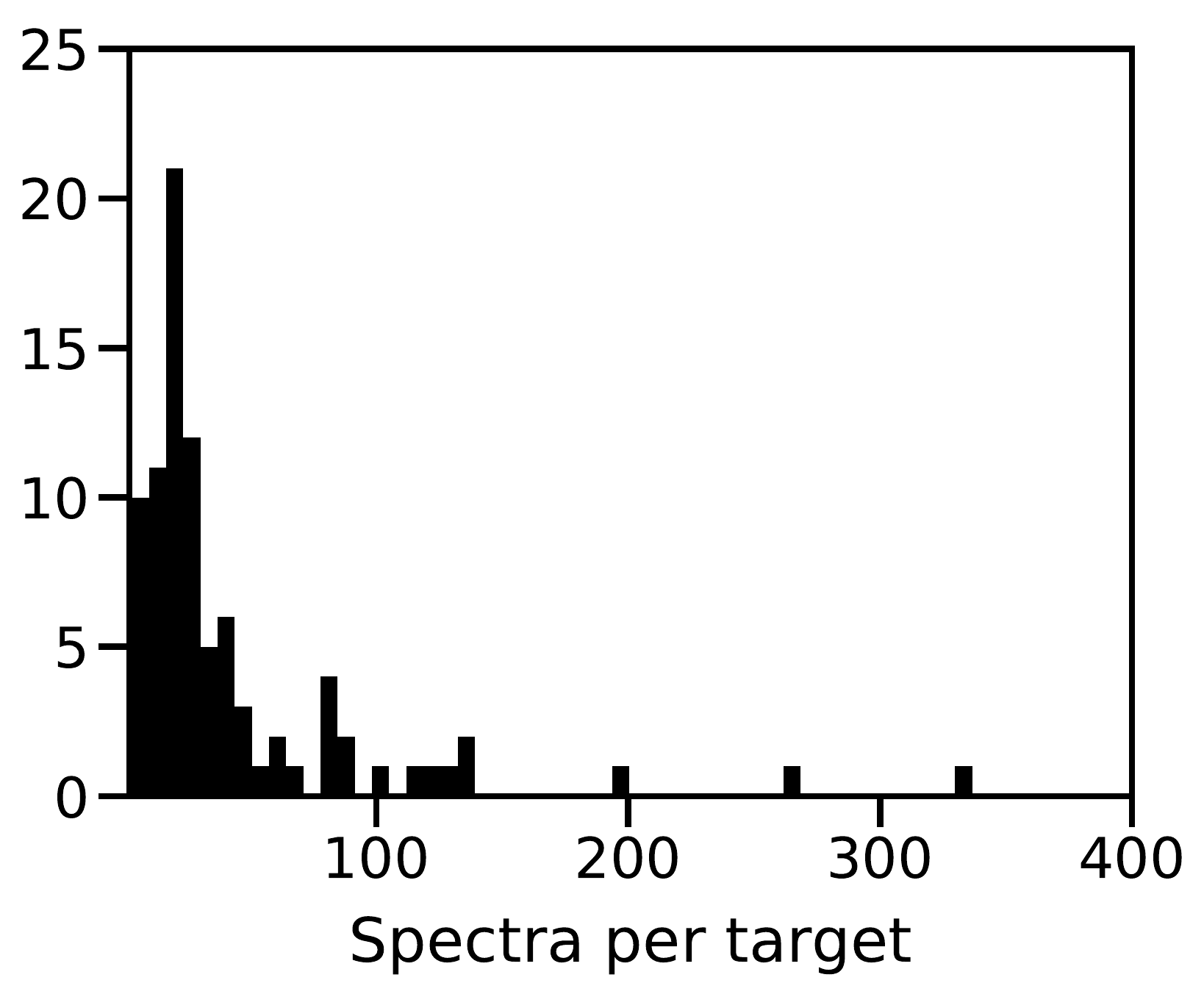}
\caption{\label{Nmes}}
\end{subfigure}
\begin{subfigure}[t]{0.32\textwidth}
\includegraphics[width=1\hsize]{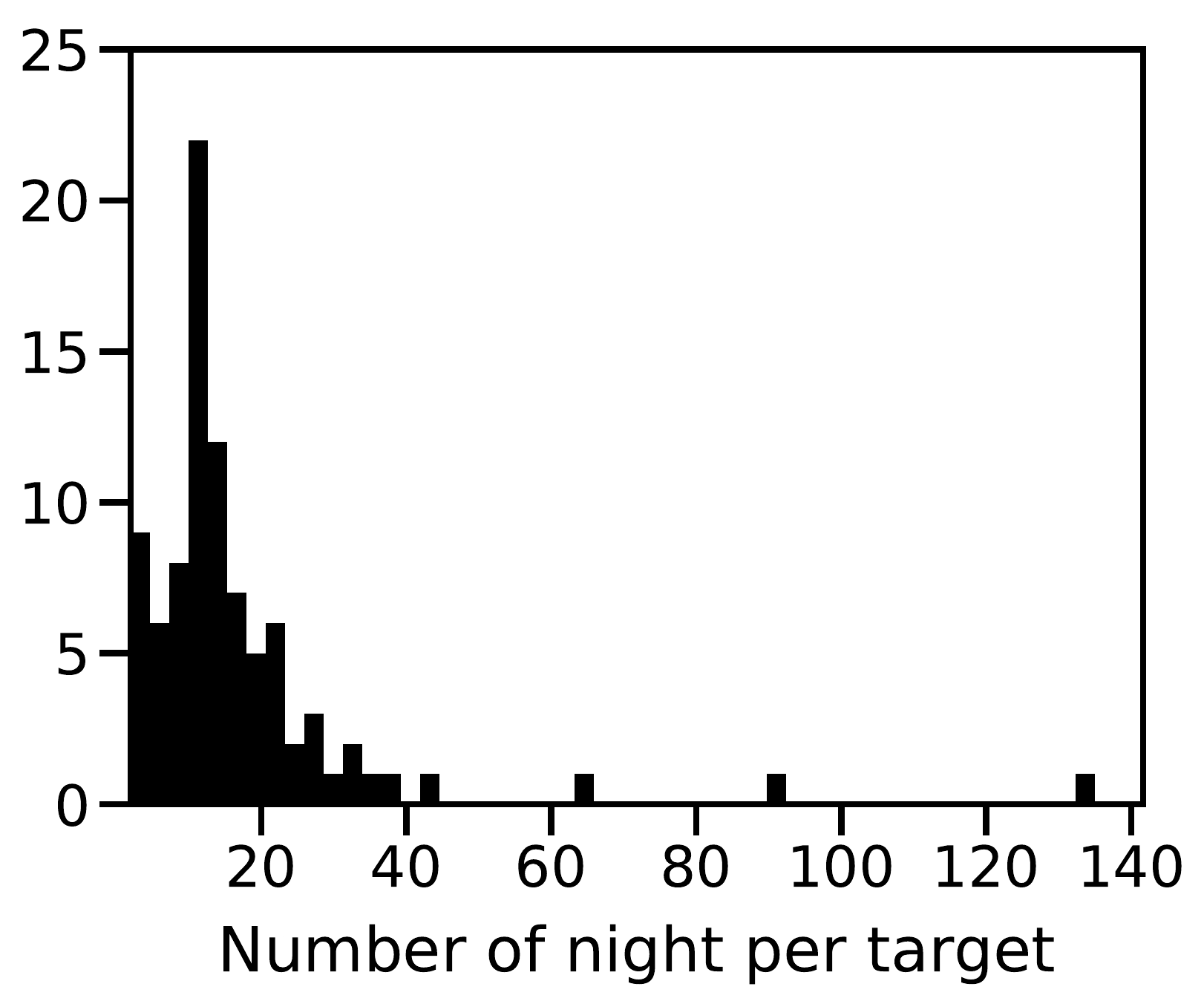}
\caption{\label{Nb_day}}
\end{subfigure}
\begin{subfigure}[t]{0.32\textwidth}
\includegraphics[width=1\hsize]{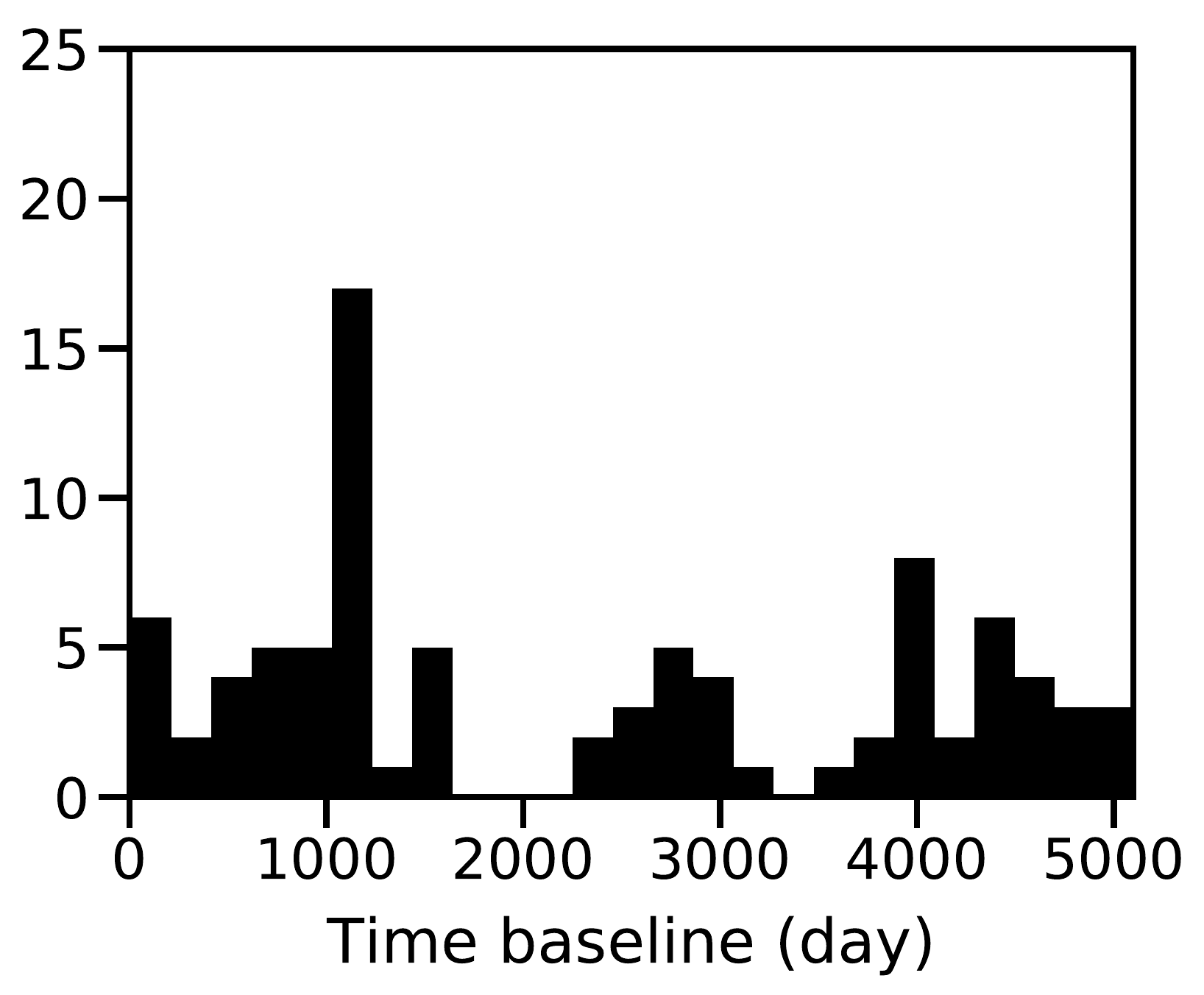}
\caption{\label{time_bsl}}
\end{subfigure}
\caption{Observation summary.
 \subref{Nmes}) Histogram of the number of spectra per target, excluding HD216956 (Fomalhaut, 834 spectra) and HD039060 ($\beta$ Pic, 5108 spectra).
 \subref{Nb_day}) Histogram of the  number of nights per target. 
\subref{time_bsl}) Histogram of the time baselines.} 
       \label{survey_carac_2}
\end{figure*}

\subsection{Observations}

We observed our $89$ targets mainly between 2013 and 2016. 
Some stars were part of previous surveys by \cite{Simon_IX} and \cite{Lagrange_2009}, which allows us to reach a time baseline up to $\SI{10}{\year}$.
Some stars had already been observed with \harps \  before, some since the \harps \ commissioning in 2003. 
Additional observations were also obtained in October  2017, December 2017, and March 2018. 

We use the observing strategy described in \cite{Simon_IX}, which consist of recording two spectra per visit and to observe each target on several consecutive nights to have a good sampling of the short-term jitter.
The median time baseline is \SI{1639}{\day} (mean time baseline of  \SI{2324}{\day}), with a  median number of spectra per target of $25$ ($52$ on average) spaced on a median number of $\SI{12}{\night}$  ($17$ on average, \Cref{survey_carac_2}).
Details can be found in \cref{tab_carac}.

\subsection{Observables}

\label{safir}

From the \harps \  spectra we derived the RV and whenever possible the cross-correlation function (CCF), bisector velocity span (hereafter BVS), and \rhk \ with our \safir \ software for Spectroscopic data via Analysis of the Fourier Interspectrum Radial velocities.
It builds a reference spectrum from the median of all spectra available on a given star and computes the relative RV in the Fourier plane.
The computed RV are then relative to the  reference spectrum. 
 The efficiency of this method was proved in the search of companions \citep{SAFIR,Galland_GP}.
We mainly use the correlation between RV and BVS  to determine the main source of RV variability : magnetic activity, pulsations or  companions \citep{Lagrange_2009,Simon_IX}. 
We excluded spectra with S/N at \SI{550}{\nano\meter} which was too high ($> 380$), to avoid saturation or too low ($<80$), to avoid bad data, as well as spectra with an air mass that was too high ($sec \ z > 3$), and spectra that were too different from the reference spectrum of the star ($\chi^2 > 10$).
For M-type stars, we used a lower limit of $40$ in S/N at  \SI{550}{\nano\meter} as it is a better compromise for these stars to provide enough spectra to perform our analysis without including bad spectra.

\section{Detected companions in the \harps \ survey}
\label{comp} 

\subsection{Long period companions, RV long-term trends, and stellar binaries}

In this section we describe the stars for which we identified a GP companion with a period higher than $\SI{1000}{\day}$, a long-term trend RV signal or a binary signal.
When possible, we characterize the companion using the \emph{yorbit} software that fits RV with a sum of keplerian or a sum of keplerians plus a drift  \citep{Segransan}.

\subsubsection{HD39060}

$\beta$ Pic is an A6V pulsating star that hosts an imaged edge-on debris \citep{Beta_pic_disc} and gas disk \citep{Beta_pic_gas_disc}, and an imaged GP at $\SI{9}{\au}$ \citep{Beta_pic_b,Lagrange_2018}.
This star presents also exocomets signatures in its spectra \citep{Beta_pic_FEB_1988,Beust_1989,Kiefer_comet}.
$\beta$ Pic was observed with \harps \  since its  commissioning in 2003, totalizing more than  $6000$ spectra with a mean S/N at $\SI{550}{\nano\meter}$ of $273$.
Until 2008, spectra were taken to study the Ca II absorption lines associated to the falling exocomets. 
Since 2008, we adopted a specific observation strategy, to properly sample the stellar pulsations and to correct the RV from them, \citep{Lagrange_2012,Beta_pic_c}.
It consisted in observing the star for continuous sequences of $1- \SI{2}{\hour}$. 
Longer sequences up to $\SI{6}{\hour}$ were obtained in $2017-2018$.
This allowed to detect a new GP companion within the pulsations signal of $\beta$ Pic. The discovery of this \SI{10}{\MJ}, \SI{1200}{\day} period companion is detailed in \cite{Beta_pic_c}. 

\subsubsection{HD106906}
HD106906 is a F5V star in the Sco-Cen young association.  \cite{Bailey} Imaged a giant planet companion at $\SI{7.1}{\arcsecond}$ ($\SI{650}{\au}$) in 2014.
\harps \ data from this survey together with the PIONER interferometer data allowed to detect a close low mass stellar companion to HD106906 with a period of \SI{10}{\day}, \citep{Lagrange_106b}. 
The presence of the binary could explain the wide orbit of HD106906b under some circumstances \citep{Rodet_106}.

\subsubsection{HD131399}

HD131399 is member of a complex hierarchical system. HD131399A forms a binary with the tight binary HD131399BC.
A GP companion was discovered with \sphere \ around HD131399A by \cite{Wagner_16} but is now identified as a background star with similar proper motion \citep{Nielsen_17}.
We detected in this survey the presence of a close stellar companion to HD131399A, with a period of $\SI{10}{\day}$ ($\SI{0.1}{\au}$), and an $M\sin{i}$ of $\SI{450}{\MJ}$ \citep{Lagrange_131}.

\begin{figure}[h!]
  \centering
\includegraphics[width=0.46\hsize,valign=t]{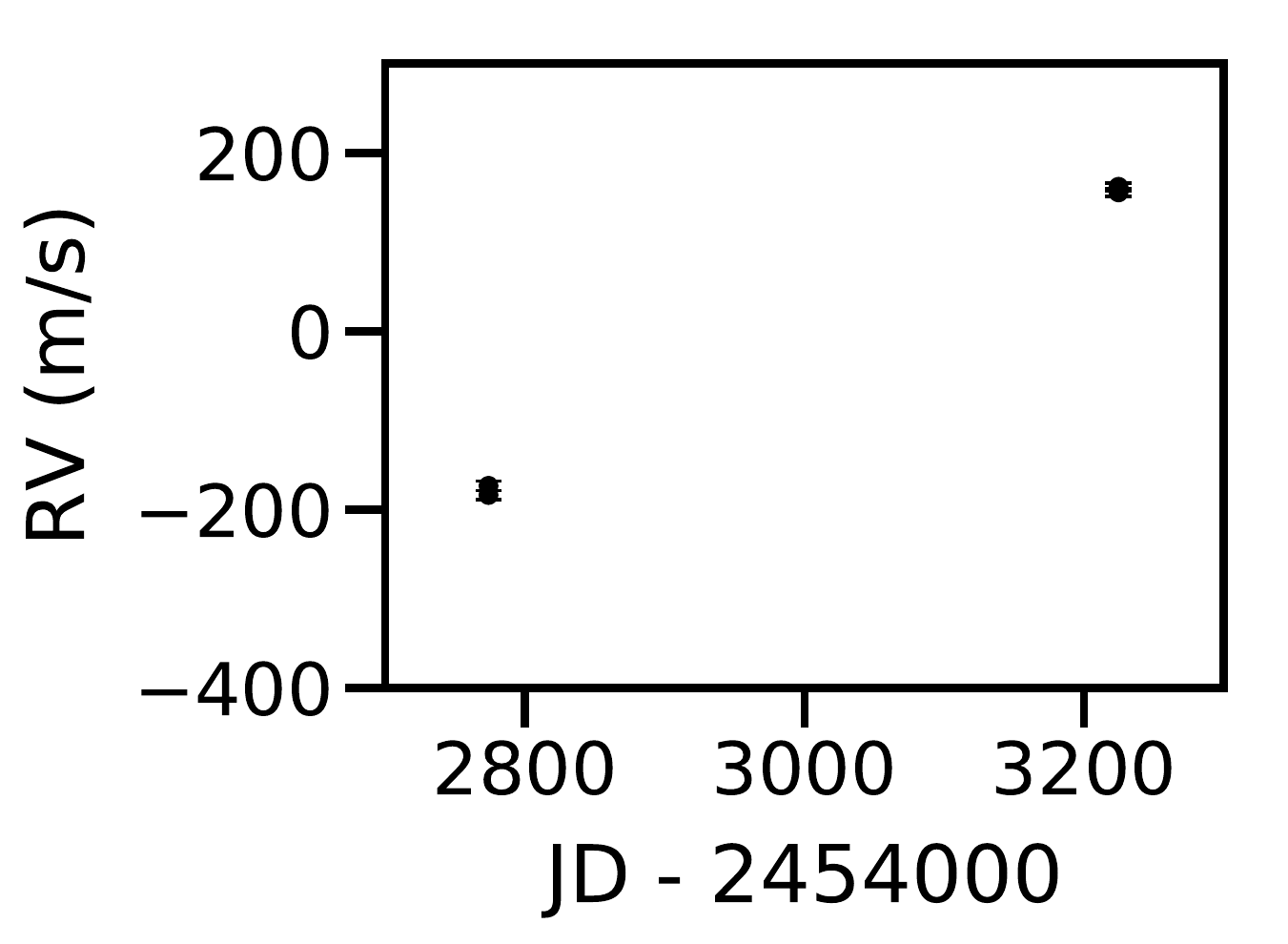}
\includegraphics[width=0.46\hsize,valign=t]{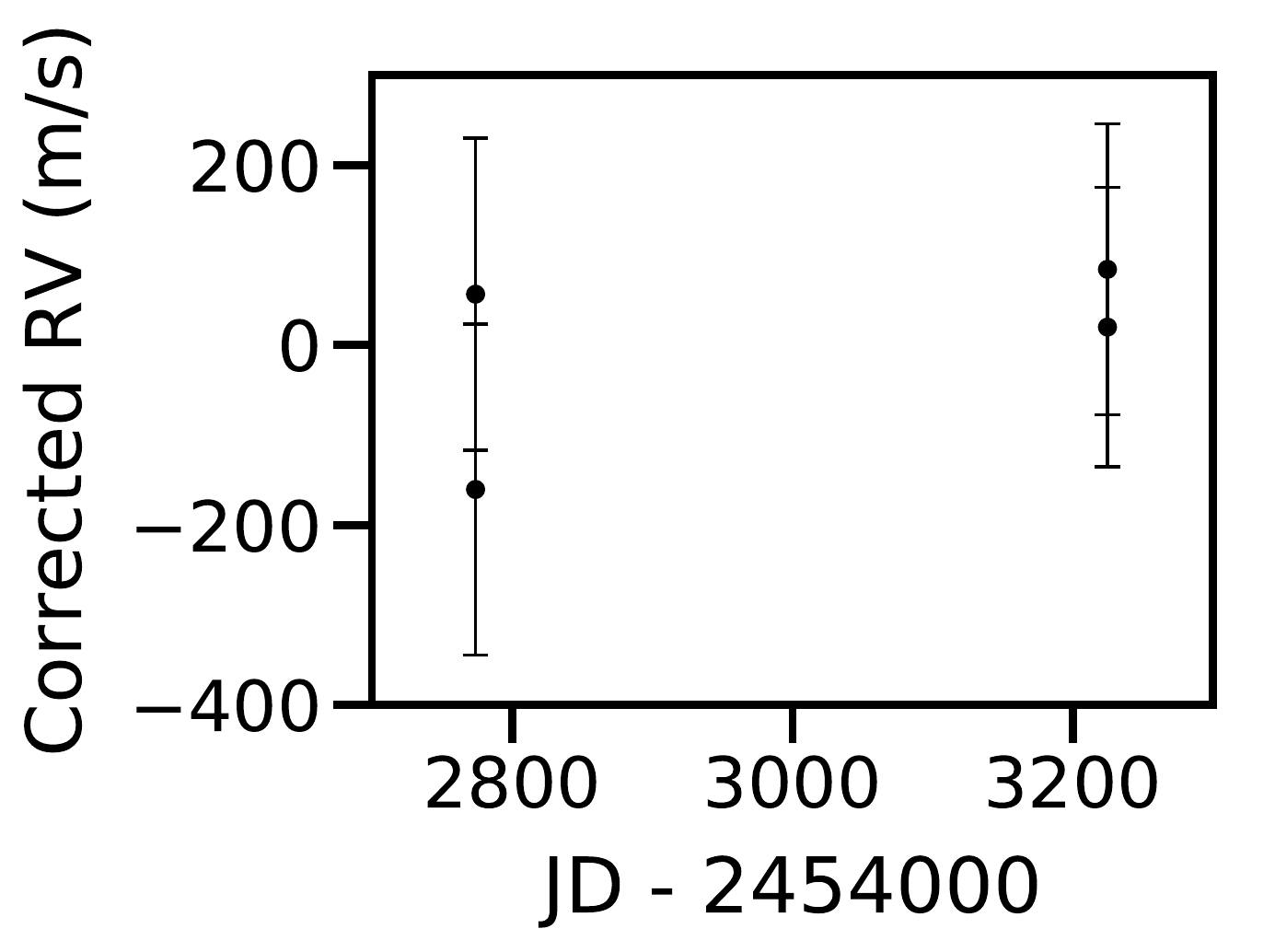}
\caption{HD186704 RV long term trend.   ({\it Left}) RV time variations,  ({\it ight}) RV corrected from the RV-BVS correlation time variations.} 
       \label{HD186704}
\end{figure}

\subsubsection{HD177171}

HD 177171 ($\rho$ Tel) was reported as a spectroscopic binary by \cite{Nordstrom_181} and as an astrometric binary by \cite{Frankowski} from the \hipp \ data. This is confirmed by \cite{Lagrange} in the feasibility study of this survey.
We measure an amplitude of at least \SI{20}{\kilo\meter\per\second} in the RV. Our time sampling does not allow for the estimation the period of this stellar companion.

\begin{figure*}[h!]
  \centering
\includegraphics[width=0.45\hsize,valign=t]{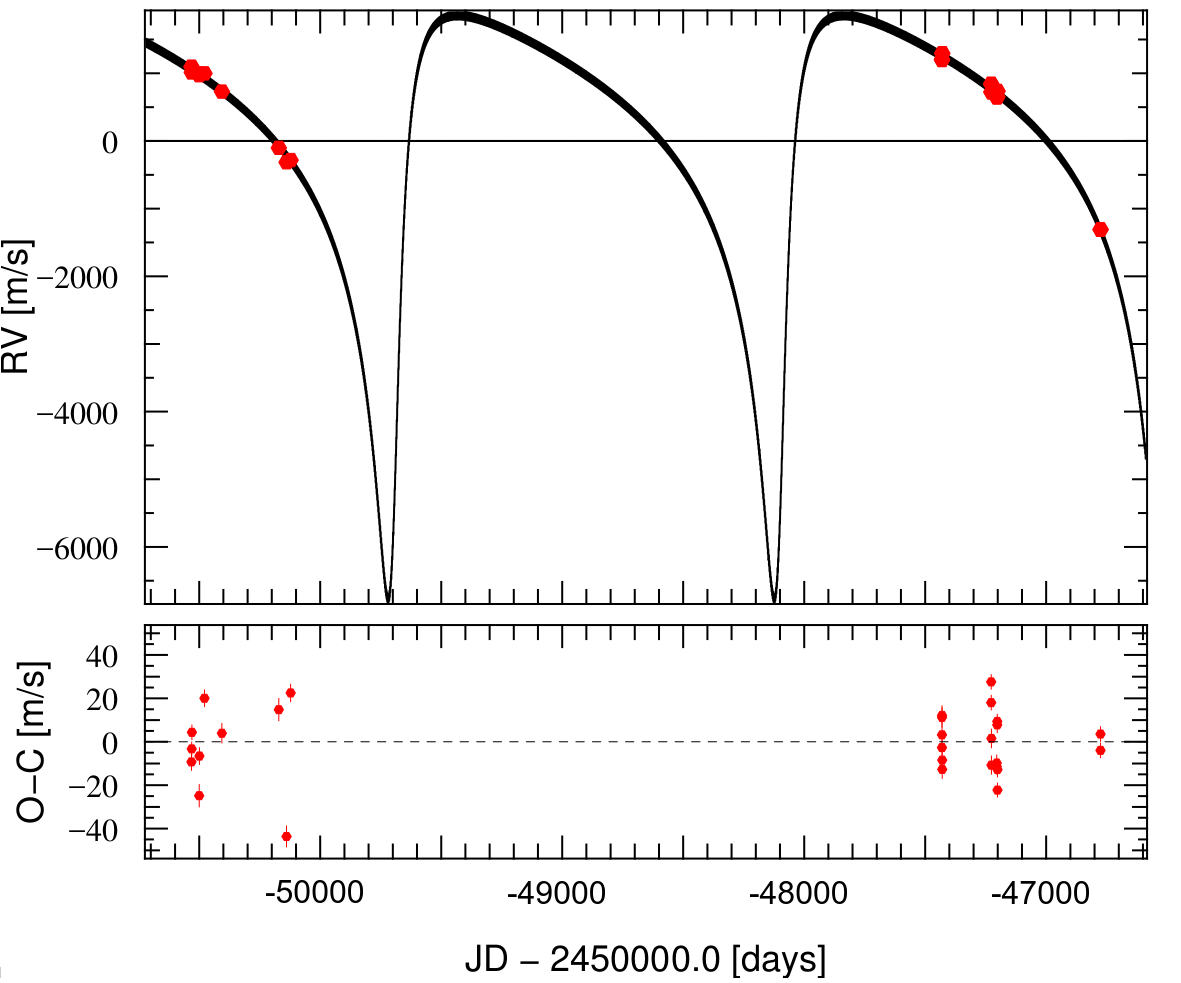}
\includegraphics[width=0.45\hsize,valign=t]{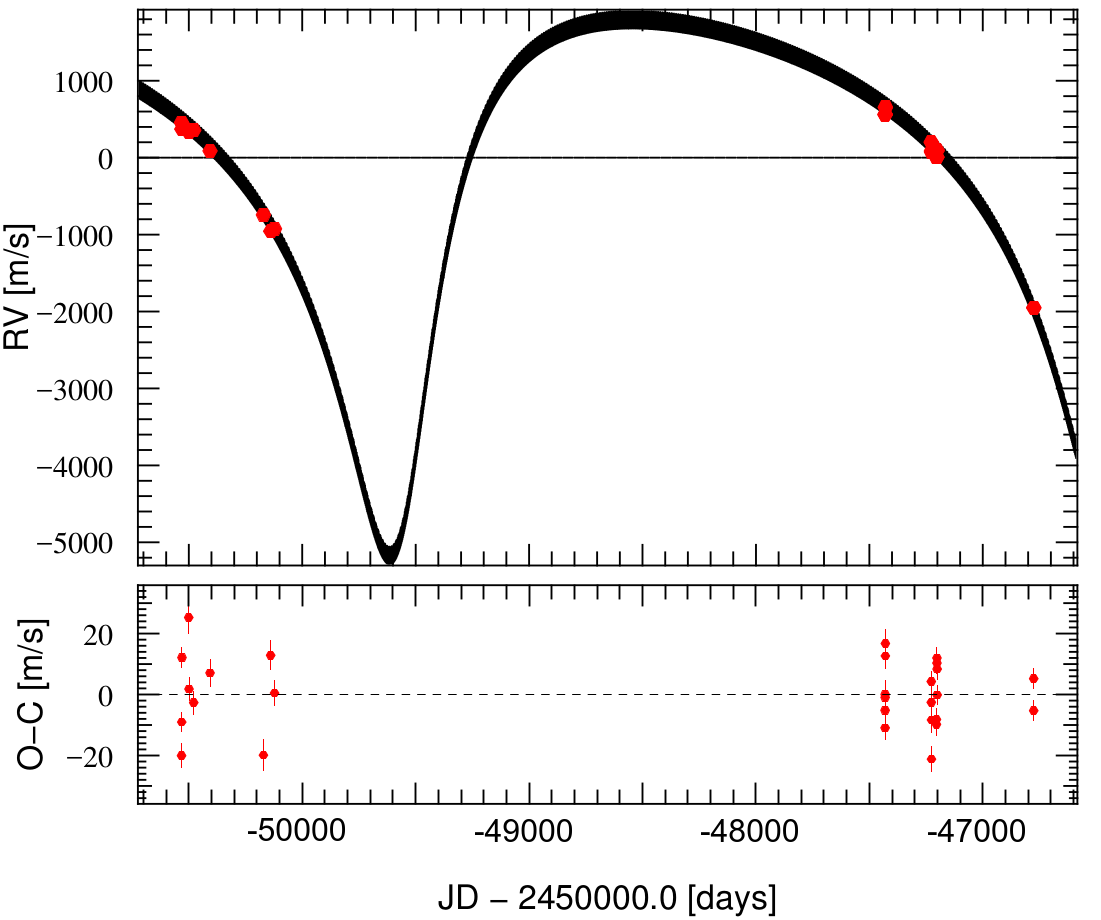}

\caption{Solutions of HD181321 RV fit by the sum of two keplerian ({\it Top}) and their residuals ({\it Bottom}) . {\it Left}: $\SI{1600}{\day}$ solution. {\it Right}: $\SI{3200}{\day}$ solution.} 
       \label{HD181321}
\end{figure*}

\subsubsection{HD186704A}

HD186704 is a known binary system with a companion at \SI{10}{\arcsecond} \citep{Zuckerman_13}.
\cite{Nidever} report a trend in the RV of $88 \pm 8$ $\si{\meter\per\second\per\jdb}$ with a negative curvature based on $4$ observations spaced on \SI{70}{\day} for HD186704AB.
\cite{Tremko} observe a change in the RV of \SI{4200}{\meter\per\second} in \SI{8682}{\day} on HD186704A.
Finally, \cite{Tokovinin} reporte HD186704A as hosting a spectroscopic binary (SB) companion with a \SI{3990}{\day} period.
We observe a trend of \SI{340}{\meter\per\second} over a duration of \SI{450}{\day} in the RV of HD186704A based on $4$ spectra.
This corresponds to a slope of $\SI{0.75}{\meter\per\second\per\jdb}$.
The star present signs of activity, we therefore corrected the RV using the RV-BVS correlation using \cite{Melo_BVS_RV_corr} method  (see \cref{appendix_corr_BVS}). The trend is still visible with a lower slope $\SI{0.23}{\meter\per\second\per\jdb}$. We present the RV and corrected RV of HD186704A in \Cref{HD186704}.
The difference of one order of magnitude on the slope of the trend between \cite{Nidever} observations and ours can be explained by the fact that our observations were made when the SB was closer to the periastron or apoastron of its orbit.

\subsubsection{HD181321}
We confirm that HD181321 is an SB. \cite{Nordstrom_181} report a variation of  \SI{2.3}{\kilo\meter\per\second} over \SI{9}{\year}, \cite{Guenther_181} report a trend with a slope of \SI{1.4}{\kilo\meter\per\second\per\year}.
Our observations spread on \SI{3757}{\day} (\SI{10.3}{\year})  cover at least two orbital periods.
The star is active and show  BVS variations with a period of $2-\SI{2.5}{\day}$ that should correspond to the rotational period of the star.
We use \emph{yorbit} to fit the RV with two Keplerian models, one to fit the stellar activity variation and an another to fit the binary variations.
We find two possible solutions for the companion, one with a period of \SI{1600}{\day} (\SI{2.7}{\au}) and a minimum mass of  $\SI{0.1}{\msun}$, and a second with a period of \SI{3200}{\day} (\SI{4.4}{\au}) and a minimum mass of $\SI{0.18}{\msun}$.
In both cases, the eccentricity is $\sim 0.5$.
We present those two solutions in \Cref{HD181321}.
More data are needed in order to distinguish between those two solutions.

\subsubsection{HD206893}

HD206893 is a F5V star that hosts a directly imaged BD companion at a separation of \SI{270}{\milli\arcsecond}  \citep{Milli,Delorme}. 
We recently reported in \cite{Grandjean} a long-term trend in the star RV coupled with pulsations with periods slightly less than one day.
We performed an MCMC on both the RV trend, imaging data, and \hipp-Gaia proper motion measurements to constrain the orbit and dynamical mass of the BD.
We concluded that the trend can not be attributed to the BD as it leads to dynamical masses incompatible with the object's spectra. 
The presence of an inner companion that contributes significantly to the RV trend is suggested with a mass of $\SI{15}{\MJ}$, and a period between $1.6$ and $\SI{4}{\year}$.

\subsubsection{HD217987}

\begin{figure*}[t!]
  \centering
\begin{subfigure}[t]{0.32\textwidth}
\includegraphics[width=1\hsize,valign=m]{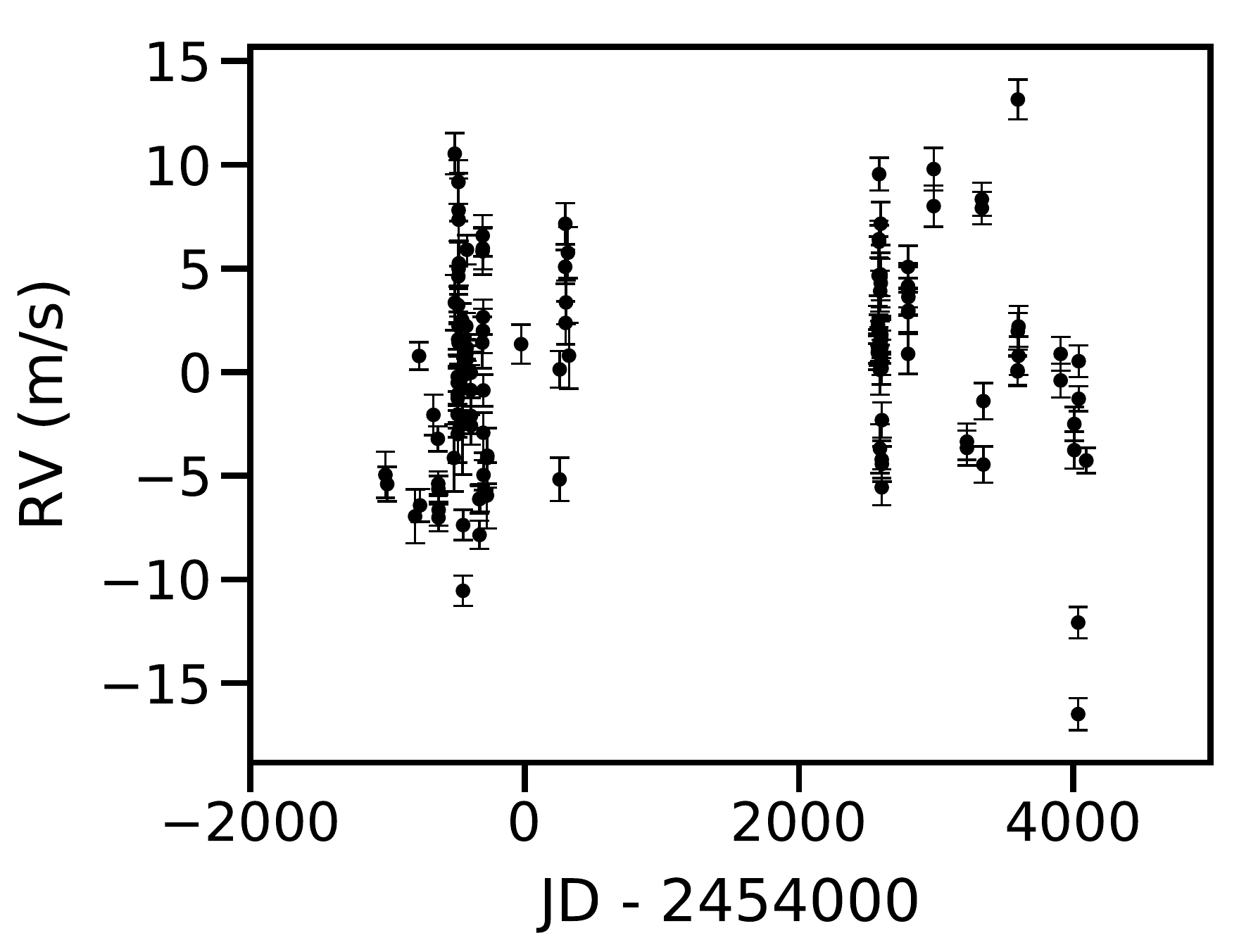}
\caption{\label{RV_217}}
\end{subfigure}
\begin{subfigure}[t]{0.32\textwidth}
\includegraphics[width=1\hsize,valign=m]{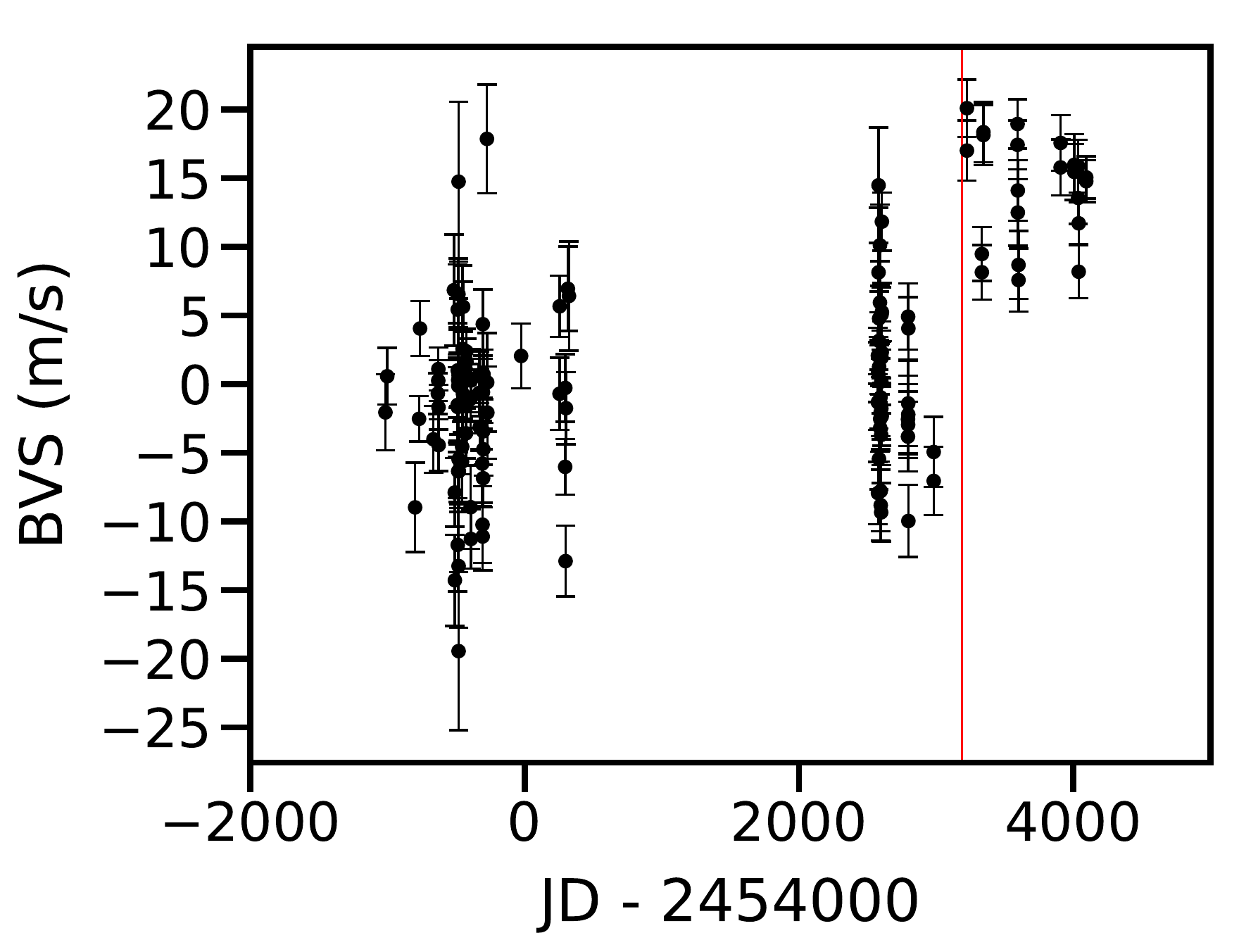}
\caption{\label{span_217}}
\end{subfigure}
\begin{subfigure}[t]{0.32\textwidth}
\includegraphics[width=1\hsize,valign=m]{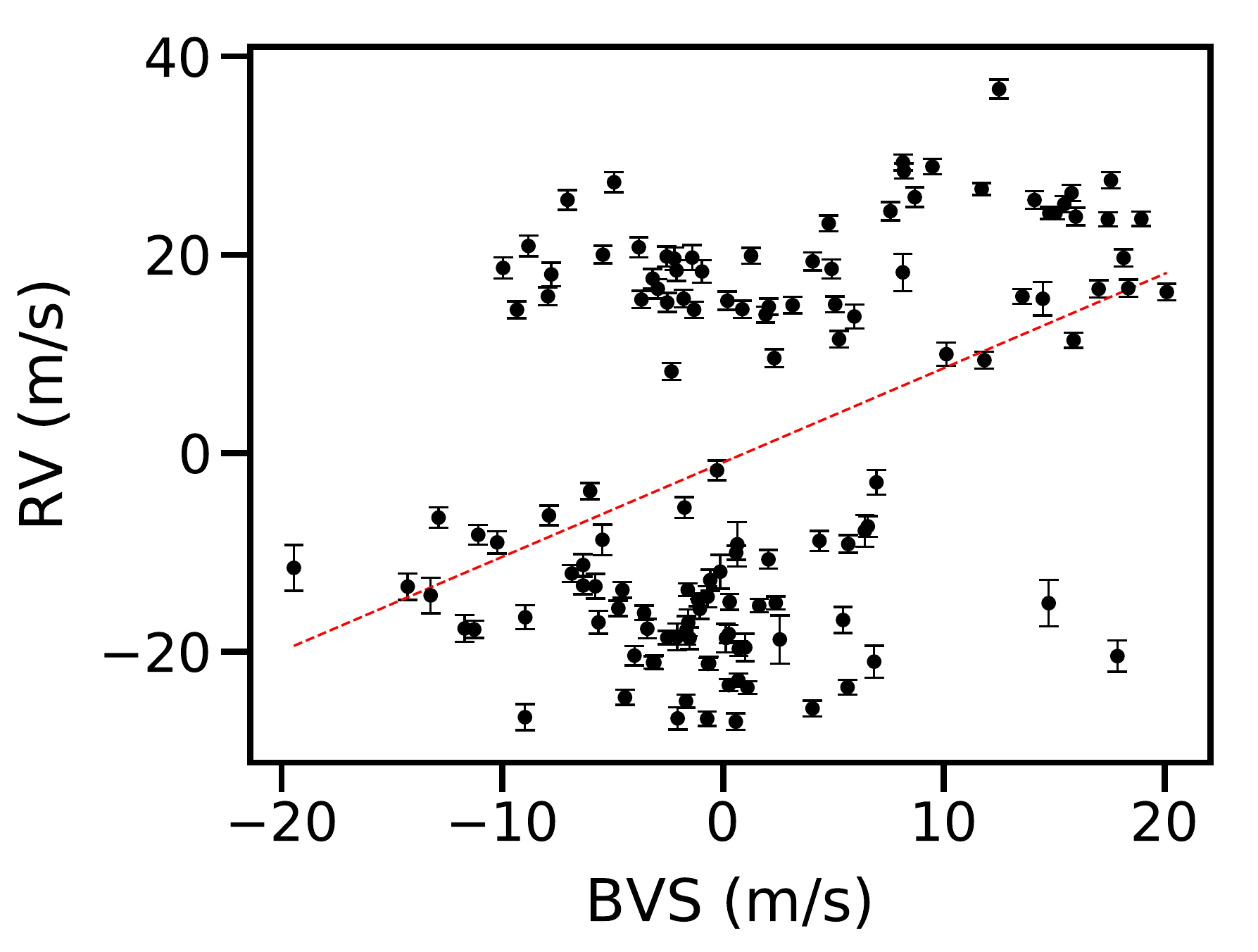}
\caption{\label{BVS_RV_217}}
\end{subfigure}

\begin{subfigure}[t]{0.32\textwidth}
\includegraphics[width=1\hsize]{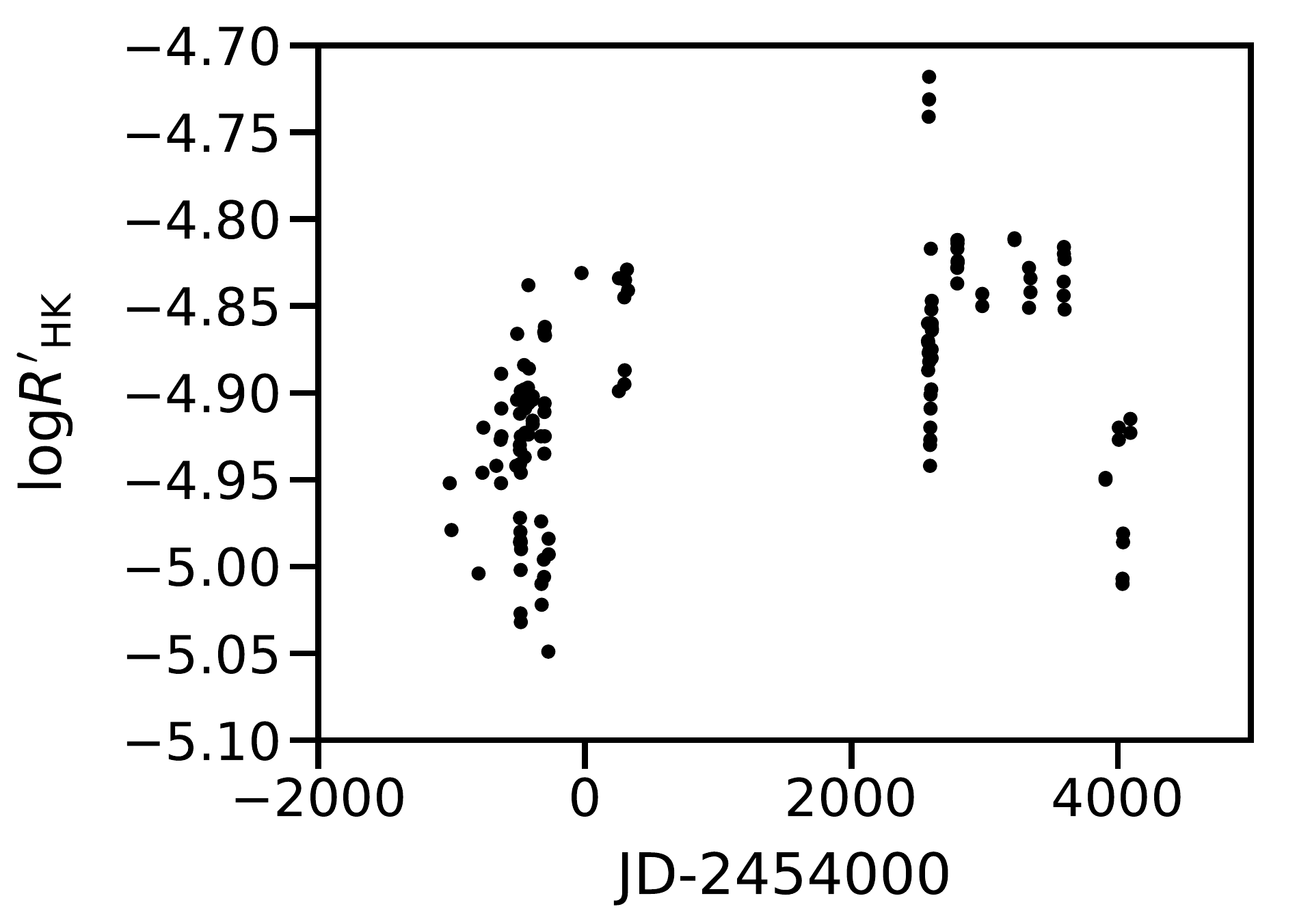}
\caption{\label{rhk_217}}
\end{subfigure}
\begin{subfigure}[t]{0.32\textwidth}
\includegraphics[width=1\hsize]{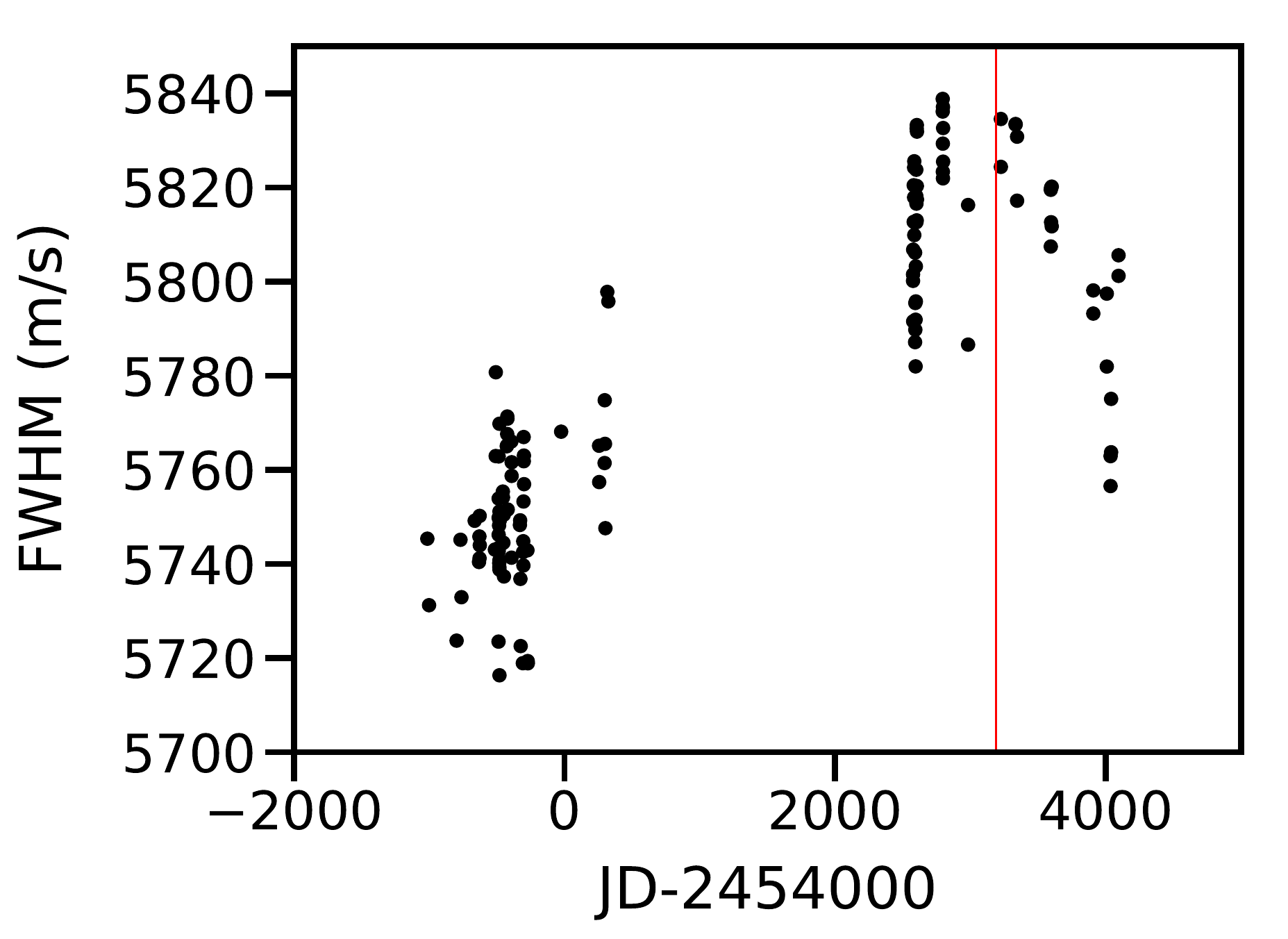}
\caption{\label{fwhm_217}}
\end{subfigure}
\caption{HD217987 summary. \subref{RV_217}) RV time variations corrected from the secular acceleration drift.
 \subref{span_217}) BVS time variations, \harps \  fiber change is shown  with a vertical red line.
  \subref{BVS_RV_217}) RV corrected from the secular acceleration drift vs BVS. The best linear fit is presentend in red dashed line.
  \subref{rhk_217}) \rhk \ time variations.
  \subref{fwhm_217}) FWHM time variations.} 
       \label{RV_BVS_corr}
\end{figure*}

HD217987 is a M2V, high proper motion star and exhibits a long-term RV trend induced by its secular acceleration.
We therefore corrected the RV from the star secular acceleration in its proper motion we deduced from its paralax and proper motion reported in \cite{DR2A1} (see \Cref{RV_BVS_corr}).
The corrected RV presents short period variations. 
The RV and BVS are correlated which indicate that the signal might be dominated by spots or plages (see Figure \ref{RV_BVS_corr}).
We observe a long-term variation in \rhk, FWHM, and RV which are better fitted by a second degree polynomial model than with a linear model (see \Cref{RV_BVS_corr}).
This variation could be then attributed to a magnetic cycle of the star with a period greater than $\SI{5000}{\day}$.
An analysis on a large number of M dwarf was made by \cite{Mignon}, including this star and we should expect an offset of $\sim \SI{-10}{\meter\per\second}$ in the RV due to the \harps \  fiber change of the 15th of June 2015 for this star.
We observe an offset in the star's BVS at this date date (see \Cref{RV_BVS_corr}), but no significant offset in the RV.
We use one template on all the data (see \cref{safir}), the impact of the \harps \ fiber change is then averaged. This can explain why we do not see a significant RV offset for this star.

\subsubsection{HIP36985}

\begin{figure}[h!]
  \centering
\includegraphics[width=1\hsize,valign=t]{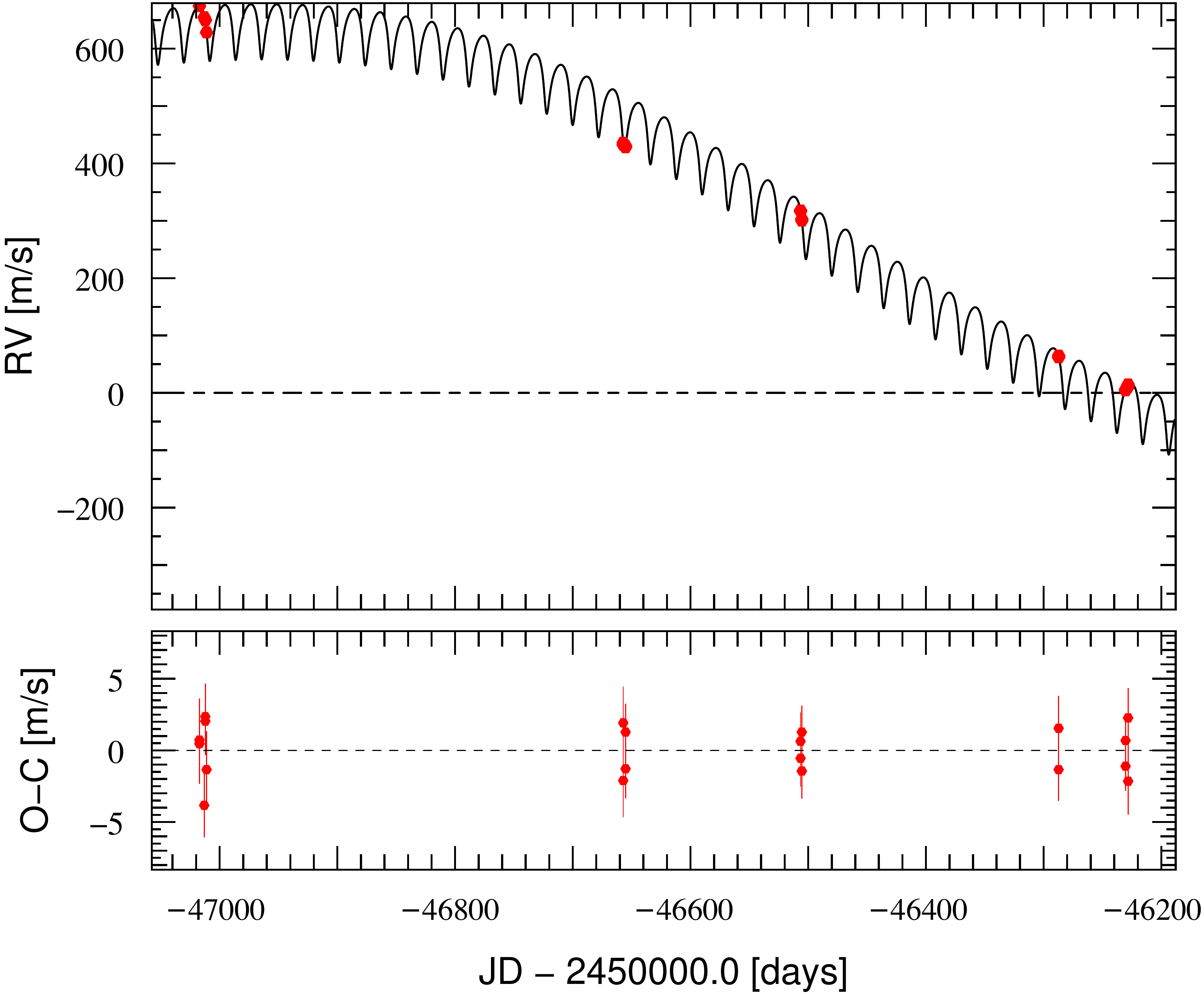}

\caption{Solution of HIP36985 RV's fit by a sum of two keplerians and its residuals} 
       \label{86985}
\end{figure}

HIP36985 is a M2V-type star reported as a wide binary with the system GJ282AB at a separation of $\ang[angle-symbol-over-decimal]{1.09}$ \citep{Poveda}.
We observe an RV long-term variation with an amplitude of \SI{700}{\meter\per\second}  on a time baseline of \SI{1400}{\day} in addition to a short-term variation with an amplitude of $\SI{50}{\meter\per\second}$.
No correlation between RV and BVS is seen, which excludes stellar activity or pulsations as the origin of the long-term variations.
The mean \rhk \ of our spectra of $-4.3$ indicates that the short-term variations come from magnetic activity (spots).
We fit the RV with \emph{yorbit} using two Keplerian models, one to fit the long-term variations and an another to fit the companion variations. We present our best solution in \Cref{86985} and the corresponding companion parameters in \cref{tab_36985}.
The companion minimum mass deduced from the present data is \SI{29}{\MJ} and the period is \SI{8400}{\day} (\SI{23}{\year}), corresponding to a semi-major axis of \SI{7}{\au} and a projected separation of \SI{497}{\milli\arcsecond}.
We observed HIP36985 with \sphere \  in 2018 and confirm the presence of a however low-mass star companion \citep{36985}.
We find a period of $\SI{22}{\day}$ for the short-term variations, while the rotation period is of $12 \pm 0.1 \ \si{\day}$ \citep{Alonso}.
The poor sampling of those short-term variations probably explains the strong difference with the rotation period.

\begin{table}[h!]
\center
\begin{tabular}[h!]{|c|c|}
Parameters & Value\\ \hline
$P \ (\si{\day})$& $8500^{+1900}_{-1900}$\\
$a \ (\si{\au})$& $7^{+1}_{-1.1}$\\
$Separation \ (\si{\milli\arcsecond})$& $500^{+70}_{-80}$\\
$e$& $0.55 \pm 0.04$\\
$ \omega \ (\si{\degree})$& $50 \pm 67$\\
$ K\ (\si{\meter\per\second})$& $480 \pm 230$\\
$ M \sin i\ (\si{\MJ})$& $29^{+18}_{-16}$\\
\end{tabular}
\caption{HIP36985B's orbital parameters.}
\label{tab_36985}
\end{table}

\subsection{Giant planets}

\begin{figure*}[h!]
  \centering
\includegraphics[width=1\hsize]{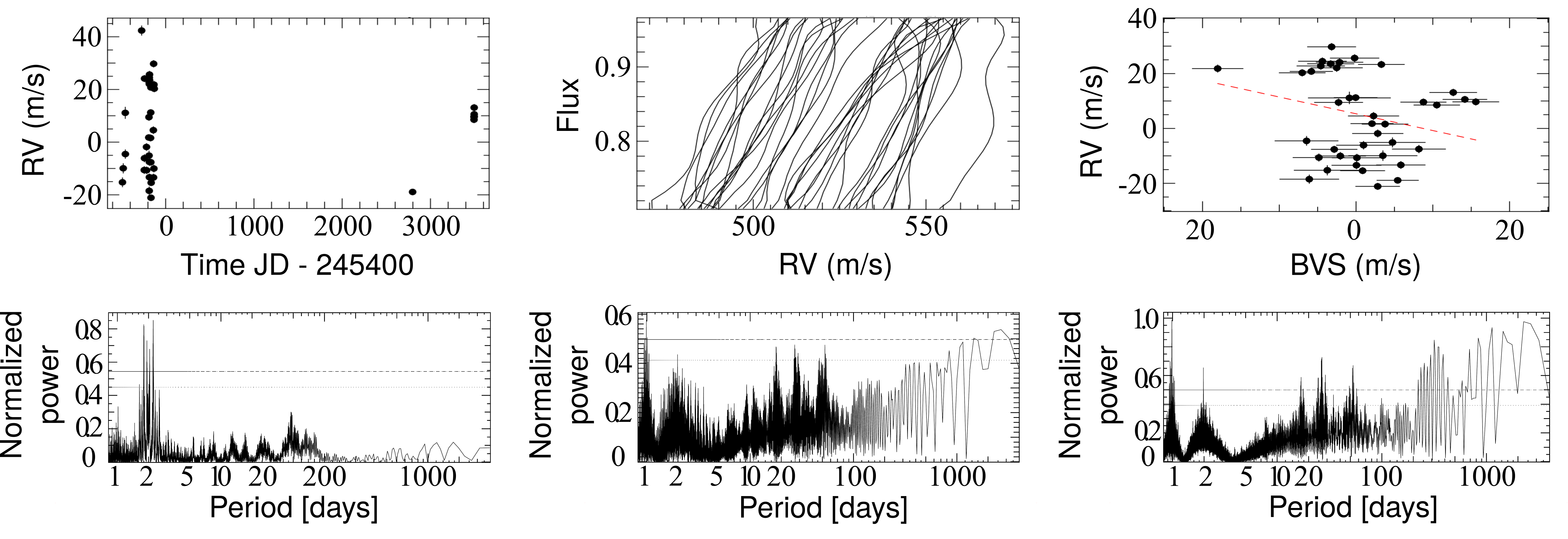}
\caption{BD$+$20 2465 summary. ({\it Top row}) RV time variations ({\it Left}), bisectors ({\it Middle}), RV vs BVS ({\it Right}). 
({\it Bottom row}) RV periodogram  ({\it Left}), BVS periodogram ({\it Middle}), time window peridogram ({\it Right}). The FAP at  $1 \%$ and $10 \%$ are presented respectively in dashed lines and in dotted lines.} 
       \label{901741}
\end{figure*}

No giant planet companion of period less than \SI{1000}{\day} is detected.
In addition to the stars presented in \cref{comp}, some stars presented RV variations without a significant correlation between RV and BVS. For these stars we compared the RV periodogram with the BVS periodogram and the time window periodogram.
In all cases, putative periods associated with the RV were also present in the BVS or time window periodograms, or in both. 
This implies that these RV variations are most likely due to stellar activity or pulsations.
We present as an example BD$+$20 2465 in  \Cref{901741}. This star shows RV variations with periods near $\SI{2}{\day}$ which are \text{both} present in the time window periodogram and BVS  periodogram.

\section{Analysis}
\label{occ_rate}

\subsection{Stellar intrinsic variability}

\label{jitter}

\begin{figure*}[t!]
  \centering
\includegraphics[width=0.32\hsize]{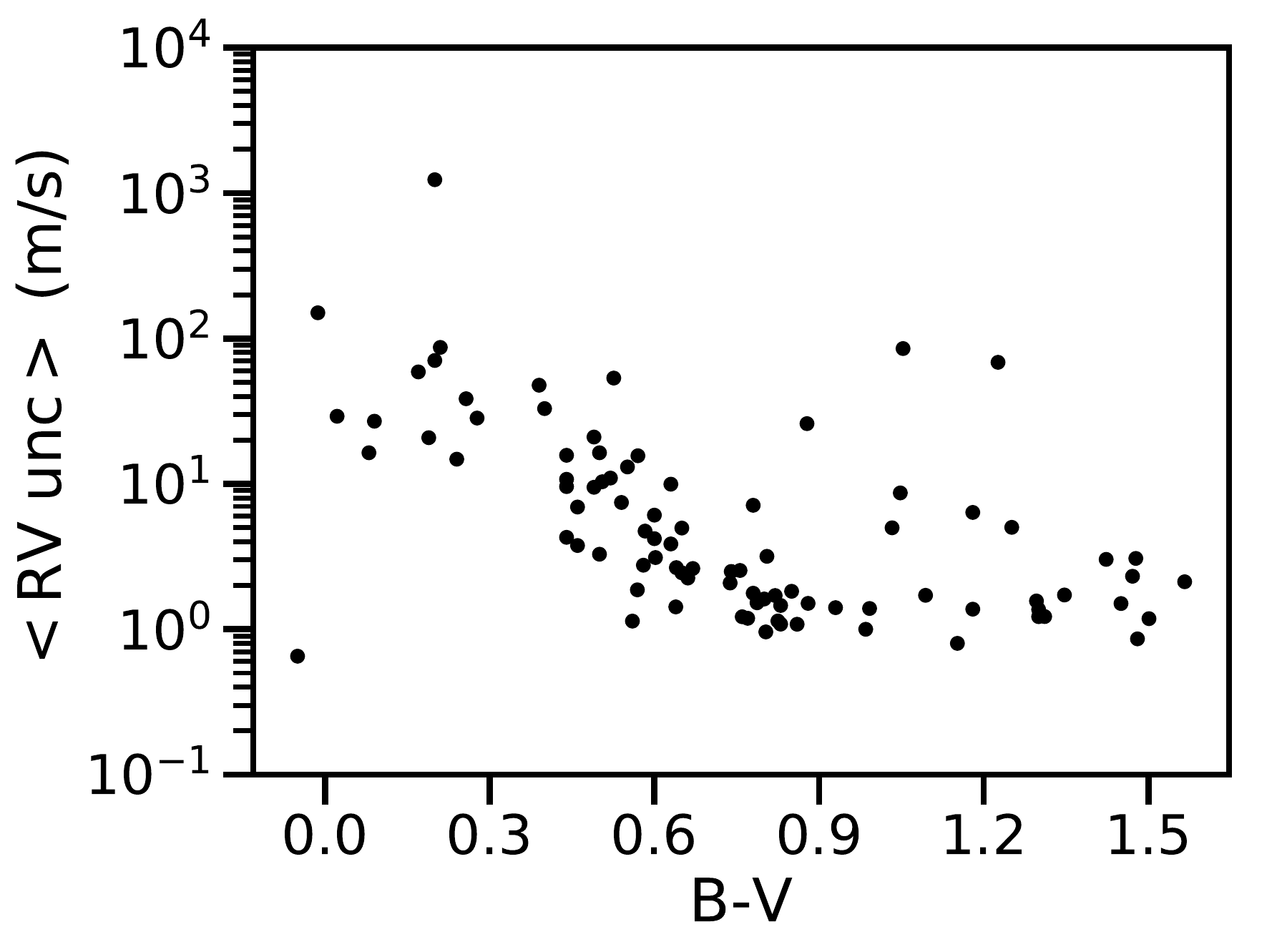}
\includegraphics[width=0.32\hsize]{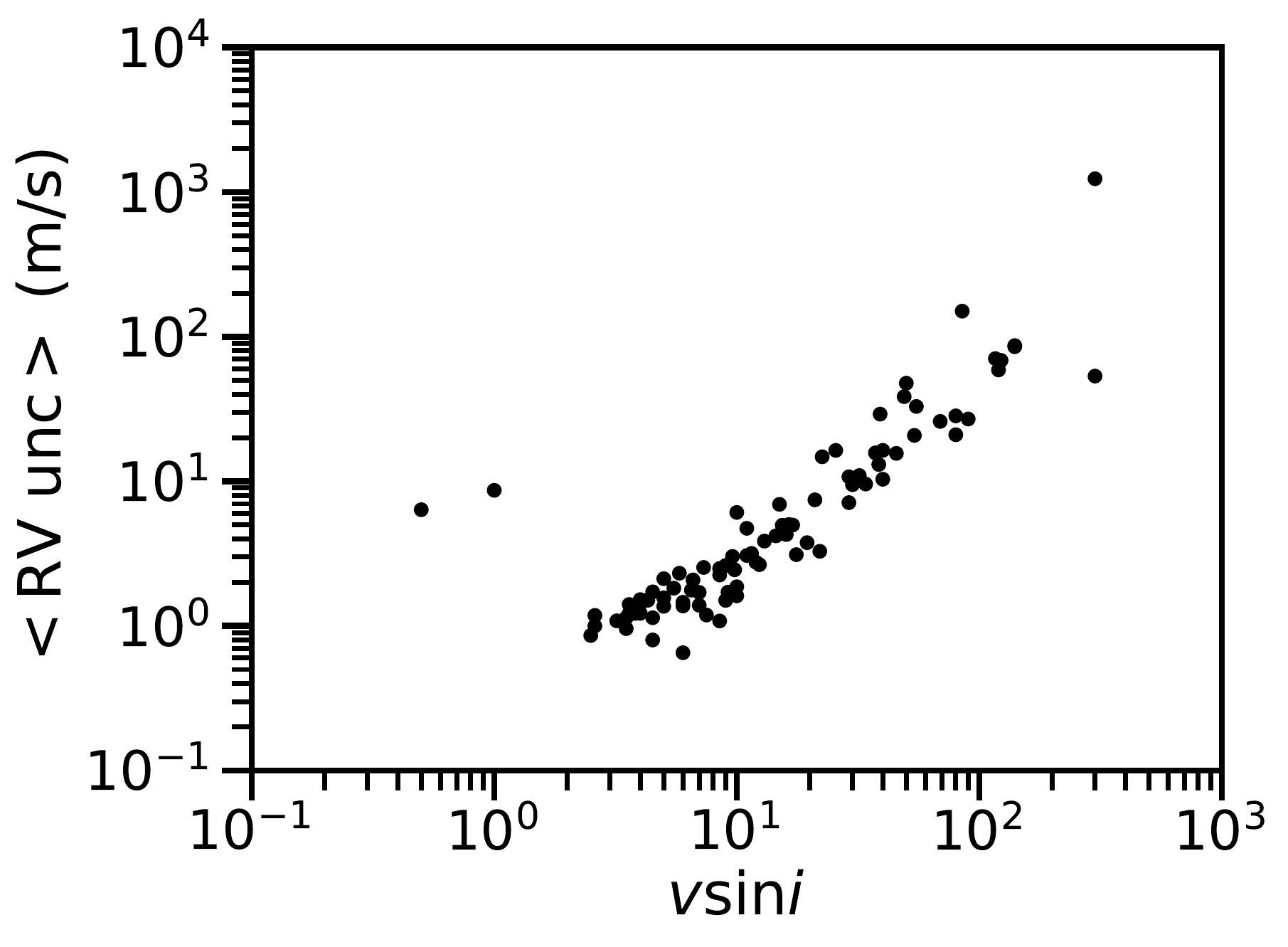}
\includegraphics[width=0.32\hsize]{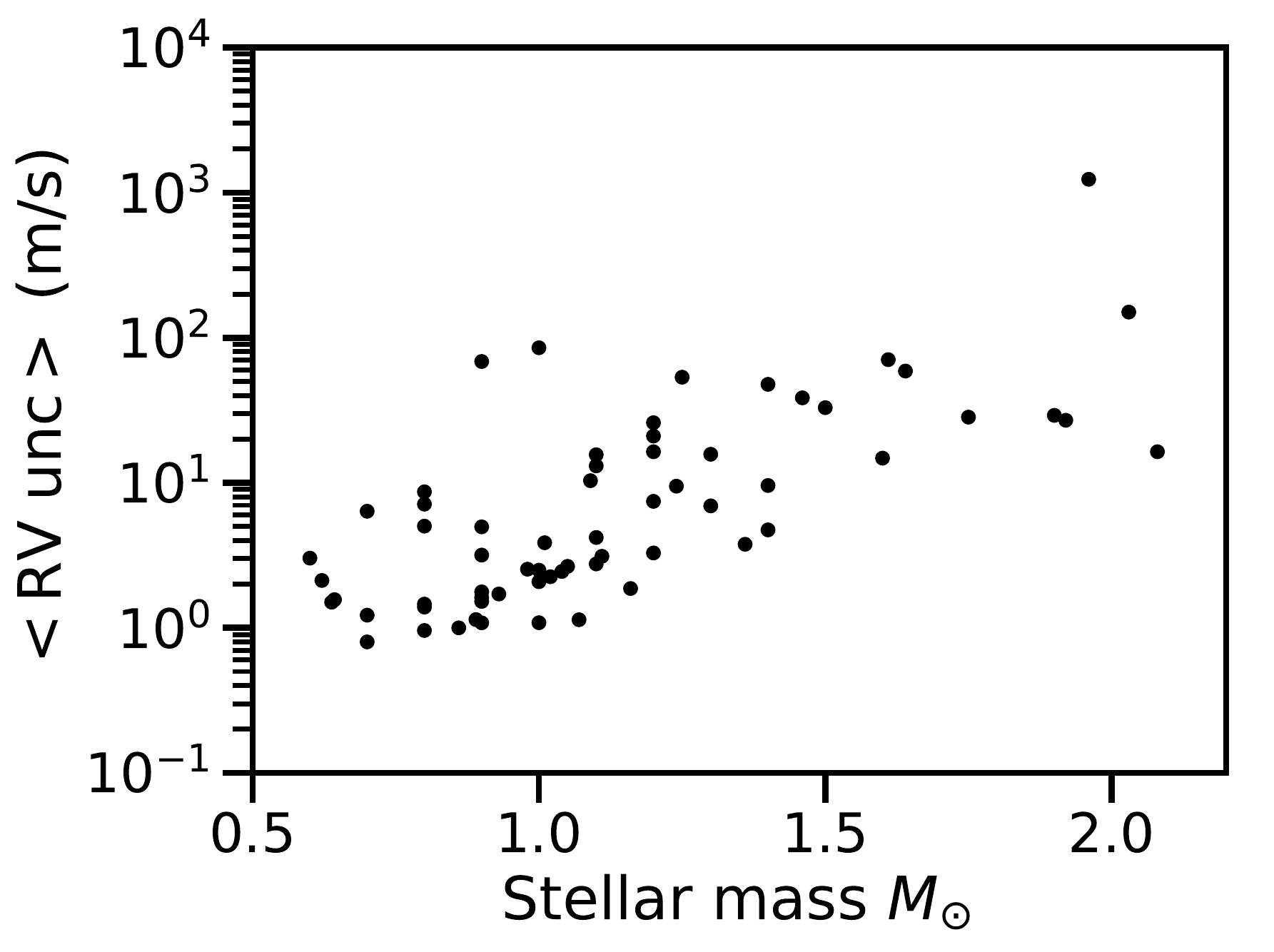}
\caption{Summary of the survey RV uncertainties. Mean RV uncertainty (accounting for the photon noise only) vs \bv ({\it Left}), vs \vsini \ ({\it Middle}) and vs \Mstar \ (in \Msun, {\it Right}).}
       \label{stell_var_1}
\end{figure*}

\begin{figure*}[t!]
  \centering

\includegraphics[width=0.32\hsize,valign=m]{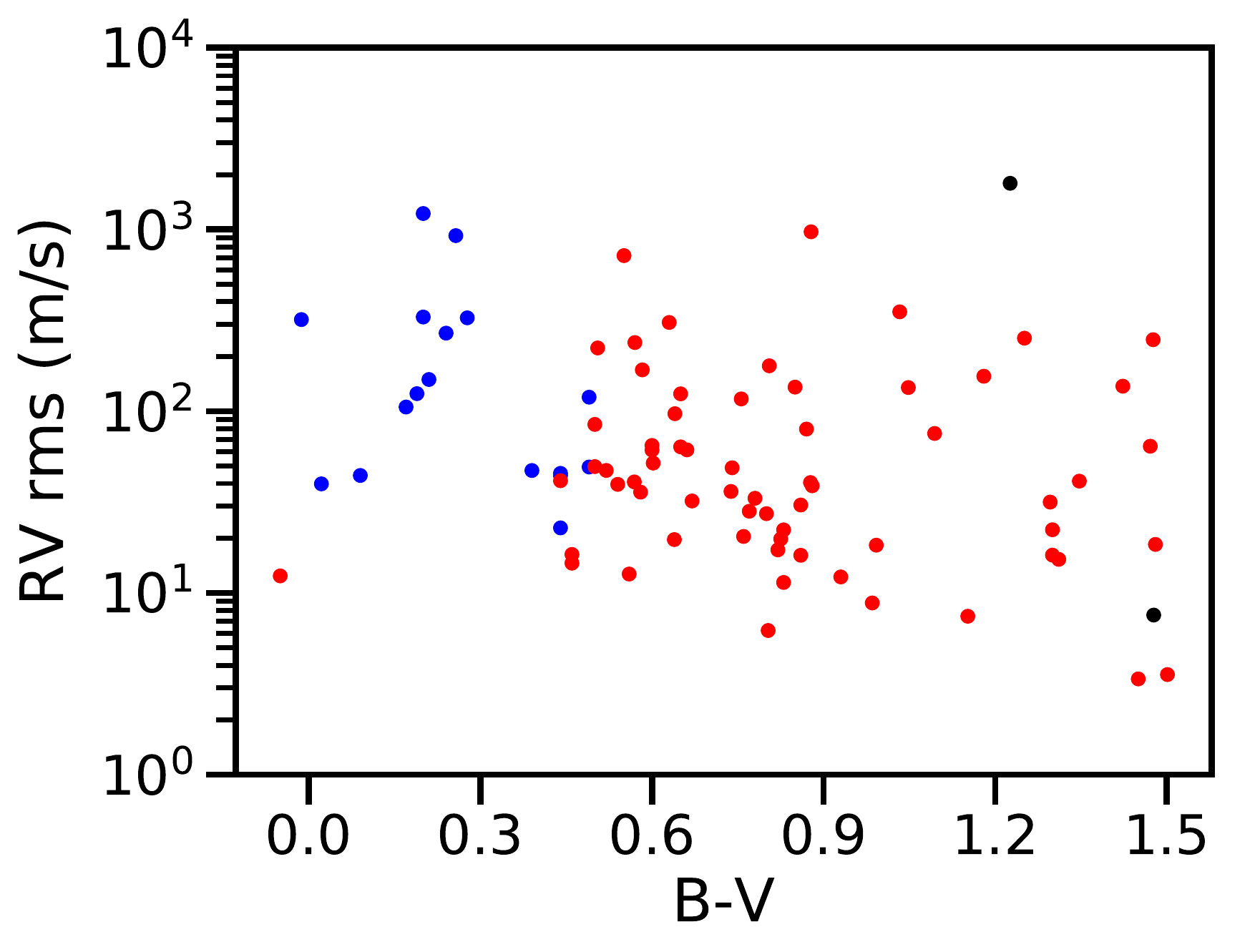}
\includegraphics[width=0.49\hsize,valign=m]{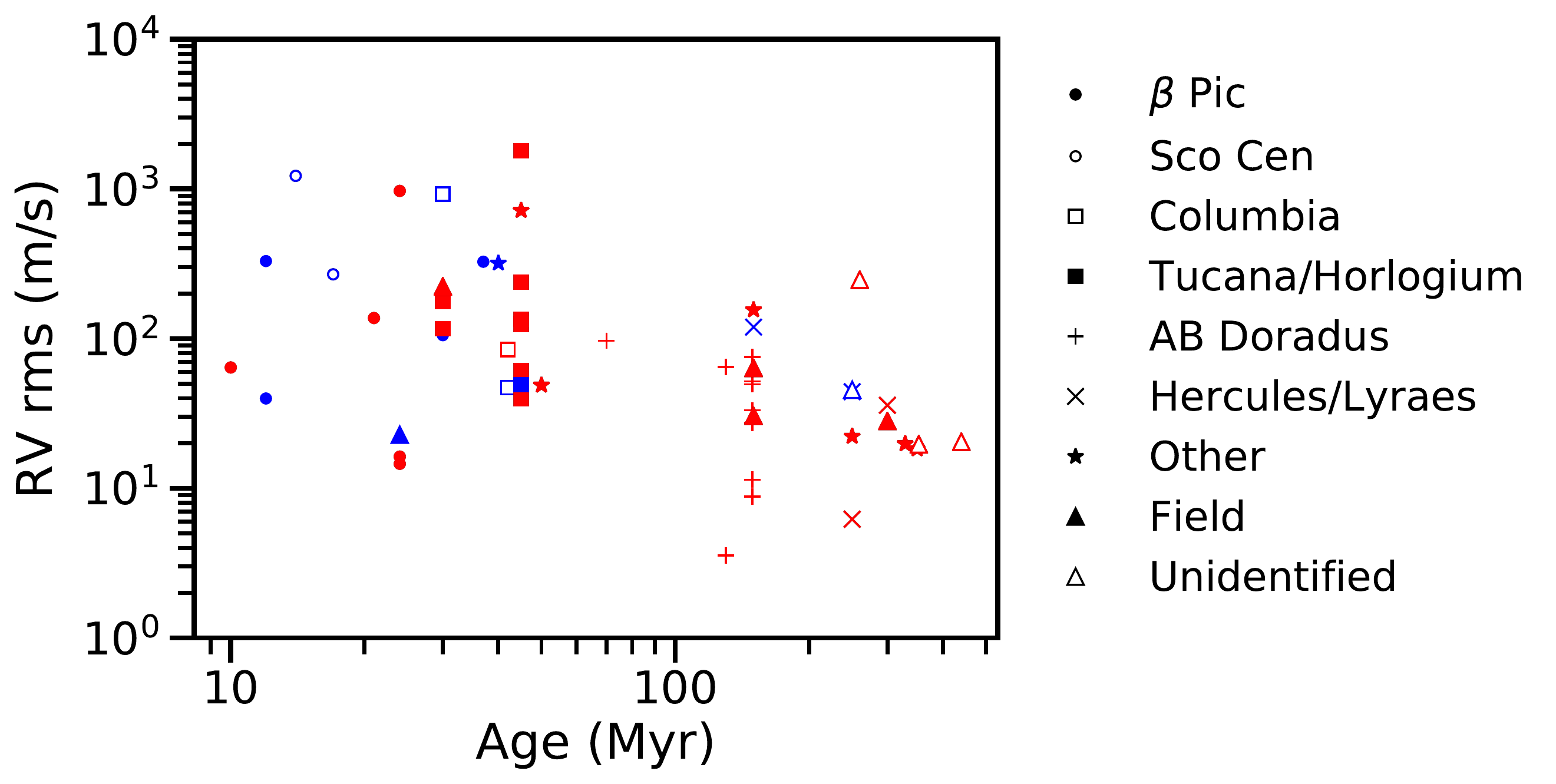}

\caption{Survey RV rms summary. ({\it Left}) RV rms vs  $B-V$, ({\it Right}) RV rms and  vs age. 
Pulsating stars are plotted in blue, stars with RV dominated magnetic activity (spots) are plotted in red and stars with undetermined main source of RV are plotted in black.
Stars with SB signature are not displayed (HD106906, HD131399, HD177171). 
HD197890 is not considered due to a too small set of data available.} 
       \label{stell_var_2}
\end{figure*}

\Cref{stell_var_1} displays the mean RV uncertainty vs $B-V$, \vsini \ and $M_*$.
We also display the RV rms vs $B-V$, and age in \Cref{stell_var_2}.
We observe that the mean RV uncertainty is correlated to the $v\sin{i}$  ($Pearson=-0.65$, $p_{value} \ll 1 \%$).
This is consistent with what is observed on older, AF main sequence stars by \cite{Simon_IX,Simon_X}.
We observe a strong jitter for most of the stars. The ratio between RV rms and the mean RV uncertainty is between $390$ and $1$ with a median at $16$.

The median RV rms is \SI{49}{\meter\per\second} (\SI{300}{\meter\per\second} on average).
This jitter is mainly caused by pulsations for early type stars (from A to F5V), and by spots and faculae for late type stars ($>F5V$). 
Those two regimes can be distinguished, as stars with pulsations shows a vertical spread of BVS(RV) diagram, whereas stars with spots present a correlation between RV and BVS \citep{Lagrange_2009}.
The main origin of RV jitter is reported in \cref{tab_result} for each target.

\begin{figure}[h!]
  \centering

\includegraphics[width=0.75\hsize]{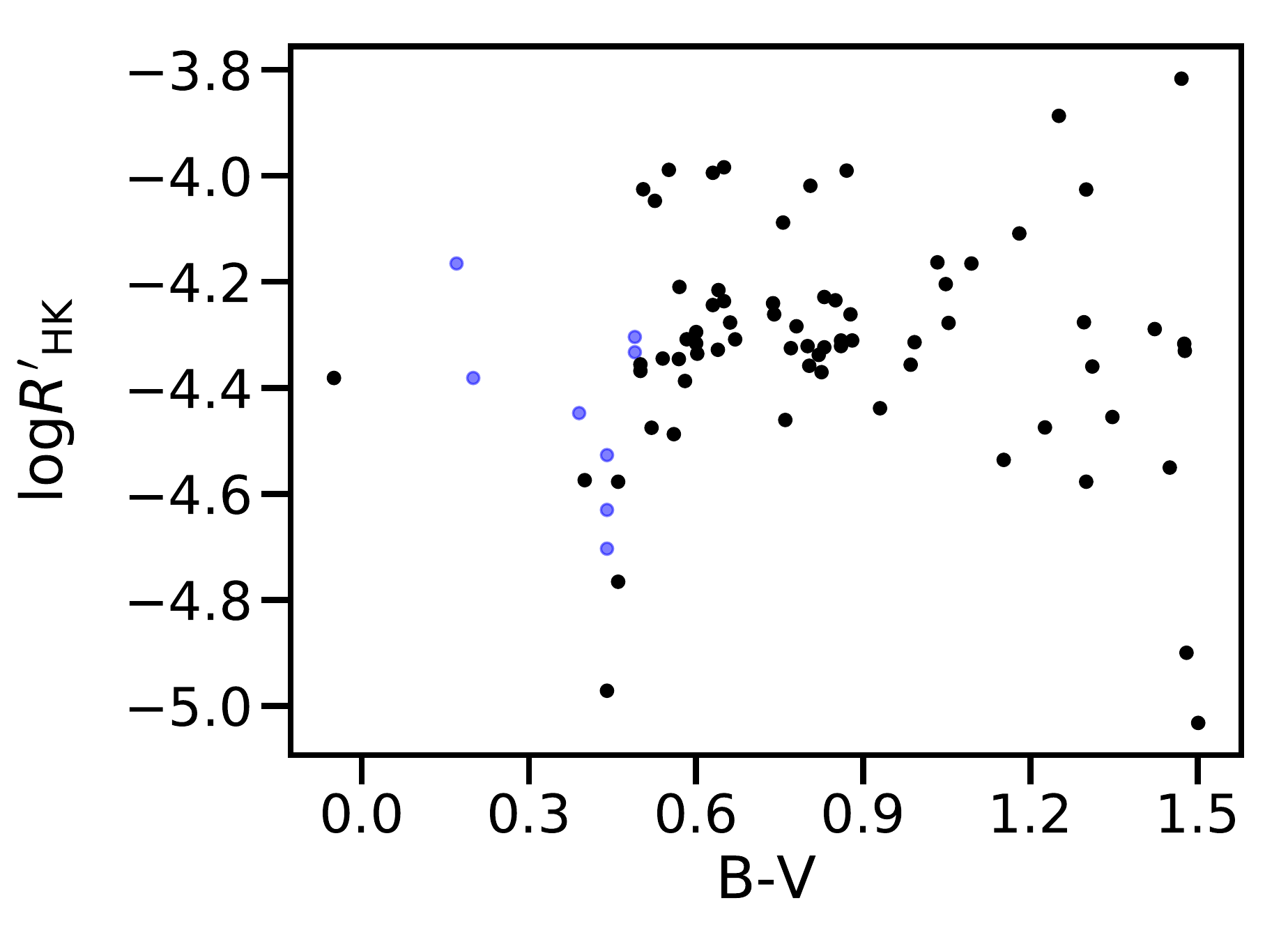}

\caption{Survey \rhk vs $B-V$.
Pulsating stars are plotted in blue.} 
       \label{stell_rhk}
\end{figure}

$78$ stars out of $89$ of our sample present variations in their Ca lines. The median \rhk \ of our sample is $-4.3$ with a standard deviation of $0.2$. 
$4$ stars present signs of low activity (\rhk $<-4.75$), $59$ are active ($-4.75<$ \rhk $< -4.2$) and $15$ stars present signs of high activity (\rhk $>-4.2$). 
We present in \Cref{stell_rhk} \rhk \ vs $B-V$. 
Some early F-type pulsating stars also present signs of actity when all late type stars present signs of activity.

\subsection{\harps \ fiber change}
\label{harps_fiber_change}
\begin{figure}[h!]
  \centering

\includegraphics[width=0.75\hsize]{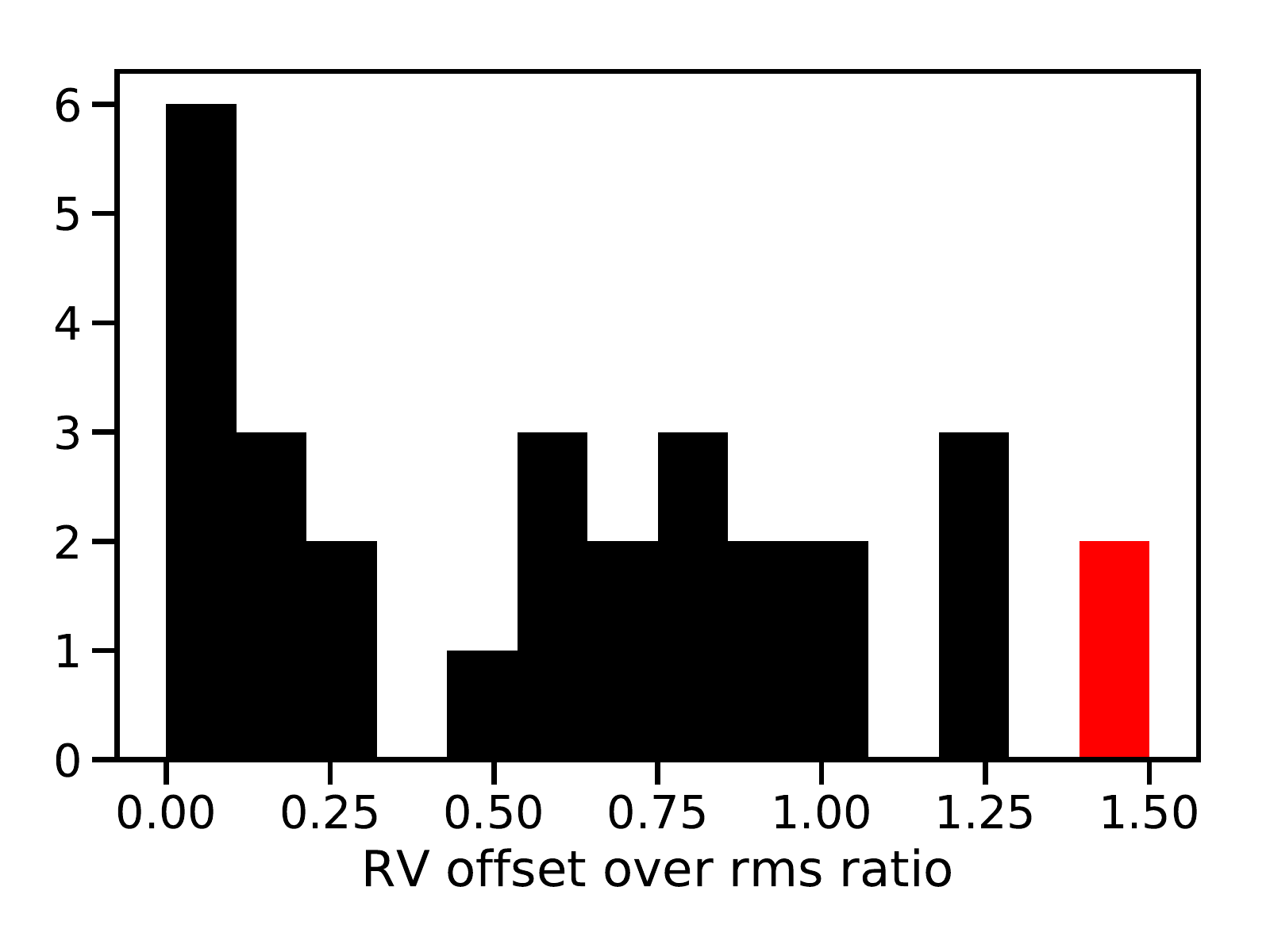}

\caption{Histogram of the ratio between the difference in the RV mean between before and after \harps' fiber change, and the maximum of the rms before and after the fiber change. The stars that present a ratio greater than $1.3$ are plotted in red.} 
       \label{fig_fac_gap}
\end{figure}

\begingroup
\renewcommand\arraystretch{1.2}
\begin{table}[h!]
\center
\resizebox{0.45\textwidth}{!}{ 
\begin{tabular}[h!]{|lccccc|}
Name & ST & $B-V$ & $<$FWHM$>$ & Offset & Ratio\\ 
&&& $(\si{\kilo\meter\per\second})$ & $(\si{\meter\per\second})$ &\\ \hline
HD7661 & K0V & 0.77 & 9.2 & 42 & 1.4 \\ 
HD218860 & G8V & 0.74 & 11.7 & 57 & 1.4 \\
\end{tabular}}
\caption{Parameters of the stars that present a significant offset. The offset is the difference between the RV mean before and after \harps' fiber change, and the ratio is the ratio between the offset and the maximum of the rms before and after the fiber change.}
\label{tab_gap}
\end{table}
\endgroup

In June 2015, the fiber of the \harps \ instrument  was changed in order to increase its stability  \citep{fiber_change}.  
\cite{fiber_change} show that it leads to a change in the instrument profile which impacts the CCF computation and therefore the RV and BVS computation.
They observed an offset in the RV between the datasets taken before and after this change of the order of $\SI{15}{\meter\per\second}$ for old F to K-type stars, based on $19$ stars \citep{fiber_change}.
A more detailed analysis is currently underway for a large sample of close M-type main sequence stars \citep{Mignon}. In this analysis, two reference spectra are computed, one before the change and one after.
Our current set of spectra for each star is not big enough to build reference spectra as done in \cite{Mignon}.
The offset is different from one star to the other and its correlation with stellar parameters is not yet determined (spectral type, \vsini \ etc.).
This offset has not been estimated on young stars before and the impact of high \vsini \ and strong jitter is not known.
One should be careful before trying to correct this offset in order to not remove signal.

The difference between the mean of the RV before and after the fiber change can be measured, however, it won't probe the offset alone, as long-term variations (star magnetic cycle, unseen companions, etc.) or a bad sampling of the jitter can also induce a difference in the mean of the RV.

We then decided to correct the offset only when it is significant enough. 
We compared the difference between the RV mean before and after the fiber change to the maximum of the rms before and after the fiber change to determine it.
First, we selected the stars for which computing rms before and after the fiber change is relevant : we excluded the stars for which we have less than $10$ spectra, less than $6$ spectra before the fiber change, and less than $6$ spectra after.
We excluded the stars identified as SB (marked "B" in \cref{tab_result}) or presenting a long-term variation due to a companion (marked "T" in \cref{tab_result}).
We also excluded the stars for which the RV amplitude is more than $\SI{900}{\meter\per\second}$ as the offset should be negligible  compared to the jitter.
We finally excluded HD169178 that presents a trend due to a magnetic cycle.
Finally, $29$ stars were not excluded. We present the histogram of the ratio between the difference in mean of RV between before and after the fiber change and the maximum of the rms before and after the fiber change in \Cref{fig_fac_gap}.
We chose to correct the RV offset for the stars that present a ratio larger than $1.3$ a threshold  that ensure a significant offset. Two stars correspond to this criterion. 
We present the characteristics of their offset in \cref{tab_gap}, and we present the correction of the offset for one of them in \cref{218}.
They have G to K-types and present an offset of the order of $\SI{50}{\meter\per\second}$. Such offsets are three times larger than those founds on old stars \citep{fiber_change}. 

\subsection{Exclusion of peculiar stars and RV correction for further analysis}

In order to better estimate the number of potentially missed planets in our survey, we excluded some stars from our analyze and made some corrections on others before computing the detection limits.
We excluded SB stars (HD106906, HD131399, HD177171, HD181321). 
Further, we excluded HD116434 since its high value of $v\sin{i}$ ($>\SI{200}{\kilo\meter\per\second}$) prevents the measurements of BVS. 
We also excluded HD197890 for which our data are too sparse to reliably quantify the detection limits ($3$ spectra). Thus leads to $83$ stars,  $23$ A-F stars, $52$ F-K stars, and $8$ K-M stars.

For stars with RV dominated by spots (marked A in the \cref{tab_result}) we corrected their RV from the RV-BVS correlation using \cite{Melo_BVS_RV_corr} method (see \cref{appendix_corr_BVS}). 
 We corrected HD217987 RV from its proper motion and HD186704 RV from its long-term trend with a linear regression.  
For $\beta$ Pic, we considered the RV corrected from its pulsations  as well as $\beta$ Pic b and c contributions \citep{Beta_pic_c}. 

For the stars for which we identified a non ambiguous offset in the RV, we corrected their RV from this offset (see \cref{tab_gap}).

\subsection{Detection limits}
\label{detlim}

We compute $m_p\sin{i}$ detection limits for periods between $1$ and \SI{1000}{\day} in the GP domain (between $1$ and $\SI{13}{\MJ}$), and in the BD domain (between $13$ and  $\SI{80}{\MJ}$).
We use the Local Power Analysis (LPA) \citep{Meunier,Simon_IX} which determines, for all periods $P$, the minimum $m_p\sin{i}$ for which a companion on a circular orbit with a period $P$ would lead to a signal consistent with the data, by comparing the synthetic companion maximum power of its periodogram to the maximum power of the data periodogram within a small period range around the period $P$.

\begin{figure}[t!]
  \centering
\includegraphics[width=1\hsize]{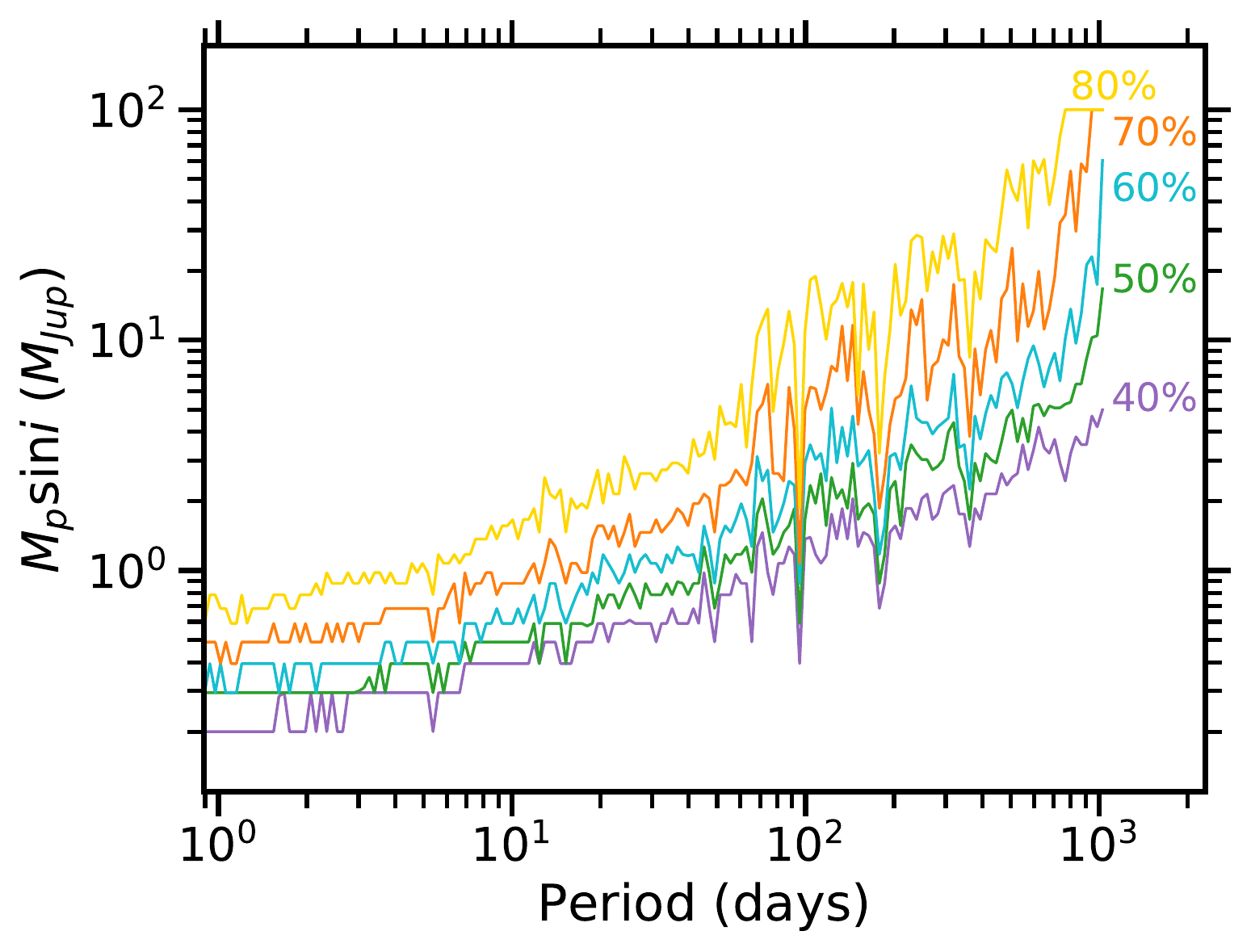}
\caption{Search completeness of our survey, corresponding to the lower $m_p \sin{i}$ for which $X\%$ of the star of the survey have detection limits below this $m_p \sin{i}$ at a given period $P$. From bottom to top $40\%$ to $80\%$.\label{Completeness}} 
       \label{sample}
\end{figure}

We then compute the completeness function $C(m_p\sin{i},P)$ which corresponds for a given couple $(m_p\sin{i},P)$ to the fraction of stars for which the companion is excluded by the detection limits \citep{Simon_IX}. 
We present the $40$ to $80\%$ search completeness in \Cref{Completeness}.

\subsection{Companion occurrence rates}

We compute the upper limits of companions occurrence rates for our $83$ stars in the GP ($1$ to $\SI{13}{\MJ}$) and BD ($13$ to $\SI{80}{\MJ}$) domains for AF ($B-V \in [-0.05:0.52[$), FK ($B-V \in [0.52:1.33[$), KM ($B-V \geq 1.33$) type stars for different ranges of periods : $1$-$10$, $10$-$100$, $100$-$1000$, and $1$-\SI{1000}{\day}.
We use the method described in \cite{Simon_IX} to compute the occurrence rates and to correct them from the estimated number of missed companions $n_{miss}$ derived from the search completeness.

We present the upper limits of the occurrence rates for all stars  in \Cref{sample} and for AF, FK, M, and all stars in \cref{tab_occur}. 
The GP occurrence rate is below $2_{-2}^{+3} \ \%$ ($1 \sigma$) and the BD occurrence rate is below $1_{-1}^{+3}\ \%$ ($1 \sigma$)  for periods under \SI{1000}{\day}.

\begin{figure}[h!]
  \centering
\includegraphics[width=1\hsize]{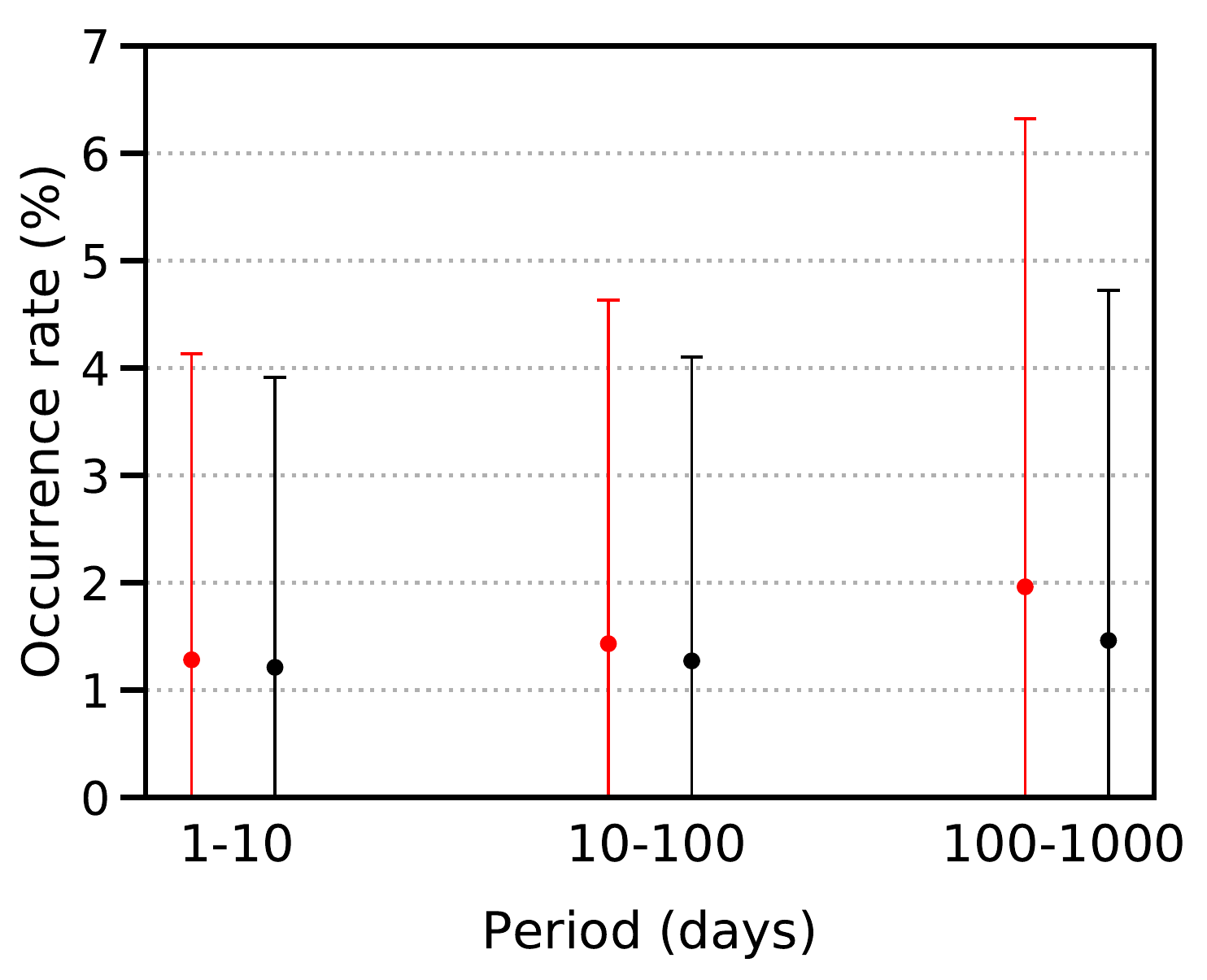}
\caption{Upper limits ($1 \sigma$) on the occurrence rates for our survey for period ranges of $1$-$10$, $10$-$100$, and $100$- \SI{1000}{\day} in the GP domain ($1-13$ \Mjup,  \emph{Red}) and BD domain ($13-80$ \Mjup,  \emph{Black}).} 
       \label{sample}
\end{figure}

\renewcommand{\arraystretch}{1.25}
\begin{table*}[t!]
\caption{GP ($m_p\sin{i} \in [1,13] \ \si{\MJ}$) and BD  ($m_p\sin{i} \in [13,80] \ \si{\MJ}$) occurrence rates around young nearby stars. The parameters are displayed in normal, bold, italic or bold and italic fonts when considering the full star sample, the early type AF stars, FK-type stars or KM-type stars, respectively.}
\label{tab_occur}
\begin{center}
\resizebox{1\textwidth}{!}{ 
\begin{tabular}{l c c c c c c l l}\\
\hline
\hline
\msini    & Orbital period  & $B-V$ & Search       & Detected    &  Missed       & GP occurrence rate    & \multicolumn{2}{c}{Confidence intervals}\\
interval  & interval        && completeness & GP systems  & GP systems    & upper limit                & $1\sigma$ & $2\sigma$        \\
(\Mjup)   & (day)           & & $C$  (\%)   &             &     \textbf{upper limit}          &                (\%)         & (\%)      &  (\%)            \\

\hline 
1-13        &  1-10      & all  & 94             &  0     &  0.1         & 1.3     & 0 - 4.1      & 0 - 7.0    \\ 
(GP)        &     &  $[-0.05:0.52[$  & {\bf 89}             &  {\bf 0}     &  {\bf 0.1}         & {\bf 4.9}     & {\bf 0 - 14.7}      & {\bf 0 - 24.0}    \\ 
       &     & $[0.52:1.33[$   & {\it 95}             &  {\it 0}     &  {\it 0.1}         & {\it2.0}     & {\it 0 - 6.4}      & {\it 0 - 10.8}    \\ 
       &     & $\geq 1.33$  & \textbf{\textit{98}}             & \textbf{\textit{0}}     &  \textbf{\textit{0.0}}         & \textbf{\textit{12.8}}     & \textbf{\textit{0 - 32.9}}      & \textbf{\textit{0 - 50.0}}    \\ 
\hline 
       &  1-100     & all   & 89             &  0     &  0.1         & 1.4     & 0 - 4.4      & 0 - 7.4    \\ 
        &       & $[-0.05:0.52[$ & {\bf 81}             &  {\bf 0}     &  {\bf 0.2}         & {\bf 5.4}     & {\bf 0 - 16.3}      & {\bf 0 - 26.7}    \\ 
       &      & $[0.52:1.33[$ & {\it 91}             &  {\it 0}     &  {\it 0.1}         & {\it2.1}     & {\it 0 - 6.7}      & {\it 0 - 11.3}    \\ 
       &      &  $\geq 1.33$  & \textbf{\textit{92}}             & \textbf{\textit{0.1}}     &  \textbf{\textit{0.0}}         & \textbf{\textit{13.6}}     & \textbf{\textit{0 - 35.0}}      & \textbf{\textit{0 - 53.1}}    \\ 
\hline 
      &  1-1000       & all & 80             &  0     &  0.3         & 1.5     & 0 - 4.9     & 0 - 8.3    \\ 
        &       & $[-0.05:0.52[$ & {\bf 71}             &  {\bf 0}     &  {\bf 0.4}         & {\bf 6.2}     & {\bf 0 - 18.6}      & {\bf 0 - 30.5}    \\ 
       &      & $[0.52:1.33[$ & {\it 82}             &  {\it 0}     &  {\it 0.2}         & {\it2.4}     & {\it 0 - 7.5}      & {\it 0 - 12.6}    \\ 
       &      &  $\geq 1.33$  & \textbf{\textit{86}}             & \textbf{\textit{0}}     &  \textbf{\textit{0.2}}         & \textbf{\textit{14.5}}     & \textbf{\textit{0 - 37.4}}      & \textbf{\textit{0 - 56.6}}    \\ 
\hline 
\hline 
13-80        &  1-10       & all & 99             &  0     &  0.0         & 1.2     & 0 - 3.9      & 0 - 6.6    \\ 
(BD)        &      &  $[-0.05:0.52[$ & {\bf 99}             &  {\bf 0}     &  {\bf 0.0}         & {\bf 4.4}     & {\bf 0 - 13.2}      & {\bf 0 - 21.6}    \\ 
       &     &  $[0.52:1.33[$ & {\it 99}             &  {\it 0}     &  {\it 0.0}         & {\it1.9}     & {\it 0 - 6.2}      & {\it 0 - 10.4}    \\ 
       &      &  $\geq 1.33$  & \textbf{\textit{99}}             & \textbf{\textit{0}}     &  \textbf{\textit{0.0}}         & \textbf{\textit{12.6}}     & \textbf{\textit{0 - 32.6}}      & \textbf{\textit{0 - 49.5}}    \\ 
\hline 
       &  1-100    &  all  & 97             &  0     &  0.0         & 1.2     & 0 - 4.0      & 0 - 6.8    \\ 
        &       & $[-0.05:0.52[$ & {\bf 95}             &  {\bf 0}     &  {\bf 0.0}         & {\bf 4.6}     & {\bf 0 - 13.7}      & {\bf 0 - 22.5}    \\ 
       &       & $[0.52:1.33[$ & {\it 97}             &  {\it 0}     &  {\it 0.0}         & {\it2.0}     & {\it 0 - 6.3}      & {\it 0 - 10.6}    \\ 
       &     & $\geq 1.33$    & \textbf{\textit{93}}             & \textbf{\textit{0}}     &  \textbf{\textit{0.1}}         & \textbf{\textit{13.4}}     & \textbf{\textit{0 - 34.7}}      & \textbf{\textit{0 - 52.6}}    \\ 
\hline 
        &  1-1000     & all   & 92             &  0     &  0.1         & 1.3     & 0 - 4.2      & 0 - 7.2    \\ 
        &       & $[-0.05:0.52[$ & {\bf 89}             &  {\bf 0}     &  {\bf 0.1}         & {\bf 4.9}     & {\bf 0 - 14.7}      & {\bf 0 - 24.0}    \\ 
       &       & $[0.52:1.33[$ & {\it 93}             &  {\it 0}     &  {\it 0.1}         & {\it2.1}     & {\it 0 - 6.6}      & {\it 0 - 11.1}    \\ 
       &     &  $\geq 1.33$  & \textbf{\textit{90}}             & \textbf{\textit{0}}     &  \textbf{\textit{0.1}}         & \textbf{\textit{13.8}}     & \textbf{\textit{0 - 35.7}}      & \textbf{\textit{0 - 54.1}}    \\ 
\hline 
\hline

\end{tabular}}
\end{center}
\end{table*}
\renewcommand{\arraystretch}{1.}

\subsection{Comparaison to surveys on main sequence stars}

No GP companion with periods lower than $\SI{1000}{\day}$ was detected in the present survey.
This non-detection is robust for HJ as $70\%$ of the star of the survey have detection limits lower than $\SI{1}{\MJ}$ for period lower than $\SI{10}{\day}$.
The completeness of the survey is over $71 \ \%$ for AF and FK stars and $n_{miss} < 0.4$ for AF and FK stars in the $1$- \SI{1000}{\day} domain (\emph{cf} \cref{tab_occur}).
However, we may have missed some planets with low masses and long period as only $40\%$ of the stars of the survey have detection limits lower than $\SI{5}{\MJ}$ between $100$ and $\SI{1000}{\day}$.

We statistically tested if our GP non-detection implies GP occurrence rates around young stars significantly lower than around older main sequence (MS) stars, using the p-value formalism.
The p-value is the probability to get the observed results given a null hypothesis, if it is lower than $10 \%$ then the null hypothesis can be rejected.
We tested the following null hypothesis : GP occurrence rates are identical around young  and main sequence stars in the same mass and period range.

We first applied this formalism in the  $1$- \SI{1000}{\day} domain with a survey that have a completeness similar to ours.
For AF MS stars the GP occurrence rate in this domain was estimated at $2.5^{+2.5}_{-0.5} \ \%$  \citep{Simon_X}. 
We detected $0$ companion out of $23$ stars, the corresponding p-value is $56^{+7}_{-21} \ \%$. 
The null hypothesis can not be rejected.

Then, we applied this formalism in the HJ ($P<\SI{10}{\day}$) domain.
For FK MS stars, the occurrence rate for HJ in this domain was estimated at : $0.46^{+0.3}_{-0.3} \ \%$ by  \cite{Cumming} and at $1.2 \pm 0.38 \ \%$ by  \cite{Wright_2012}.
We detected $0$ companion out of $52$ stars, the corresponding  p-value are respectively $79^{+13}_{-11} \ \%$ and  $53^{+12}_{-9} \ \%$.
The null hypothesis can not be rejected.

There is no evidence for a difference in occurrence rates of HJ between young and MS stars.

BD companions occurrence rates around MS stars is low : $f \leq 1.1^{+4.1}_{-1.1} \%$ for periods less than $\SI{1000}{\day}$ \citep{Simon_IX}, our non detection is not surprising. A bigger sample is needed to determine if BD occurrence rates around young and MS stars are different or not.

\section{Conclusions}
\label{conc}

We observed $89$ young A- to M-type stars over $3$ years or more with \harps \  to search for  GP and BD with periods less than \SI{1000}{\day}.
This survey allowed to detect close binaries around HD106906 and around HD131399 \citep{Lagrange_106b,Lagrange_131}.
We constrained the period of HD181321B and confirmed the RV trend in HD186704's RV. 
We also discovered a low-mass star companion to HIP36985.
No GP companion was detected in this survey. 
We obtain upper limits on the GP and BD occurrence rates, they are respectively $2_{-2}^{+3} \ \%$ and $1_{-1}^{+3}\ \%$ for periods of less than \SI{1000}{\day}.
Our comparison of these occurrence rates to those derived for MS stars \citep{Simon_X,Cumming,Wright_2012} indicates that there is no evidence for a difference in occurrence rates of HJ between young and MS stars.

The forthcoming analysis of our SOPHIE survey around young stars and of our on-going \harps \ survey on Sco-Cen stars will add $60$ and $80$ stars, respectively, in our analysis. 
This will allow for the derivation of more accurate occurrence rates and it will help in the search for the possible impact of system ages on occurrence rates.

\begin{acknowledgements}
   We acknowledge support from the French CNRS and from the Agence Nationale de la Recherche (ANR grant GIPSE ANR-14-CE33-0018). 
This work has been supported by a grant from Labex OSUG@2020 (Investissements d’avenir – ANR10 LABX56).
These results have made use of the SIMBAD database, operated at the CDS, Strasbourg, France.  ESO
SD acknowledges the support by INAF/Frontiera through the "Progetti
Premiali" funding scheme of the Italian Ministry of Education,
University, and Research
Based on observations collected at the European Southern Observatory under ESO programme(s) 060.A-9036(A),072.C-0488(E),072.C-0636(A),072.C-0636(B),073.C-0733(A),073.C-0733(C),073.C-0733(D),073.C-0733(E),074.C-0037(A),074.C-0364(A),075.C-0202(A),075.C-0234(A),075.C-0234(B),075.C-0689(A),075.C-0689(B),076.C-0010(A),076.C-0073(A),076.C-0279(A),076.C-0279(B),076.C-0279(C),077.C-0012(A),077.C-0295(A),077.C-0295(B),077.C-0295(C),077.C-0295(D),078.C-0209(A),078.C-0209(B),078.C-0751(A),078.C-0751(B),079.C-0170(A),079.C-0170(B),079.C-0657(C),080.C-0032(A),080.C-0032(B),080.C-0664(A),080.C-0712(A),081.C-0034(B),081.C-0802(C),082.C-0308(A),082.C-0357(A),082.C-0412(A),082.C-0427(A),082.C-0427(C),082.C-0718(B),083.C-0794(A),083.C-0794(B),083.C-0794(C),083.C-0794(D),084.C-1039(A),085.C-0019(A),087.C-0831(A),089.C-0732(A),089.C-0739(A),090.C-0421(A),091.C-0034(A),094.C-0946(A),098.C-0739(A),099.C-0205(A),0104.C-0418(A),1101.C-0557(A),183.C-0437(A),183.C-0972(A),184.C-0815(A),184.C-0815(B),184.C-0815(C),184.C-0815(E),184.C-0815(F),191.C-0873(A),192.C-0224(B),192.C-0224(C),192.C-0224(G),192.C-0224(H).
\end{acknowledgements}

\section{References}

\bibliography{yns}

\bibliographystyle{aa}

%
%


\begin{appendix}

\onecolumn
\section{Sample}

\begingroup
\renewcommand\arraystretch{1.3}
\begin{longtable}{lcccccccc}
\caption{Stars characteristics for the $89$ stars of our \harps~RV survey.
Spectral type (ST) and \bv~values are taken from the CDS database. 
The \vsini~values are taken from the CDS if present, or are computed with \safir~based on the CCF width.
The IR/D column report if either an IR excess is reported ($y$) or not ($n$) and if a disc has been imaged ($y$) or not ($n$) in the literature.
\label{tab_carac} }
 \\ \hline
Name & HIP & ST & \bv & Mass & Age  &\vsini &Rotation          &  IR/D\\ 
  HD/BD/CD & &&&(\Msun) & (\si{\mega\year}) &(\kms) & period ($\si{\day}$)   & \\
\hline \hline
\endfirsthead
\caption{Continued.}\\
 \\ \hline
Name & HIP & ST & \bv & Mass & Age  &\vsini &Rotation          &  IR/D\\ 
  HD/BD/CD & &&&(\Msun) & (\si{\mega\year})  &(\kms) & period ($\si{\day}$)   & \\
 \\ \hline
  \hline
\endhead
\hline
\endfoot
HD105 & 490 &  G0V & 0.600 & 1.1\footnote{\label{Vigan}\cite{Vigan}} & $45_{10}^{5}$\footnoteref{Vigan}  & 14.5 & - & y\footnote{\label{Meyer}\cite{Meyer_105}}/y\footnoteref{Meyer} \\ 
HD984 & 1134 &  F7V & 0.500 & 1.2\footnoteref{Vigan} & $42_{7}^{8}$\footnoteref{Vigan}  & 40.0 & - & -/n\footnote{\label{Weise}\cite{Weise}} \\ 
HD987 & 1113 &  G8V & 0.756 & 0.98\footnote{\label{Lagrange}\cite{Lagrange}} & $30_{15}^{15}$\footnoteref{Weise}  & 7.3 & $3.72\pm0.01$\footnote{\label{Messina}\cite{Messina}} & n\footnote{\label{Rebull}\cite{Rebull}}/y\footnoteref{Lagrange} \\ 
HD1466 & 1481 &  F8V & 0.540 & 1.2\footnoteref{Vigan} & $45_{10}^{5}$\footnoteref{Vigan}  & 21.0 & - & y\footnote{\label{Sierchio}\cite{Sierchio}}/y\footnote{\label{Mamajek_1466}\cite{Mamajek_1466}} \\ 
HD3221 & 2729 &  K4V & 1.226 & 0.9\footnoteref{Messina} & $45_{10}^{5}$\footnoteref{Vigan}  & 123.0 & $0.370\pm0.002$\footnoteref{Messina} & n\footnote{\label{Donaldson}\cite{Donaldson}}/y\footnoteref{Mamajek_1466} \\ 
HD6569 & 5191 &  K1V & 0.830 & 0.8\footnoteref{Vigan} & $149_{49}^{31}$\footnoteref{Vigan}  & 6.0 & $7.13\pm0.05$\footnoteref{Messina} & -/- \\ 
HD7661 & 5938 &  K0V & 0.770 & - & $300_{50}^{50}$\footnoteref{Weise}  & 7.5 & 7.46\footnote{\label{Wright_prot}\cite{Wright_prot}} & n\footnote{\label{Lawler}\cite{Lawler}}/n\footnoteref{Mamajek_1466} \\ 
HD10008 & 7576 &  K0V & 0.803 & 0.8\footnoteref{Vigan} & $250_{50}^{50}$\footnoteref{Vigan}  & 3.5 & $7.15\pm0.10$\footnote{\label{Folsom}\cite{Folsom}} & y\footnote{\label{Plavchan}\cite{Plavchan}}/y\footnote{\label{Patel}\cite{Patel}} \\ 
HD16765  & 12530 & F71V & 0.520 & - & - &  32.0 & - & -/- \\ 
HD17925 & 13402 &  K1V & 0.860 & 0.9\footnoteref{Vigan} & $150_{80}^{150}$\footnoteref{Vigan}  & 8.5 & 6.76\footnoteref{Wright_prot} & y\footnote{\label{Hallenbrand}\cite{Hallenbrand}}/- \\ 
HD18599  & 13754 & K2V & 0.880 & - & - &  4.3 & - & -/- \\ 
HD19668 & 14684 &  K0V & 0.780 & 0.9\footnoteref{Vigan} & $149_{49}^{31}$\footnoteref{Vigan}  & 6.5 & $5.46\pm0.08$\footnoteref{Messina} & y\footnote{\label{Carpenter}\cite{Carpenter}}/y\footnoteref{Weise} \\ 
HD24916  & 18512 & K4V & 1.152 & 0.7\footnote{\label{Ammler}\cite{Ammler}} & - &  4.5 & - & -/- \\ 
HD25457 & 18859 &  F6V & 0.500 & 1.2\footnoteref{Vigan} & $149_{49}^{31}$\footnoteref{Vigan}  & 22.0 & 3.13\footnoteref{Wright_prot} & y\footnote{\label{Zuckerman}\cite{Zuckerman}}/- \\ 
HD26923  & 19859 & G0IV & 0.560 & 1.07\footnoteref{Ammler} & - &  3.5 & - & -/- \\ 
HD29391 & 21547 &  F0IV & 0.277 & 1.75\footnote{\label{Simon}\cite{Simon}} & $37_{9}^{9}$\footnoteref{Montet}  & 80.0 & - & y\footnoteref{Patel}/- \\ 
HD30447 & 22226 &  F3V & 0.390 & 1.4\footnoteref{Vigan} & $42_{7}^{8}$\footnoteref{Vigan}  & 50.0 & - & y\footnote{\label{Chen}\cite{Chen}}/y\footnote{\label{Soummer}\cite{Soummer}} \\ 
HD35650  & 25283 & K6V & 1.311 & 0.7\footnoteref{Messina} & - &  4.0 & $9.34\pm0.08$\footnoteref{Messina} & y\footnoteref{Zuckerman}/y\footnote{\label{Choquet}\cite{Choquet}} \\ 
HD37572 & 26373 &  K0V & 1.094 & 0.9\footnoteref{Lagrange} & $149_{49}^{31}$\footnoteref{Vigan}  & 9.2 & $4.52\pm0.02$\footnoteref{Messina} & y\footnoteref{Zuckerman}/n\footnoteref{Lagrange} \\ 
HD39060 & 27321 & A6V & 0.170 & 1.64\footnoteref{Lagrange}  & $30$\footnoteref{Lagrange} &  130.0 & - & y\footnote{\label{Beta_pic_IR}\cite{Beta_pic_IR}}/y\footnote{\label{Beta_pic_disc}\cite{Beta_pic_disc}} \\ 
HD41593 & 28954 &  K0V & 0.825 & 1.01\footnoteref{Ammler} & $329_{93}^{93}$\footnoteref{Plavchan}  & 4.5 & - & -/- \\ 
HD43989 & 30030 &  G0V & 0.570 & 1.1\footnoteref{Vigan} & $45_{10}^{5}$\footnoteref{Vigan}  & 45.5 & 1.15\footnoteref{Wright_prot} & y\footnoteref{Carpenter}/y\footnoteref{Weise} \\ 
HD44627 & 30034 &  K1V & 0.805 & 0.9\footnoteref{Messina} & $30_{15}^{15}$\footnoteref{Weise}  & 11.5 & $3.85\pm0.01$\footnoteref{Messina} & -/n\footnoteref{Weise} \\ 
HD45081  & 29964 & K4V & 1.251 & 0.8\footnoteref{Messina} & - &  16.4 & $2.67\pm0.01$\footnoteref{Messina} & n\footnote{\label{McDonald}\cite{McDonald}}/y\footnoteref{Weise} \\ 
HD45270 & 30314 &  G1V & 0.602 & 1.11\footnoteref{Lagrange} & $149_{49}^{31}$\footnoteref{Vigan}  & 17.6 & - & y\footnoteref{Carpenter}/n\footnoteref{Lagrange} \\ 
HD59967 & 36515 &  G3V & 0.639 & 1.09\footnoteref{Plavchan} & $353_{58}^{58}$\footnoteref{Plavchan}  & 4.0 & - & y\footnoteref{Plavchan}/- \\ 
HD61005 & 36948 &  G8V & 0.740 & 1.0\footnoteref{Vigan} & $50_{10}^{20}$\footnoteref{Vigan}  & 8.5 & $5.04\pm0.03$\footnote{\label{Folsom}\cite{Folsom}} & y\footnote{\label{Meyer_excess}\cite{Meyer_excess}}/y\footnote{\label{Meyer_61005}\cite{Meyer_61005}} \\ 
HD63608 & 37923 &  K0V & 0.830 & 1.0\footnoteref{Vigan} & $250_{50}^{50}$\footnoteref{Vigan}  & 3.2 & - & -/- \\ 
HD77825 & 44526 &  K2V & 0.992 & 0.8\footnoteref{Vigan} & $350_{150}^{150}$\footnoteref{Vigan}  & 7.0 & 8.64\footnote{\label{Kiraga}\cite{Kiraga}} & -/- \\ 
HD82558  & 46816 & K1V & 0.870 & 0.8\footnote{\label{Kovari}\cite{Kovari}} & - &  29.0 & 1.70\footnoteref{Wright_prot} & -/- \\ 
HD89449  & 50564 & F6IV & 0.440 & - & - &  16.0 & - & -/- \\ 
HD90905 & 51386 &  F5V & 0.569 & 1.16\footnoteref{Lagrange} & $170_{70}^{180}$\footnoteref{Vigan}  & 10.0 & 2.60\footnoteref{Wright_prot} & y\footnoteref{Carpenter}/y\footnote{\label{Morales}\cite{Morales}} \\ 
HD92945 & 52462 &  K1V & 0.877 & 0.9\footnoteref{Vigan} & $170_{70}^{130}$\footnoteref{Vigan}  & 4.0 & - & n\footnote{\label{Chen_92945}\cite{Chen_92945}}/y\footnote{\label{Golimowski}\cite{Golimowski}} \\ 
HD95086 & 53524 &  A8III & 0.240 & 1.6\footnote{\label{Rameau_95086}\cite{Rameau_95086}} & $17_{4}^{4}$\footnote{\label{Meshkat_2013}\cite{Meshkat_2013}}  & 22.5 & - & y\footnote{\label{Rhee}\cite{Rhee}}/y\footnote{\label{Moor}\cite{Moor}} \\ 
\pagebreak 
HD95650  & 53985 & M2V & 1.477 & 0.59\footnote{\label{Montet}\cite{Montet_14}} & - &  11.0 & 14.80\footnote{\label{Kiraga}\cite{Kiraga}} & -/- \\ 
HD99211  & 55705 & A7V & 0.210 & - & - &  140.0 & - & y\footnote{\label{Mannings}\cite{Mannings}}/- \\ 
HD102458  & 57524 & G4V & 0.630 & 1.70\footnote{\label{Reza}\cite{reza}} & - &  31.0 & - & y\footnoteref{Rebull}/n\footnoteref{Lagrange} \\ 
HD103743  & 58241 & G4V & 0.670 & - & - &  9.0 & - & -/- \\ 
HD105690 & 59315 &  G5V & 0.661 & 1.02\footnoteref{Lagrange} & $8_{8}^{15}$\footnoteref{Weise}  & 8.5 & - & -/- \\ 
HD106906 & 59960 &  F5V & 0.400 & 1.5\footnote{\label{Bailey}\cite{Bailey}} & $13_{2}^{2}$\footnote{\label{Pecaut_2012}\cite{Pecaut_2012}}  & 55.0 & - & y\footnote{\label{Sierchio}\cite{Sierchio}}/y\footnote{\label{Kalas}\cite{Kalas}} \\ 
HD108767  & 60965 & K0V & -0.050 & 2.74\footnote{\label{Montesinos}\cite{Montesinos}} & - &  6.0 & - & -/- \\ 
HD116434 & 65426 & A2V &  0.200 & 1.96\footnote{\label{Chauvin}\cite{Chauvin}} & $14_{4}^{4}$\footnoteref{Chauvin} &  300.0 & - & n\footnoteref{Chauvin}/- \\ 
HD118100 & 66252 &  K5V & 1.180 & 0.7\footnoteref{Vigan} & $150_{50}^{50}$\footnoteref{Vigan}  & 0.5 & 3.96\footnoteref{Wright_prot} & -/- \\ 
HD131399 & 72940 &  A1V & 0.080 & 2.08\footnote{\label{Nielsen}\cite{Nielsen}} & $21_{3}^{4}$\footnoteref{Nielsen}  & 25.6 & - & -/- \\ 
HD141943 & - &  G2V & 0.505 & 1.09\footnoteref{Lagrange} & $30_{15}^{15}$\footnoteref{Weise}  & 40.0 & 2.2\footnoteref{Kiraga} & -/n\footnoteref{Lagrange} \\ 
HD146464  & 79958 & K3V & 1.033 & - & - &  17.0 & 2.329\footnote{\label{Koen}\cite{Koen}} & -/- \\ 
HD146624  & 79881 & A0V &  0.022 & 1.9\footnoteref{Lagrange} & $12$\footnoteref{Lagrange} &  39.0 & - & -/n\footnoteref{Lagrange} \\ 
HD152555 & 82688 &  F8V & 0.600 & - & $130_{20}^{20}$\footnote{\label{Brandt_2014}\cite{Brandt_2014}}  & 10.0 & 2.77\footnoteref{Wright_prot} & n\footnoteref{Zuckerman}/n\footnoteref{Weise} \\ 
HD159492  & 86305 & A5IV & 0.189 & - & - &  54.0 & - & -/y\footnote{\label{Morales}\cite{Morales}} \\ 
HD164249 & 88399 &  F6V & 0.460 & 1.3\footnoteref{Vigan} & $24_{5}^{5}$\footnoteref{Vigan}  & 15.0 & - & y\footnoteref{Chen}/- \\ 
HD169178  & - & K0V & 0.850 & - & - &  5.5 & - & -/- \\ 
HD171488 & 91043 &  G2V & 0.551 & 1.1\footnoteref{Vigan} & $45_{25}^{35}$\footnoteref{Vigan}  & 38.5 & $1.3371\pm0.0002$\footnote{\label{Strassmeier}\cite{Strassmeier}} & -/- \\ 
HD172555  & 92024 & A7V &  0.200 & 1.61\footnoteref{Lagrange} & $12$\footnote{\label{Messina_2017}\cite{Messina_2017}} &  116.5 & - & y\footnoteref{Lagrange}/y\footnoteref{Lagrange} \\ 
HD174429 & 92680 &  G9IV & 0.878 & 1.2\footnoteref{Vigan} & $24_{5}^{5}$\footnoteref{Vigan}  & 69.0 & $0.944\pm0.001$\footnoteref{Messina_2017} & y\footnoteref{Chen}/- \\ 
HD177171 & 93815 & F6V & 0.526 & 1.25\footnoteref{Reza} & $30$\footnoteref{Lagrange} &  300.0 & 4.737\footnoteref{Koen} & -/n\footnoteref{Lagrange} \\ 
HD181321  & 95149 & G2V & 0.630 & 0.89\footnote{\label{Fuhrmann}\cite{Fuhrmann}} & - &  13.0 & 5.7\footnote{\label{Olmedo}\cite{Olmedo}} & -/- \\ 
HD181327 & 95270 &  F6V & 0.460 & 1.36\footnoteref{Lagrange} & $24_{5}^{5}$\footnoteref{Vigan}  & 19.5 & - & y\footnoteref{Mannings}/y\footnote{\label{Schneider}\cite{Schneider}} \\ 
HD183414 & 96334 &  G3V & 0.650 & 1.04\footnoteref{Lagrange} & $150_{80}^{70}$\footnoteref{Vigan}  & 9.8 & 3.924\footnoteref{Koen} & -/- \\ 
HD186704  & 97255 & G0V &  0.583 & 1.4\footnote{\label{Bonavita}\cite{Bonavita}} & $100$\footnote{\label{Zuckerman_2013}\cite{Zuckerman_2013}} &  11.0 & 3.511\footnoteref{Kiraga} & y\footnoteref{McDonald}/- \\ 
HD188228  & 98495 & A0V &  -0.013 & 2.03\footnoteref{Lagrange} & $40$\footnoteref{Lagrange} &  85.0 & - & -/n\footnoteref{Lagrange} \\ 
HD189245 & 98470 &  F7V & 0.490 & 1.2\footnoteref{Vigan} & $150_{50}^{150}$\footnoteref{Vigan}  & 80.0 & $1.88\pm0.01$\footnote{\label{Desidera}\cite{Desidera}} & -/- \\ 
HD191089 & 99273 &  F5V & 0.440 & 1.3\footnoteref{Vigan} & $24_{5}^{5}$\footnoteref{Weise}  & 37.3 & $0.488\pm0.005$\footnoteref{Desidera} & y\footnoteref{Mannings}/y\footnote{\label{Churcher}\cite{Churcher}} \\ 
HD197481 & 102409 &  M1V & 1.423 & 0.6\footnoteref{Messina} & $21_{5}^{7}$\footnoteref{Brandt_2014}  & 9.6 & $4.84\pm0.02$\footnoteref{Messina} & y\footnote{\label{Rameau}\cite{Rameau}}/y\footnote{\label{Kalas}\cite{Kalas}} \\ 
HD197890 & 102626 &  K3V & 1.053 & 1.0\footnoteref{Vigan} & $45_{35}^{55}$\footnoteref{Vigan}  & 140.0 & 0.3804\footnoteref{Kiraga} & y\footnoteref{McDonald}/- \\ 
HD202917 & 105388 &  G7V & 0.650 & 0.9\footnoteref{Vigan} & $45_{10}^{5}$\footnoteref{Vigan}  & 15.4 & $3.36\pm0.01$\footnoteref{Messina} & y\footnoteref{Patel}/y\footnoteref{Soummer} \\ 
HD206860 & 107350 &  G0V & 0.580 & 1.1\footnoteref{Vigan} & $300_{100}^{700}$\footnoteref{Vigan}  & 12.0 & 4.86\footnoteref{Wright_prot} & y\footnoteref{Chen}/y\footnote{\label{Moro-Martin}\cite{Moro-Martin}} \\ 
HD206893 & 107412 &  F5V & 0.440 & 1.24\footnote{\label{Milli}\cite{Milli}} & $250_{200}^{450}$\footnote{\label{Delorme}\cite{Delorme}}  & 29.0 & - & y\footnoteref{Sierchio}/y\footnoteref{Milli} \\ 
HD207575 & 107947 &  F6V & 0.490 & 1.24\footnoteref{Lagrange} & $45_{15}^{5}$\footnoteref{Vigan}  & 30.0 & - & y\footnoteref{Zuckerman}/n\footnoteref{Lagrange} \\ 
HD213845 & 111449 &  F7V & 0.440 & 1.4\footnoteref{Vigan} & $250_{50}^{750}$\footnoteref{Vigan}  & 34.0 & - & y\footnote{\label{Beichman}\cite{Beichman}}/- \\ 
HD215641 & 112491 &  G8V & 0.760 & - & $440_{40}^{40}$\footnoteref{Brandt_2014}  & 3.6 & - & -/- \\ 
HD216956 & 113368 & A3V & 0.090 & 1.92\footnote{\label{Mamajek}\cite{Mamajek}} & $440_{40}^{40}$\footnoteref{Mamajek}  & 90.0 & - & y\footnote{\label{Backman}\cite{Backman}}/y\footnote{\label{Holland}\cite{Holland}} \\ 
\pagebreak 
HD217343  & 113579 & G5V &  0.640 & 1.05\footnoteref{Vigan} & $70$\footnoteref{Lagrange} &  12.4 & - & -/n\footnoteref{Lagrange} \\ 
HD217987 & 114046 & M2V & 1.480 & 0.47\footnoteref{Montet}& $100-10000$\footnote{\label{Delorme_2012}\cite{Delorme_2012}} &   2.5 & - & -/- \\ 
HD218396  & 114189 & A5V &  0.257 & 1.46\footnoteref{Vigan} & $30$\footnoteref{Lagrange} &  49.0 & - & y\footnote{\label{Zuckerman_8799}\cite{Zuckerman_8799}}/y\footnoteref{Rhee} \\ 
HD218860  & 114530 & G8V & 0.738 & 1.0\footnoteref{Messina} & - &  6.6 & $5.17\pm0.02$\footnoteref{Messina} & y\footnoteref{Zuckerman}/- \\ 
HD221575  & 116258 & K2V & 0.930 & - & - &  3.6 & - & -/- \\ 
HD223340  & - & K1V & 0.820 & - & - &  7.0 & - & -/- \\ 
HD224228 & 118008 &  K2V & 0.985 & 0.86\footnoteref{Lagrange} & $149_{49}^{31}$\footnoteref{Vigan}  & 2.6 & - & y\footnoteref{Zuckerman}/n\footnoteref{Lagrange} \\ 
- & 6276 &  G9V & 0.800 & 0.9\footnoteref{Vigan} & $149_{49}^{31}$\footnoteref{Vigan}  & 10.0 & 6.40\footnoteref{Wright_prot} & y\footnoteref{Sierchio}/y\footnoteref{Weise} \\ 
-  & 116384 & K7V & 1.347 & - & - &  4.5 & - & -/- \\ 
-  & 17157 & K7V & 1.300 & - & - &  5.0 & - & -/- \\ 
- & 23309 &  M0V & 1.471 & 0.55\footnoteref{Reza} & $10_{3}^{3}$\footnoteref{Weise}  & 5.8 & $8.60\pm0.07$\footnoteref{Messina} & -/n\footnoteref{Rebull} \\ 
-  & 31878 & K7V & 1.296 & 0.643\footnote{\label{Lindgren}\cite{Lindgren}} & - &  5.0 & $9.06\pm0.08$\footnoteref{Messina} & -/- \\ 
- & 36985 &  M2V & 1.476 & 0.621\footnote{\label{Carmenes}\cite{Carmenes}} & $260_{260}^{420}$\footnote{\label{Poveda}\cite{Poveda}}  & 5.0 & 12.16\footnoteref{Kiraga} & -/- \\ 
-  & 44722 & K7V & 1.450 & 0.638\footnoteref{Lindgren} & - &  9.0 & - & -/- \\ 
-  & 46634 & G5V & 0.860 & - & - &  6.0 & $3.05\pm0.03$\footnoteref{Messina} & -/- \\ 
- & 51317 &  M2V & 1.501 & - & $130_{20}^{40}$\footnoteref{Brandt_2014}  & 2.6 & - & -/- \\ 
BD$+$20 2465  & - & M5V & 1.300 & 0.42\footnote{\label{Morin}\cite{Morin}} & - &  3.8 & 2.60\footnoteref{Wright_prot} & -/- \\ 
CD$-$46 1064 & - &  K3V & 1.048 & 0.8\footnoteref{Vigan} & $45_{10}^{5}$\footnoteref{Vigan}  & 1.0 & $3.74\pm0.04$\footnoteref{Messina} & -/- \\ 
\end{longtable}
\endgroup

\begin{landscape}
\begingroup
\begin{longtable}{lcc|cccccccccccc}
\caption{Results for the $89$ stars of our \harps~RV survey.
Spectral type (ST) are taken from the CDS database.
The survey results include the time baseline (TBL), the number of computed spectra $N_{\rm m}$, the amplitude corresponding to the difference between the maximum and the minimum of the RV({\it A}), rms and mean uncertainty $<${\it U}$>$ on the RV and BVS measurements, the RV-BVS correlation factor (slope of the best linear fit), the mean FWHM ($<$FWHM$>$), and the mean \rhk \ ($<$\rhk$>$).
V stands for the dominant source of RV variations, with A for stellar activity (spots), P for pulsations, B for binary and, T for long-term trend. 
\label{tab_result}}\\ 
\\ \hline
\multicolumn{3}{c|}{Stellar characteristics} & \multicolumn{12}{c}{Survey results.}
 \\ \hline
Name & HIP & ST &  TBL & $N_{\rm m}$   & \multicolumn{3}{c}{RV}        & \multicolumn{3}{c}{BVS}      & RV-    & $<$FWHM$>$ & $<$\rhk$>$  & V   \\ 
 HD/BD/CD & &  & & & \multicolumn{3}{c}{\raisebox{.5\baselineskip}{$\overbrace{\hspace{2.7cm}}$}} & \multicolumn{3}{c}{\raisebox{.5\baselineskip}{$\overbrace{\hspace{2.7cm}}$}} & BVS & & &\\
&& &  &  & {\it A} &rms  & $<${\it U}$>$ & {\it A} & rms &$<${\it U}$>$ & corr.    &                 &  &  \\
&&&&    & \multicolumn{3}{c}{\raisebox{.5\baselineskip}{$\underbrace{\hspace{2.7cm}}$}} & \multicolumn{3}{c}{\raisebox{.5\baselineskip}{$\underbrace{\hspace{2.7cm}}$}} & &  &\\
&& (day)&            && \multicolumn{3}{c}{($\si{\meter\per\second}$)}      & \multicolumn{3}{c}{($\si{\meter\per\second}$)}     &   &      & ($\si{\kilo\meter\per\second}$)    &\\ \hline
\endfirsthead
\caption{Continued.}\\
 \\ \hline
\multicolumn{3}{c|}{Stellar characteristics} & \multicolumn{12}{c}{Survey results.}
 \\ \hline
Name & HIP & ST &  TBL & $N_{\rm m}$   & \multicolumn{3}{c}{RV}        & \multicolumn{3}{c}{BVS}      & RV-    & $<$FWHM$>$ & $<$\rhk$>$ & V   \\ 
 HD/BD/CD & &  & & & \multicolumn{3}{c}{\raisebox{.5\baselineskip}{$\overbrace{\hspace{2.7cm}}$}} & \multicolumn{3}{c}{\raisebox{.5\baselineskip}{$\overbrace{\hspace{2.7cm}}$}} & BVS & & &\\
&& &  &  & {\it A} &rms  & $<${\it U}$>$ & {\it A} & rms &$<${\it U}$>$ & corr.    &                 &    &\\
&&&&    & \multicolumn{3}{c}{\raisebox{.5\baselineskip}{$\underbrace{\hspace{2.7cm}}$}} & \multicolumn{3}{c}{\raisebox{.5\baselineskip}{$\underbrace{\hspace{2.7cm}}$}} & & &  \\
&& (day)&            && \multicolumn{3}{c}{($\si{\meter\per\second}$)}      & \multicolumn{3}{c}{($\si{\meter\per\second}$)}     &   &      &  ($\si{\kilo\meter\per\second}$)  &\\ \hline
\endhead
\hline
\endfoot
HD105 & 490 & G0V & 4606 & 36 & 236.8 & 61.1 & 4.2 & 310.7 & 9.8 & 72.9 & -0.67 & 22.6 & -4.316 & A \\ 
HD984 & 1134 & F7V & 867 & 21 & 301.9 & 84.6 & 16.4 & 571.8 & 38.2 & 137.4 & -0.46 & 59.7 & -4.368 & A \\ 
HD987 & 1113 & G8V & 2621 & 19 & 502.6 & 116.8 & 2.5 & 393.6 & 6.4 & 103.1 & -1.08 & 13.0 & -4.088 & A \\ 
HD1466 & 1481 & F8V & 4400 & 19 & 135.8 & 39.6 & 7.4 & 189.8 & 17.6 & 55.9 & -0.66 & 32.3 & -4.344 & A \\ 
HD3221 & 2729 & K4V & 4014 & 5 & 4793.4 & 1794.3 & 68.6 & 3115.1 & 172.5 & 1447.8 & -0.54 & 203.6 & -4.474 & - \\ 
HD6569 & 5191 & K1V & 8 & 4 & 24.0 & 11.4 & 1.5 & 17.8 & 3.8 & 7.5 & -1.49 & 9.6 & -4.228 & A \\ 
HD7661 & 5938 & K0V & 1525 & 29 & 96.0 & 28.1 & 1.2 & 50.5 & 3.3 & 12.5 & -1.42 & 9.2 & -4.325 & A \\ 
HD10008 & 7576 & K0V & 4021 & 17 & 20.5 & 6.2 & 1.0 & 17.5 & 2.7 & 5.2 & 0.44 & 7.5 & -4.358 & A \\ 
HD16765 & 12530 & F71V & 926 & 27 & 173.0 & 47.2 & 11.0 & 402.9 & 27.9 & 93.1 & -0.42 & 48.5 & -4.475 & A \\ 
HD17925 & 13402 & K1V & 1470 & 40 & 113.6 & 30.5 & 1.1 & 58.7 & 3.1 & 17.4 & -0.87 & 10.0 & -4.310 & A \\ 
HD18599 & 13754 & K2V & 1055 & 16 & 115.9 & 38.9 & 1.5 & 54.3 & 3.9 & 17.6 & -1.46 & 9.8 & -4.310 & A \\ 
HD19668 & 14684 & K0V & 4402 & 20 & 143.1 & 33.2 & 1.8 & 91.9 & 4.7 & 26.0 & -0.95 & 11.8 & -4.284 & A \\ 
HD24916 & 18512 & K4V & 792 & 22 & 24.3 & 7.4 & 0.8 & 36.7 & 2.2 & 11.6 & 0.46 & 7.8 & -4.536 & A \\ 
HD25457 & 18859 & F6V & 4089 & 78 & 187.2 & 49.5 & 3.3 & 345.4 & 7.7 & 60.2 & -0.66 & 28.2 & -4.355 & A \\ 
HD26923 & 19859 & G0IV & 4010 & 47 & 50.2 & 12.7 & 1.1 & 24.7 & 3.1 & 5.6 & 1.12 & 8.0 & -4.487 & A \\ 
HD29391 & 21547 & F0IV & 3996 & 81 & 1476.1 & 326.2 & 28.4 & 2564.3 & 72.1 & 487.8 & - & 101.4 &  & P \\ 
HD30447 & 22226 & F3V & 792 & 19 & 194.1 & 47.2 & 47.7 & 7048.7 & 106.6 & 1728.3 & - & 120.7 & -4.447 & P \\ 
HD35650 & 25283 & K6V & 1195 & 13 & 48.3 & 15.3 & 1.2 & 43.6 & 3.0 & 12.3 & -0.67 & 9.4 & -4.359 & A \\ 
HD37572 & 26373 & K0V & 2826 & 34 & 347.1 & 75.4 & 1.7 & 223.6 & 4.4 & 52.8 & -1.28 & 15.6 & -4.165 & A \\ 
HD39060 & 27321 & A6V & 3702 & 5108 & 755.5 & 341.4 & 58.9 &  & &  & - &  &  & P \\ 
HD41593 & 28954 & K0V & 1194 & 15 & 69.5 & 19.8 & 1.1 & 59.4 & 2.9 & 16.7 & -0.66 & 9.1 & -4.370 & A \\ 
HD43989 & 30030 & G0V & 4433 & 17 & 819.0 & 238.6 & 15.6 & 743.7 & 34.3 & 197.6 & -0.83 & 64.2 & -4.209 & A \\ 
HD44627 & 30034 & K1V & 4892 & 23 & 670.6 & 177.7 & 3.2 & 428.6 & 8.1 & 110.9 & - & 19.1 & -4.018 & A \\ 
HD45081 & 29964 & K4V & 4020 & 17 & 852.4 & 252.1 & 5.0 & 911.0 & 13.0 & 246.9 & - & 24.8 & -3.887 & A \\ 
HD45270 & 30314 & G1V & 2827 & 19 & 137.4 & 51.8 & 3.1 & 193.3 & 8.4 & 59.2 & -0.76 & 27.6 & -4.335 & A \\ 
HD59967 & 36515 & G3V & 1065 & 25 & 65.3 & 19.7 & 1.4 & 41.6 & 3.6 & 10.5 & -0.99 & 9.0 & -4.328 & A \\ 
HD61005 & 36948 & G8V & 2369 & 33 & 186.5 & 48.8 & 2.5 & 127.4 & 6.3 & 32.7 & -1.16 & 14.5 & -4.261 & A \\ 
HD63608 & 37923 & K0V & 1069 & 36 & 80.7 & 22.2 & 1.1 & 49.1 & 3.0 & 11.7 & -0.63 & 8.0 & -4.323 & A \\ 
HD77825 & 44526 & K2V & 283 & 6 & 47.0 & 18.3 & 1.4 & 46.5 & 3.4 & 14.9 & -1.09 & 9.6 & -4.314 & A \\ 
HD82558 & 46816 & K1V & 1092 & 24 & 296.0 & 79.7 & 7.1 & 475.6 & 17.5 & 124.1 & -0.43 & 42.0 & -3.990 & A \\ 
HD89449 & 50564 & F6IV & 1227 & 28 & 129.9 & 41.4 & 4.3 & 351.6 & 9.1 & 113.6 & -0.26 & 25.3 & -4.971 & A \\ 
HD90905 & 51386 & F5V & 2551 & 24 & 125.6 & 40.8 & 1.9 & 187.7 & 5.0 & 49.0 & -0.74 & 15.0 & -4.345 & A \\ 
HD92945 & 52462 & K1V & 4533 & 38 & 182.3 & 40.5 & 1.5 & 86.2 & 3.9 & 23.8 & -1.37 & 10.3 & -4.261 & A \\ 
HD95086 & 53524 & A8III & 1532 & 103 & 1279.0 & 268.8 & 14.8 & 2594.0 & 36.3 & 511.1 & - & 45.4 &  & P \\ 
HD95650 & 53985 & M2V & 4046 & 12 & 23.8 & 7.6 & 3.1 & 46.1 & 7.3 & 13.6 & - & 8.0 & -4.330 & - \\ 
HD99211 & 55705 & A7V & 3068 & 112 & 818.0 & 149.4 & 86.8 & 49704.0 & 214.7 & 5222.3 & - & 183.4 &  & P \\ 
HD102458 & 57524 & G4V & 2964 & 26 & 988.5 & 307.7 & 10.0 & 1358.4 & 23.8 & 431.7 & - & 43.6 & -3.994 & A \\ 
HD103743 & 58241 & G4V & 1065 & 30 & 102.2 & 32.1 & 2.6 & 141.6 & 6.6 & 35.1 & -0.79 & 14.9 & -4.308 & A \\ 
HD105690 & 59315 & G5V & 2975 & 133 & 269.4 & 61.3 & 2.3 & 193.7 & 5.4 & 44.0 & -0.91 & 15.1 & -4.276 & A \\ 
HD106906 & 59960 & F5V & 1230 & 46 & 5290.4 & 1285.7 & 32.9 & 9949.8 & 83.3 & 2272.4 & - & 74.0 & -4.574 & B \\ 
HD108767 & 60965 & K0V & 1201 & 18 & 47.9 & 12.4 & 0.7 & 25.3 & 2.0 & 7.7 & 0.21 & 8.2 & -4.381 & A \\ 
HD116434 & 65426 & A2V & 439 & 58 & 6273.2 & 1223.2 & 1235.6 &  & &  & - &  &  & P \\ 
HD118100 & 66252 & K5V & 720 & 10 & 534.7 & 155.6 & 6.4 & 139.5 & 15.6 & 43.0 & -3.48 & 16.5 & -4.108 & A \\ 
HD131399 & 72940 & A1V & 189 & 87 & 19478.2 & 6381.7 & 16.4 & 1287.6 & 39.3 & 237.0 & - & 43.1 &  & B \\ 
HD141943 & - & G2V & 2648 & 58 & 861.4 & 222.8 & 10.3 & 1169.7 & 25.3 & 269.6 & -0.66 & 53.0 & -4.025 & A \\ 
HD146464 & 79958 & K3V & 51 & 5 & 872.1 & 352.1 & 5.0 & 546.2 & 12.3 & 220.6 & -1.55 & 25.8 & -4.163 & A \\ 
HD146624 & 79881 & A0V & 4762 & 335 & 232.7 & 39.8 & 29.2 & 70472.6 & 72.9 & 6163.4 & - & 65.1 &  & P \\ 
HD152555 & 82688 & F8V & 1135 & 22 & 225.0 & 64.8 & 6.1 & 283.2 & 15.9 & 69.9 & -0.74 & 26.3 & -4.294 & A \\ 
HD159492 & 86305 & A5IV & 4751 & 90 & 612.4 & 124.9 & 20.8 & 2413.2 & 52.1 & 359.7 & - & 79.4 &  & P \\ 
HD164249 & 88399 & F6V & 1113 & 25 & 56.9 & 14.6 & 6.9 & 134.2 & 17.0 & 36.7 & -0.24 & 29.8 & -4.765 & A \\ 
HD169178 & - & K0V & 1123 & 19 & 354.7 & 135.7 & 1.8 & 85.6 & 4.7 & 25.0 & - & 10.7 & -4.234 & A \\ 
HD171488 & 91043 & G2V & 1111 & 18 & 2190.4 & 717.7 & 13.1 & 2249.5 & 33.2 & 680.4 & -1.0 & 58.2 & -3.988 & A \\ 
HD172555 & 92024 & A7V & 2975 & 262 & 1883.8 & 329.6 & 70.6 & 94415.3 & 185.8 & 6148.5 & - & 172.4 &  & P \\ 
HD174429 & 92680 & G9IV & 4572 & 42 & 3362.3 & 970.2 & 26.0 & 1167.7 & 60.0 & 320.7 & -1.77 & 109.9 &  & A \\ 
HD177171 & 93815 & F6V & 144 & 21 & 21233.2 & 5931.1 & 53.4 &  & &  & - &  & -4.047 & B \\ 
HD181321 & 95149 & G2V & 3757 & 28 & 2610.2 & 683.2 & 3.9 & 244.0 & 9.6 & 59.1 & - & 19.5 & -4.243 & B \\ 
HD181327 & 95270 & F6V & 3496 & 56 & 63.1 & 16.3 & 3.8 & 156.8 & 9.0 & 30.6 & -0.26 & 28.8 & -4.577 & A \\ 
HD183414 & 96334 & G3V & 3097 & 68 & 247.9 & 63.6 & 2.4 & 214.0 & 6.2 & 58.3 & -1.01 & 16.0 & -4.236 & A \\ 
HD186704 & 97255 & G0V & 451 & 4 & 345.8 & 168.8 & 4.7 & 77.1 & 11.9 & 29.3 & - & 22.3 & -4.308 & T \\ 
HD188228 & 98495 & A0V & 4315 & 194 & 1398.7 & 319.2 & 150.2 & 211414.9 & 368.1 & 18294.0 & - & 148.0 &  & P \\ 
HD189245 & 98470 & F7V & 4285 & 46 & 646.2 & 119.4 & 21.0 & 5149.4 & 43.1 & 757.5 & - & 104.2 & -4.304 & P \\ 
HD191089 & 99273 & F5V & 1025 & 26 & 107.1 & 22.8 & 15.7 & 300.0 & 37.3 & 78.4 & - & 57.0 & -4.526 & P \\ 
HD197481 & 102409 & M1V & 5619 & 55 & 667.6 & 144.0 & 3.2 & 470.1 & 8.0 & 101.2 & -1.37 & 16.2 & -4.289 & A \\ 
HD197890 & 102626 & K3V & 13 & 3 & 922.1 & 376.6 & 85.2 & 3975.6 & 217.1 & 1713.8 & - & 216.2 & -4.277 & - \\ 
HD202917 & 105388 & G7V & 4438 & 20 & 436.1 & 124.6 & 5.0 & 438.1 & 12.4 & 134.8 & -0.84 & 23.0 & -3.984 & A \\ 
HD206860 & 107350 & G0V & 1332 & 22 & 122.1 & 35.8 & 2.8 & 180.9 & 6.7 & 51.4 & - & 16.1 & -4.387 & A \\ 
HD206893 & 107412 & F5V & 542 & 15 & 185.6 & 45.5 & 10.8 & 361.3 & 21.3 & 68.2 & - & 50.7 & -4.630 & P \\ 
HD207575 & 107947 & F6V & 2391 & 39 & 227.6 & 49.3 & 9.5 & 389.5 & 19.9 & 97.6 & - & 49.1 & -4.332 & P \\ 
HD213845 & 111449 & F7V & 4436 & 79 & 220.4 & 44.2 & 9.6 & 454.3 & 25.8 & 82.6 & - & 52.3 & -4.703 & P \\ 
HD215641 & 112491 & G8V & 1639 & 78 & 101.5 & 20.5 & 1.2 & 84.4 & 3.4 & 12.2 & 0.22 & 8.3 & -4.460 & A \\ 
HD216956 & 113368 & A3V & 4929 & 834 & 382.4 & 44.3 & 27.0 &  & &  & - &  &  & P \\ 
HD217343 & 113579 & G5V & 2727 & 26 & 276.6 & 96.9 & 2.7 & 298.7 & 6.6 & 85.6 & -1.08 & 19.3 & -4.215 & A \\ 
HD217987 & 114046 & M2V & 5109 & 130 & 63.8 & 18.5 & 0.9 & 39.5 & 2.3 & 8.2 & 0.95 & 5.8 & -4.899 & A \\ 
HD218396 & 114189 & A5V & 2727 & 124 & 3616.3 & 924.5 & 38.5 & 7376.0 & 95.5 & 2191.3 & - & 62.3 &  & P \\ 
HD218860 & 114530 & G8V & 1137 & 18 & 137.1 & 36.2 & 2.1 & 82.8 & 5.4 & 18.6 & -1.3 & 11.7 & -4.240 & A \\ 
HD221575 & 116258 & K2V & 869 & 16 & 45.7 & 12.2 & 1.4 & 33.4 & 3.7 & 12.0 & -0.54 & 8.8 & -4.438 & A \\ 
HD223340 & - & K1V & 868 & 10 & 54.4 & 17.3 & 1.7 & 42.2 & 4.4 & 10.7 & -1.34 & 9.8 & -4.337 & A \\ 
HD224228 & 118008 & K2V & 2815 & 31 & 35.4 & 8.8 & 1.0 & 31.5 & 2.8 & 7.5 & 0.32 & 7.8 & -4.356 & A \\ 
- & 6276 & G9V & 1201 & 20 & 95.6 & 27.3 & 1.6 & 91.0 & 4.2 & 27.6 & -0.65 & 9.4 & -4.321 & A \\ 
- & 116384 & K7V & 733 & 8 & 104.3 & 41.2 & 1.7 & 60.2 & 4.5 & 19.2 & -1.18 & 8.6 & -4.455 & A \\ 
- & 17157 & K7V & 1457 & 6 & 67.5 & 22.3 & 1.4 & 37.1 & 3.6 & 14.9 & - & 8.1 & -4.577 & A \\ 
- & 23309 & M0V & 4030 & 16 & 245.4 & 64.1 & 2.3 & 116.1 & 5.9 & 33.0 & -1.83 & 11.7 & -3.816 & A \\ 
- & 31878 & K7V & 435 & 11 & 84.3 & 31.6 & 1.6 & 48.3 & 4.1 & 16.1 & -1.86 & 8.5 & -4.276 & A \\ 
- & 36985 & M2V & 789 & 20 & 670.1 & 247.4 & 2.1 & 58.8 & 5.4 & 14.7 & - & 8.1 & -4.316 & B \\ 
- & 44722 & K7V & 90 & 4 & 8.0 & 3.4 & 1.5 & 10.7 & 3.8 & 4.0 & -0.49 & 7.6 & -4.550 & A \\ 
- & 46634 & G5V & 281 & 3 & 38.2 & 16.1 & 1.4 & 32.6 & 3.5 & 13.9 & -1.0 & 9.8 & -4.321 & A \\ 
- & 51317 & M2V & 4541 & 139 & 17.7 & 3.6 & 1.2 & 24.5 & 3.2 & 3.5 & -0.38 & 4.8 & -5.032 & A \\ 
BD$+$20 2465 & - & M5V & 3983 & 40 & 63.5 & 16.1 & 1.2 & 33.6 & 3.1 & 6.5 & -0.61 & 6.7 & -4.026 & A \\ 
CD$-$46 1064 & - & K3V & 1185 & 12 & 477.8 & 134.9 & 8.7 & 211.2 & 21.6 & 68.7 & -1.53 & 17.4 & -4.204 & A \\ 
\end{longtable}
\endgroup
\end{landscape}

\clearpage

\twocolumn

\section{RV correction from the RV-BVS correlation}

\label{appendix_corr_BVS}

Solar to late-type stars present spots on their surfaces. These spots induce a quasi-periodic variations in the line profiles, which cause a signal in the RV and in the BVS. 
When the lines are resolved, there is at first order a correlation between the RV and the BVS (see \cite{Desort} for a detailed analysis)

To correct the RV signal induced by spots, we use \cite{Melo_BVS_RV_corr} method. It consists in correcting the RV from the linear regression of the RV vs BVS dataset. The new RV at a given date $t$ is :

\begin{displaymath}
RV_{corr}(t) = RV(t) -(a*BVS(t)+b)
\end{displaymath}

where $a$ and $b$ are the slope and the intercept of the best linear fit of the RV vs BVS dataset.

\begin{figure}[t!]
  \centering
\begin{subfigure}[t]{0.32\textwidth}
\includegraphics[width=1\hsize,valign=m]{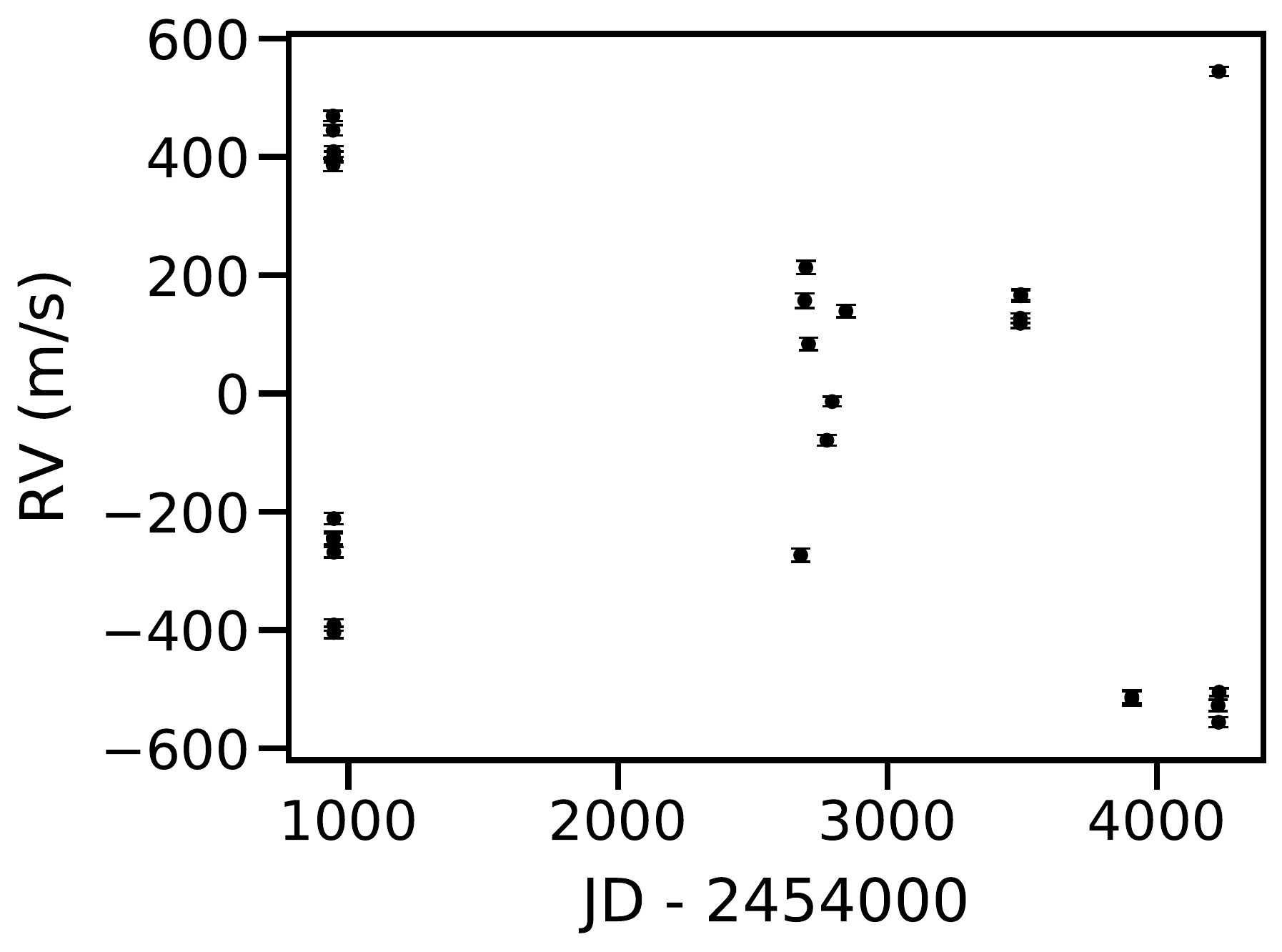}
\caption{\label{RV_102}}
\end{subfigure}
\begin{subfigure}[t]{0.32\textwidth}
\includegraphics[width=1\hsize,valign=m]{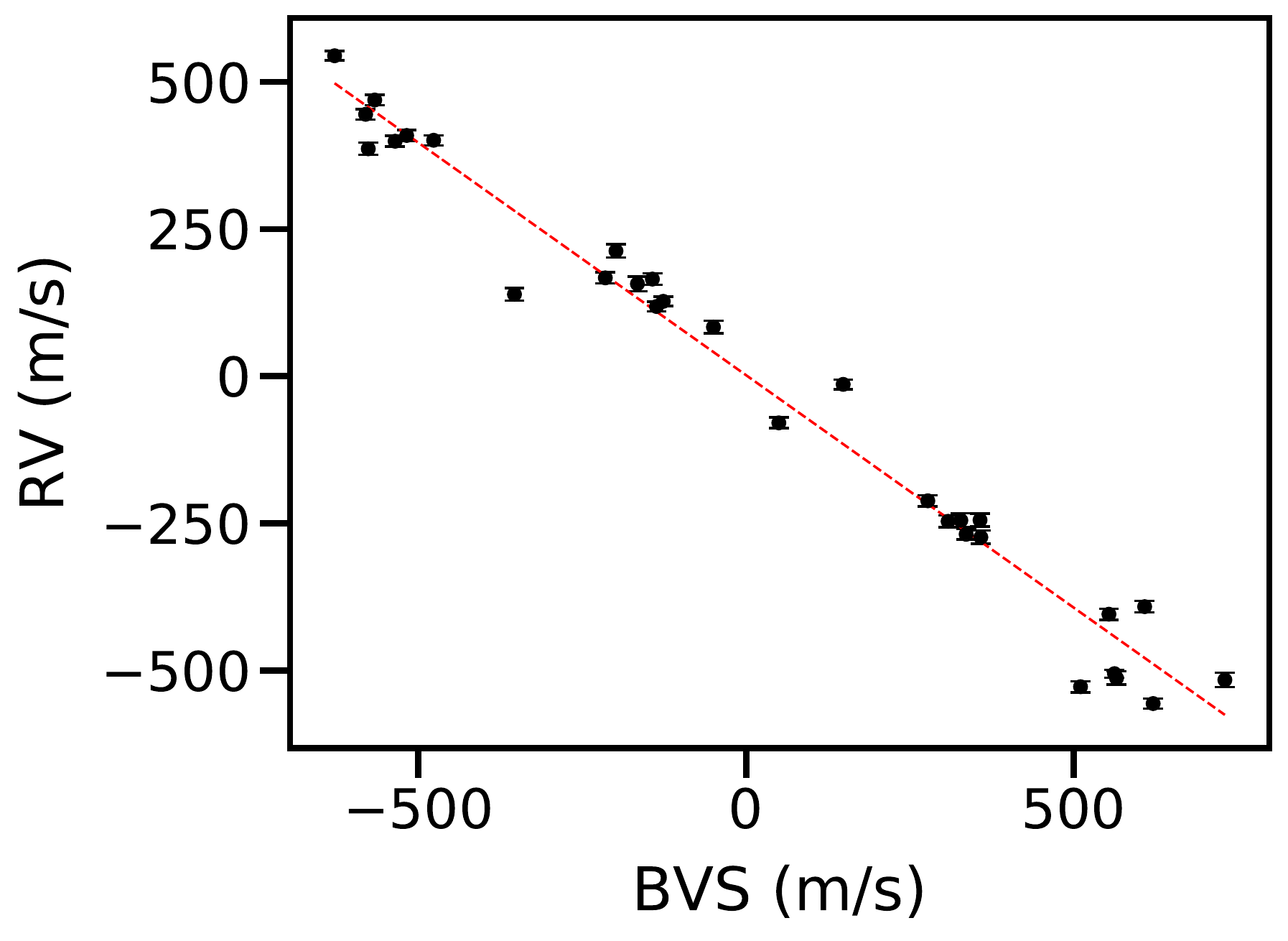}
\caption{\label{RV_BVS_102}}
\end{subfigure}
\begin{subfigure}[t]{0.32\textwidth}
\includegraphics[width=1\hsize,valign=m]{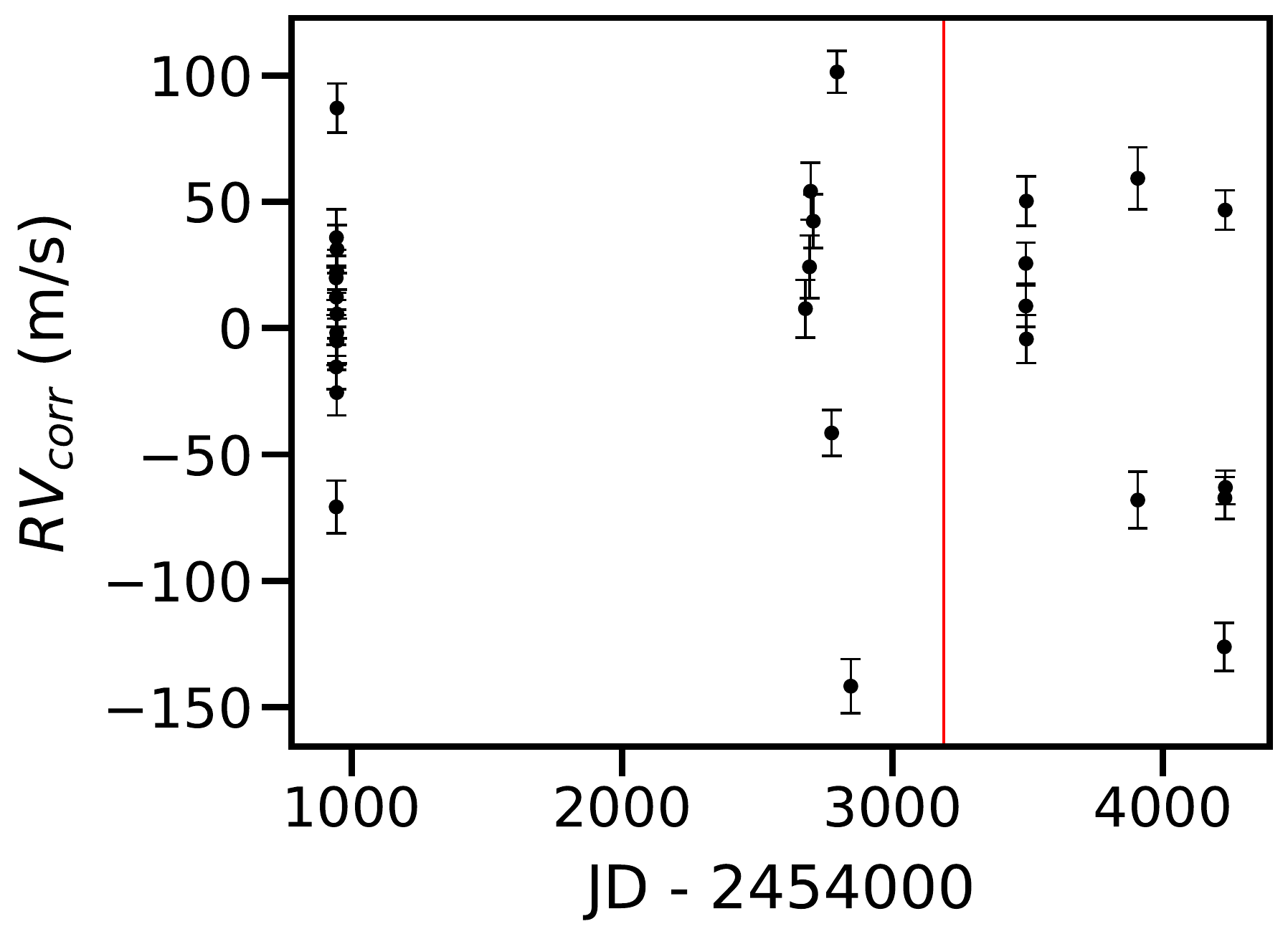}
\caption{\label{RV_corr_102}}
\end{subfigure}
\caption{HD102458 RV jitter correction.   \subref{RV_102}) RV time variations.
 \subref{RV_BVS_102}) RV vs BVS. The best linear fit is presentend in red dashed line.
 \subref{RV_corr_102}) RV corrected from the RV-BVS correlation. \harps \  fiber change is shown  with a vertical red line.
}
       \label{102}
\end{figure}

We present in \cref{102} an example of such correction for a star with $v\sin{i} = \SI{31}{\kilo\meter\per\second}$. 
The mean RV rms of  $\SI{300}{\meter\per\second}$ are reduced to  $\SI{60}{\meter\per\second}$ after correction.
This method was used in precedent works that use \safir \  \citep{Lagrange,Simon_IX,Simon_X}.

\begin{figure*}[h!]
  \centering
\begin{subfigure}[t]{0.32\textwidth}
\includegraphics[width=1\hsize,valign=m]{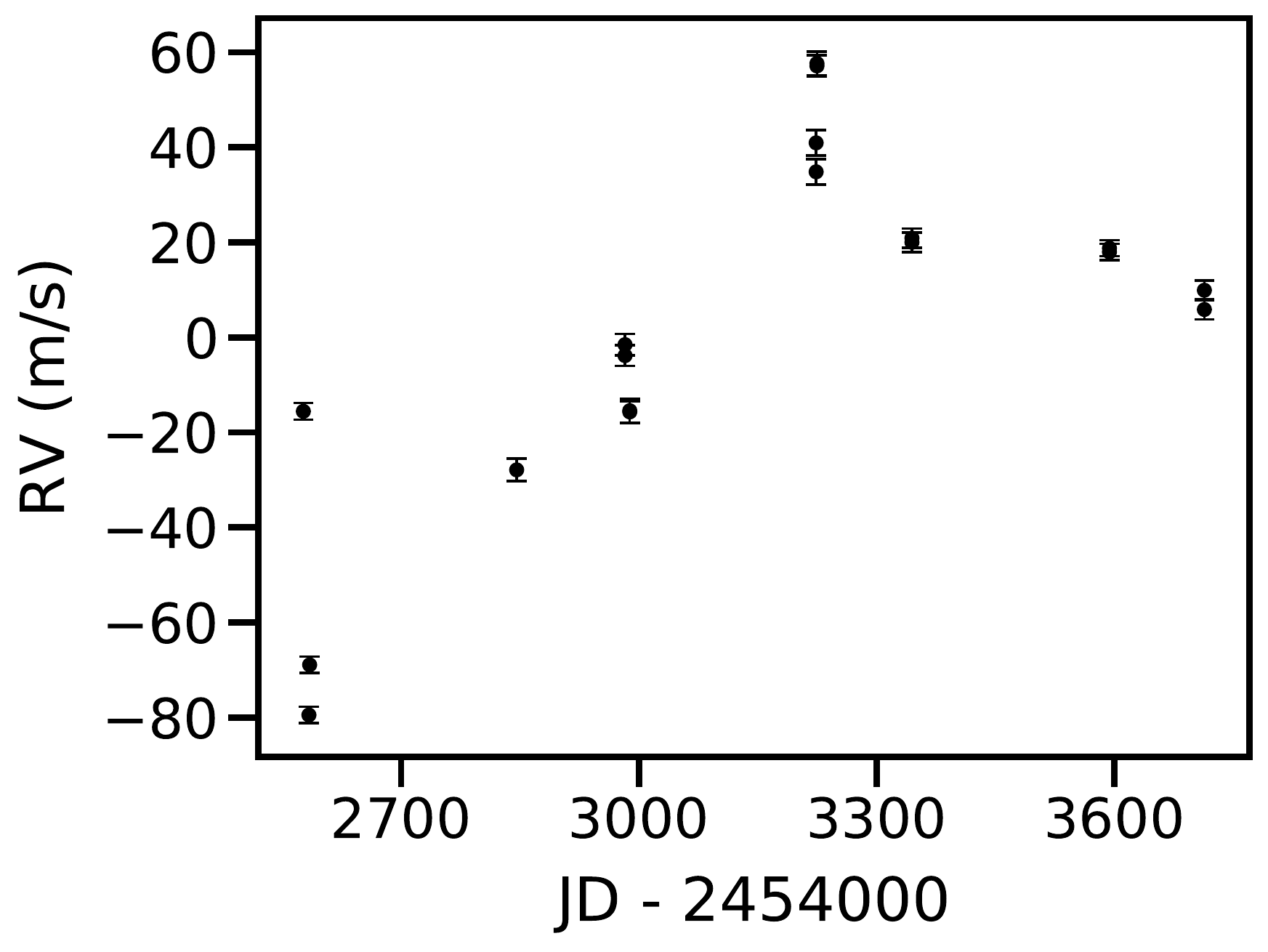}
\caption{\label{RV_218}}
\end{subfigure}
\begin{subfigure}[t]{0.32\textwidth}
\includegraphics[width=1\hsize,valign=m]{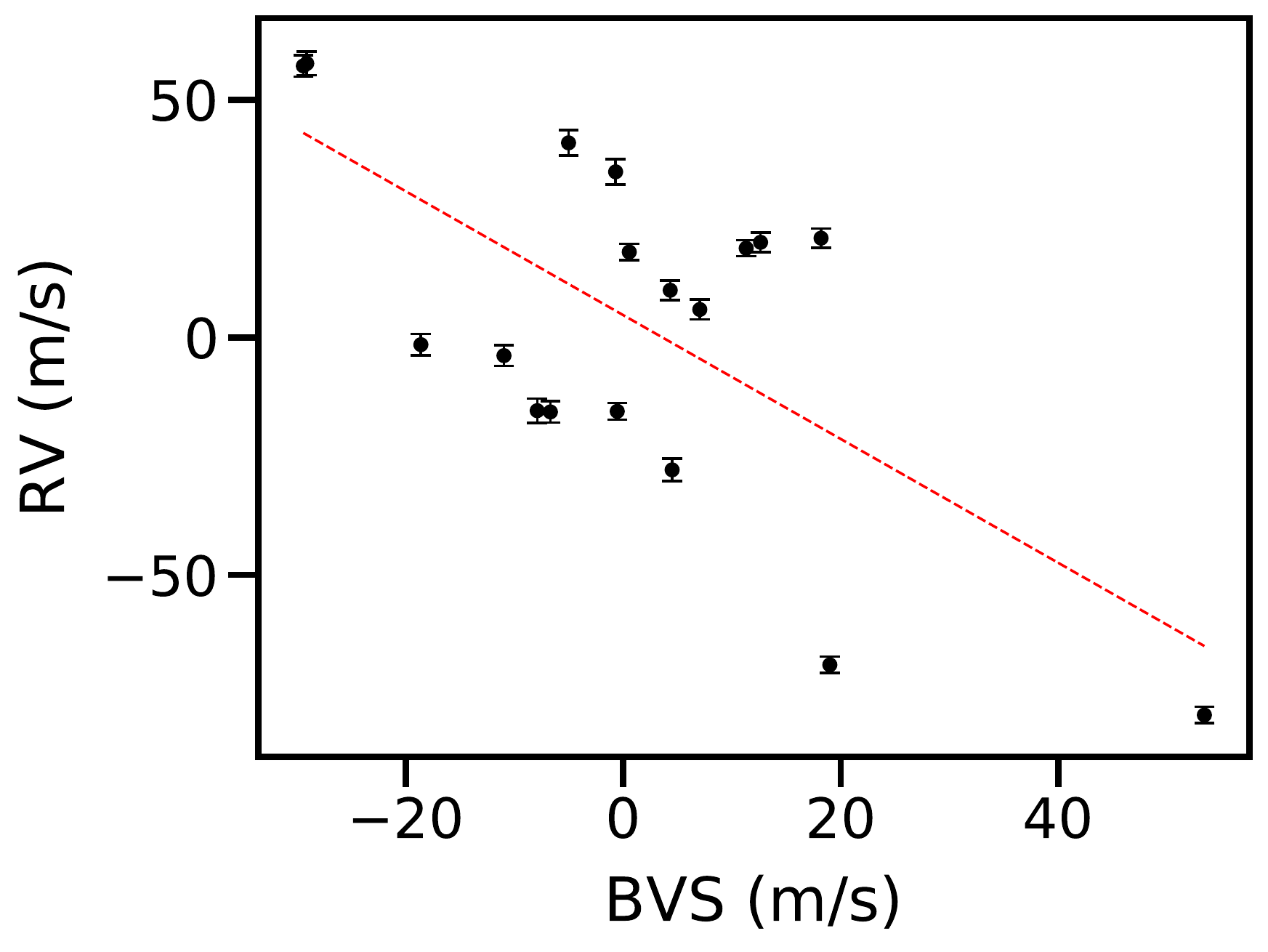}
\caption{\label{RV_BVS_218}}
\end{subfigure}
\begin{subfigure}[t]{0.32\textwidth}
\includegraphics[width=1\hsize,valign=m]{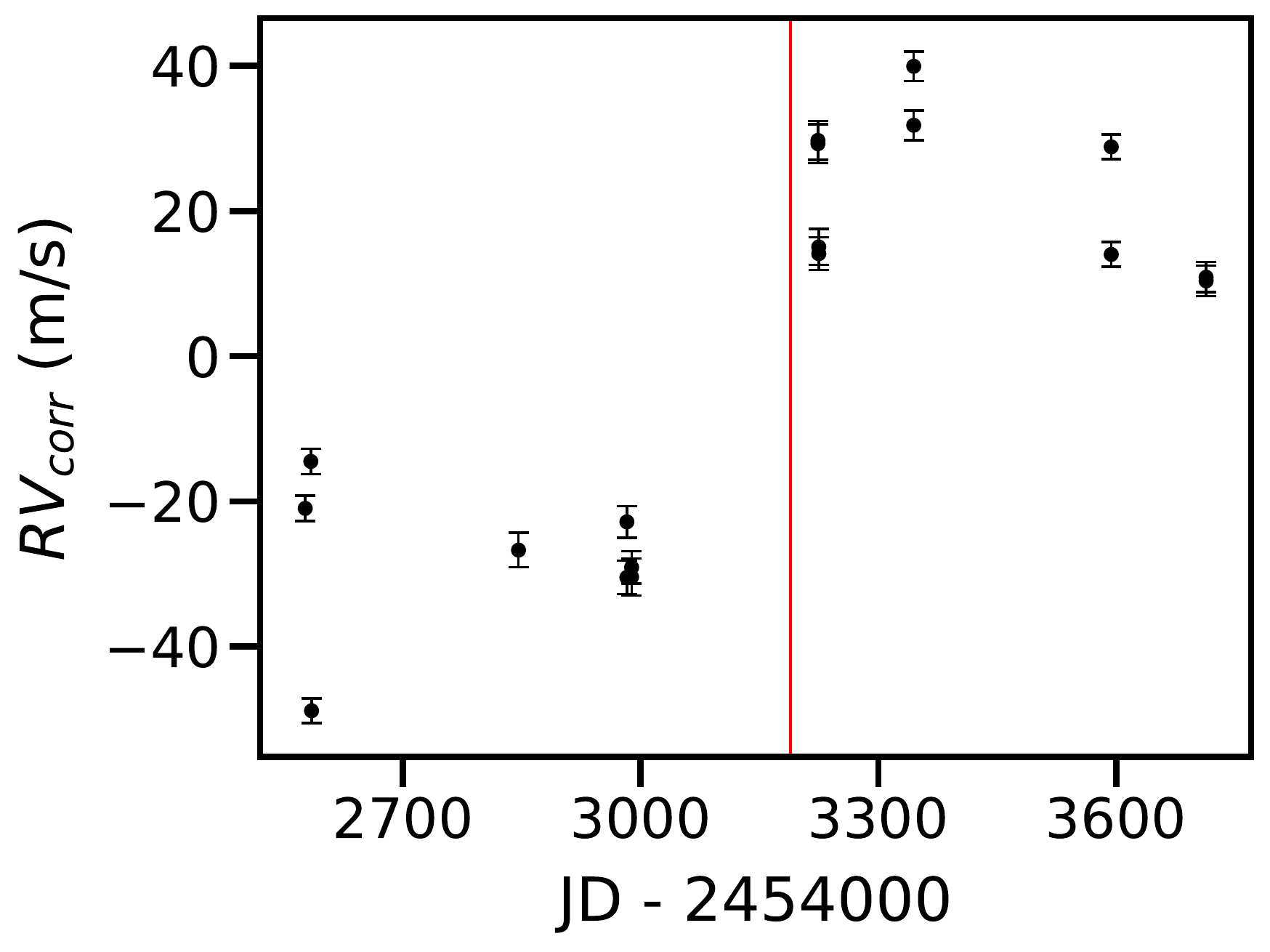}
\caption{\label{RV_corr_218}}
\end{subfigure}

\begin{subfigure}[t]{0.32\textwidth}
\includegraphics[width=1\hsize,valign=m]{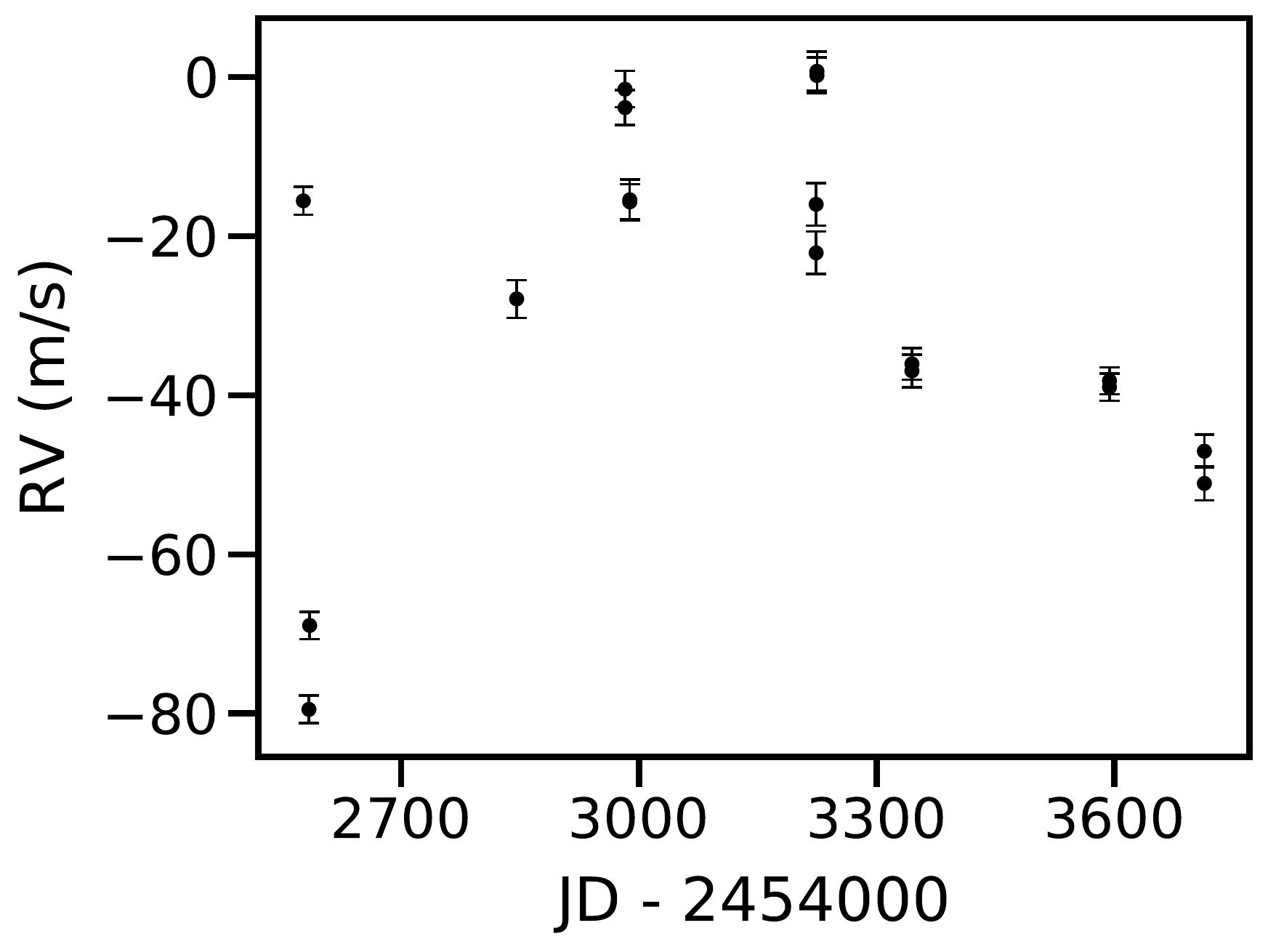}
\caption{\label{RV_corr_gap_218}}
\end{subfigure}
\begin{subfigure}[t]{0.32\textwidth}
\includegraphics[width=1\hsize,valign=m]{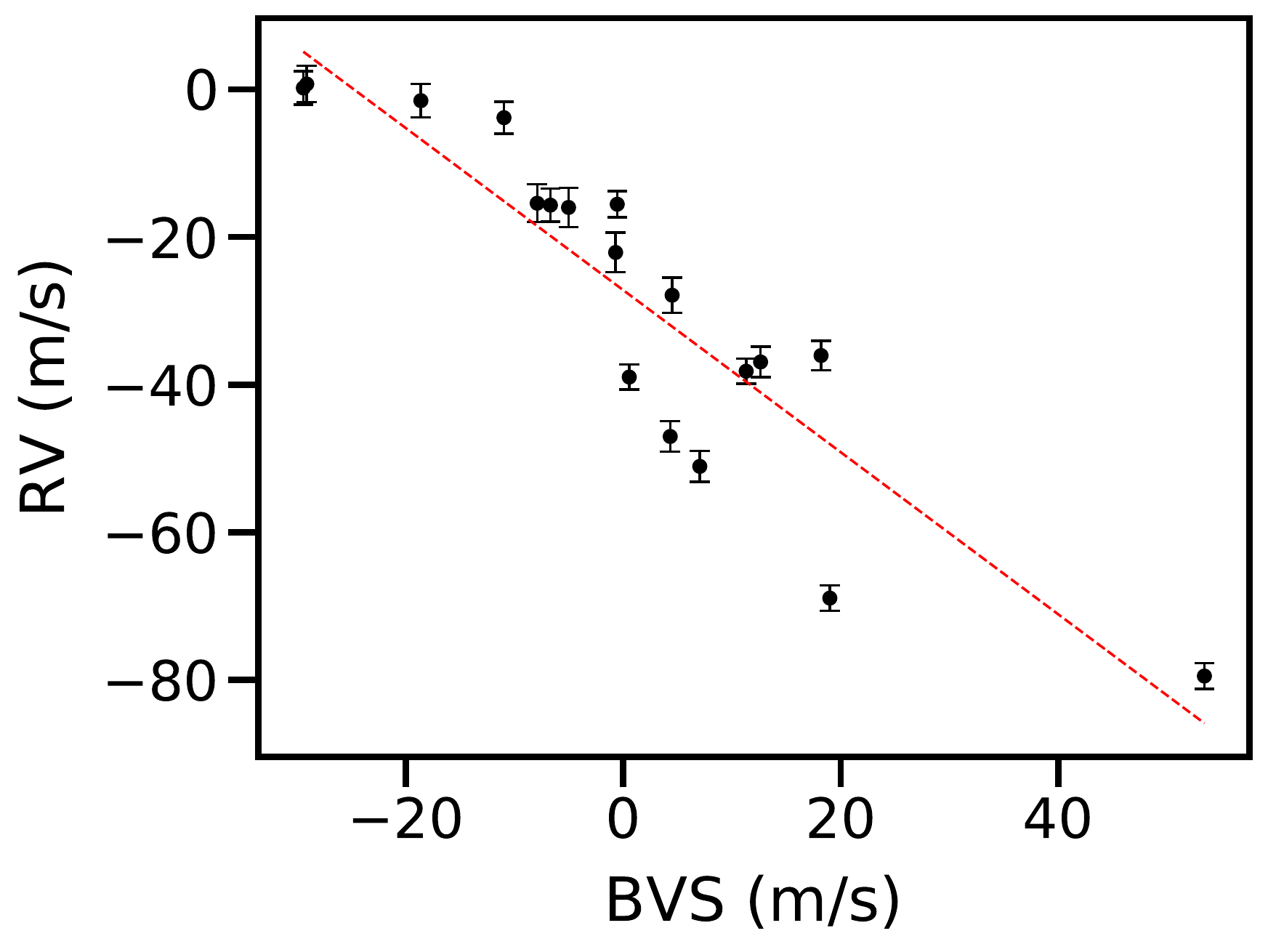}
\caption{\label{RV_BVS_corr_gap_218}}
\end{subfigure}
\begin{subfigure}[t]{0.32\textwidth}
\includegraphics[width=1\hsize,valign=m]{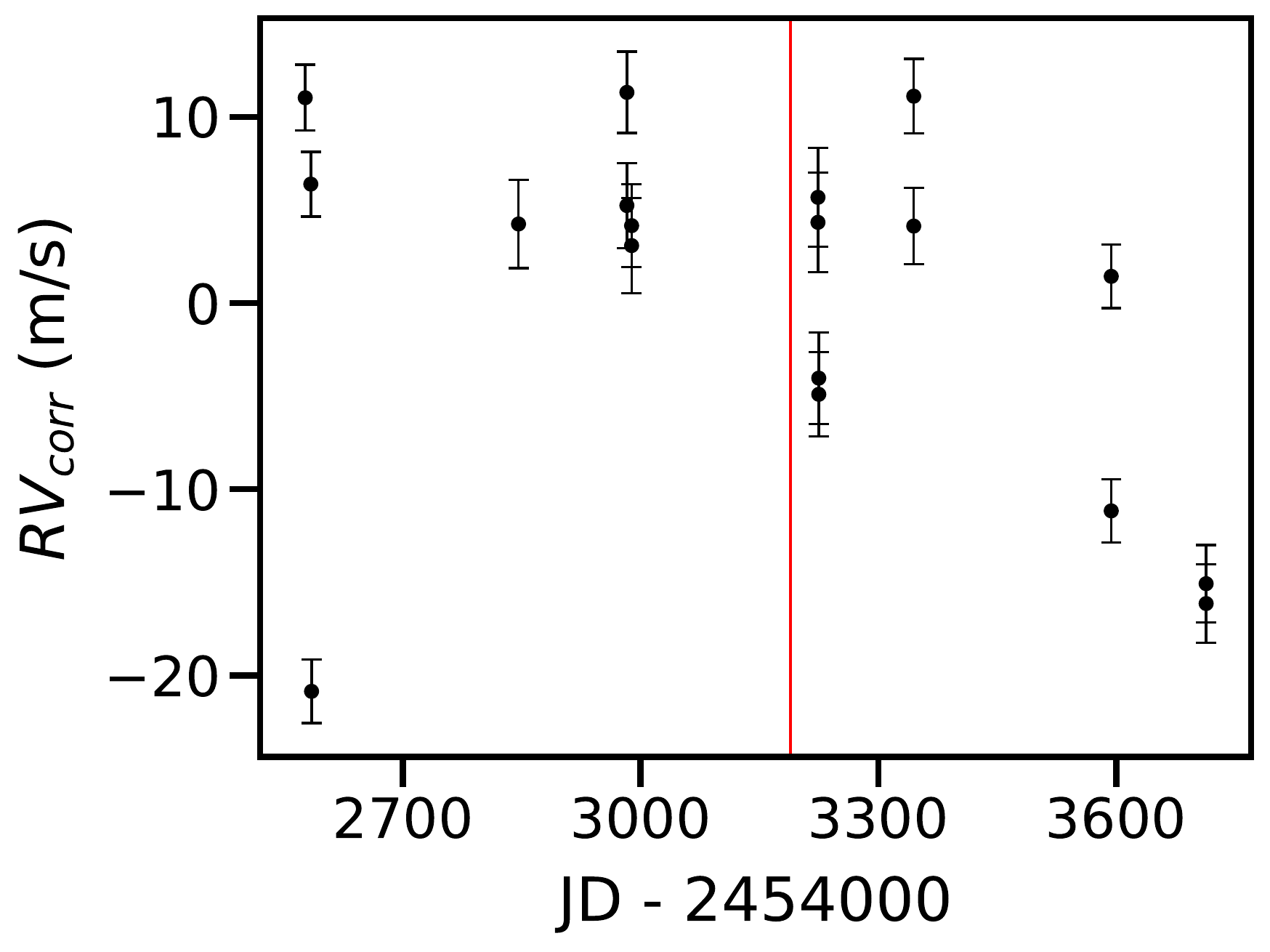}
\caption{\label{RV_corr_gap_corr_218}}
\end{subfigure}
\caption{HD218860 jitter and offset correction. \subref{RV_218})  RV time variations.
  \subref{RV_BVS_218}) RV vs BVS .  The best linear fit is presented in red dashed line.
  \subref{RV_corr_218}) RV corrected from the RV-BVS correlation. \harps \  fiber change is shown  with a vertical red line.    \subref{RV_corr_gap_218})  RV time variations corrected from the offset due to the \harps \ fiber change.
  \subref{RV_BVS_corr_gap_218}) RV corrected from offset. vs BVS .  The best linear fit is presented in red dashed line.
  \subref{RV_corr_gap_corr_218}) RV corrected from offset, corrected from their correlation to the BVS. \harps \  fiber change is shown  with a vertical red line.
}
       \label{218}
\end{figure*}

We present a star for which a significant offset due to the  \harps \ fiber change is present in the corrected RV in \cref{218}. The initial mean RV rms are $\SI{36}{\meter\per\second}$. After correction of the offset and of the RV-BVS correlation, the mean RV rms  are $\SI{10}{\meter\per\second}$.

\end{appendix}

\end{document}